\documentclass[12pt,a4paper]{article}
\usepackage[hmargin=2.5cm,vmargin=3cm]{geometry}
\usepackage[english]{babel}
\usepackage[utf8]{inputenc}
\usepackage[bbgreekl]{mathbbol}
\usepackage{amsmath,amssymb,amsfonts,amsthm}
\usepackage{mathrsfs}

\usepackage{bbold}
\DeclareSymbolFontAlphabet{\mathbb}{AMSb}
\DeclareSymbolFontAlphabet{\mathbbl}{bbold}

\numberwithin{equation}{section}
\usepackage[bottom]{footmisc}
\usepackage{cite}

\usepackage{xcolor}
\usepackage[
  bookmarksnumbered,
  linktocpage=true,
  colorlinks=true,
  citecolor={green!50!black}
]{hyperref}

\usepackage{multirow}
\usepackage{array}
\newcolumntype{C}[1]{>{\centering\let\newline\\\arraybackslash\hspace{0pt}}m{#1}}

\DeclareMathOperator{\sign}{sign}

\newcommand{\ie}{\emph{i.e.}}
\newcommand{\eg}{\emph{e.g.}}

\newcommand{\cf}{\emph{cf.}}

\newcommand{\ZZ}{\mathbb{Z}}
\newcommand{\RR}{\mathbb{R}}

\newcommand{\spindle}{\mathbbl{\Sigma}}
\newcommand{\riemann}{\Sigma_\mathrm{g}}
\newcommand{\hemi}{\mathbbl{S}^4}
\newcommand{\dd}{\mathrm{d}}
\newcommand{\e}{\mathrm{e}}
\newcommand{\ii}{\mathrm{i}}
\newcommand{\AdS}{\mathrm{AdS}}
\newcommand{\vol}[1]{\mathrm{vol}(#1)}

\newcommand{\mc}[1]{\mathcal{#1}}

\newcommand{\A}{y_N}
\newcommand{\B}{y_S}

\newcommand{\LA}{q_1}
\newcommand{\LB}{q_2}
\newcommand{\Li}{q_i}
\newcommand{\LAB}{q_{1,2}}
\newcommand{\Cd}{w}
\newcommand{\x}{x}
\newcommand{\z}{\mathtt{z}}
\newcommand{\p}[1]{p_{#1}}
\newcommand{\Fm}{\hat{F}_0}
\newcommand{\ts}{\mathfrak{n}}
\newcommand{\tr}{\mathfrak{s}}
\newcommand{\di}{\mathfrak{d}}

\newcommand{\Ispindle}{F}
\newcommand{\kk}{b}

\begin{document}

\begin{titlepage}
\vskip 2cm

\begin{center}


\vspace*{2cm}

{\Large \bf D4-branes wrapped on a spindle}

\vskip 1cm
{Federico Faedo$^{\mathrm{a,b}}$ and Dario Martelli$^{\mathrm{a,b,c}}$}

\vskip 0.5cm

${}^{\mathrm{a}}$\textit{Dipartimento di Matematica ``Giuseppe Peano'', Universit\`a di Torino,\\
Via Carlo Alberto 10, 10123 Torino, Italy}

\vskip 0.2cm

${}^{\mathrm{b}}$\textit{INFN, Sezione di Torino \&}   ${}^{\mathrm{c}}$\textit{Arnold--Regge Center,\\
 Via Pietro Giuria 1, 10125 Torino, Italy}

\end{center}

\vskip 2.5cm

\begin{abstract}
\noindent  
We construct supersymmetric $\AdS_4\times\spindle$ solutions of $D=6$ gauged supergravity, where $\spindle$ is a two-dimensional orbifold known as a spindle. These  uplift to solutions of massive type IIA  supergravity using a general prescription, that we describe. We argue that these solutions correspond to the near-horizon limit of a system of $N_f$ D8-branes, together with $N$ D4-branes wrapped on a spindle, embedded as a holomorphic curve inside a Calabi-Yau three-fold. The dual field theories are $d=3$, ${\cal N }= 2$ SCFTs that arise from a twisted compactification of the $d=5$, ${\cal N}=1$ $USp(2N)$ gauge theory. We show that the holographic  free energy associated to these solutions is reproduced by extremizing an off-shell free energy, that we conjecture to arise in the large $N$ limit of the localized partition function of the $d=5$ theories on $S^3\times\spindle$. We formulate a universal proposal for a class of off-shell free energies, whose extremization reproduces all previous results for branes wrapped on spindles, as well as on genus~$\mathrm{g}$ Riemann surfaces $\Sigma_{\mathrm{g}}$. We further illustrate this proposal discussing D4-branes wrapped on $\spindle \times \Sigma_{\mathrm{g}}$, for which we present a supersymmetric $\AdS_2\times\spindle\times \Sigma_{\mathrm{g}}$ solution of $D=6$ gauged supergravity along with the associated entropy function.
\end{abstract}

\end{titlepage}

\tableofcontents

\section{Introduction}

A plethora of examples of AdS/CFT dualities have been constructed following the idea of  \cite{Maldacena:2000mw} of wrapping branes on supersymmetric cycles.
On the field theory side, these constructions realise supersymmetric 
lower-dimensional theories as ``twisted'' compactifications of the 
theories living on the branes. Here the twisting refers to the coupling of the field theory to a background $R$-symmetry gauge field that gets identified with a connection on the tangent bundle of the manifolds on which the theory is compactified, so that supersymmetry 
can be preserved simply taking constant spinors. This is referred to as a topological twist.  On the gravity side, one generically expects to find supersymmetric solutions incorporating the backreaction of the large number of branes wrapped, and when the dual theory is a SCFT, the solutions will comprise an AdS factor. Focussing on compactifications on two-dimensional manifolds, these constructions have been realised for M2, D3, D4 and M5-branes wrapping constant curvature Riemann surfaces, which include the round two-sphere as the genus $\mathrm{g}=0$ case. The references presenting these solutions, along with a discussion of the field theory duals, are summarised in the first row of Table~\ref{minispindles}.

These solutions usually have been constructed in some $U(1)^{\di}$ gauged supergravity in $D=p+2$ dimensions, where $p$ is the world-volume dimension of the brane, and then lifted to $D=10$ or $D=11$ supergravities, which is a necessary step in order to compare gravity computations with calculations performed in the dual field theory. An exception to this is the solution corresponding to D4-branes wrapped on $\Sigma_\mathrm{g}$, that was obtained directly in massive type IIA supergravity~\cite{Bah:2018lyv}.
Below we will show that, in fact, that solution can also be  obtained in a $D=6$, $U(1)^2$  gauged supergravity and uplifted to massive type IIA, provided we take due care of the flux quantization conditions. 
 The case of M2-branes, corresponding to supersymmetric AdS$_2\times \Sigma_\mathrm{g}$ solutions, is particularly interesting, because on general grounds it corresponds to the near-horizon limit of BPS black holes in AdS$_4$.
In this case,  
one can also add rotation to the AdS$_2\times S^2$ solutions.  The black holes are  interpreted as ``flows'' across dimensions, with the AdS$_4$ conformal boundary representing the parent three-dimensional SCFT in the UV and the AdS$_2$ near-horizon region corresponding to the one-dimensional IR theory. 
Such flows have also been constructed for higher-dimensional AdS solutions, although usually they are known only numerically.
 \begin{table}[h]
\centering
\begin{tabular}{|c||c |c| c| c|}
\hline
 & M2 & D3 & D4 & M5 \\
\hline\hline
$\Sigma_\mathrm{g}$ & \cite{Cacciatori:2009iz} & \cite{Benini:2013cda} & \cite{Bah:2018lyv} & \cite{Bah:2012dg} \\[1pt]
\hline
$\spindle $ & \cite{Ferrero:2020twa} & \cite{Ferrero:2020laf} &  here & \cite{Ferrero:2021wvk} \\[1pt]
\hline
\end{tabular}
\caption{In the first row, the references discussing supersymmetric $\AdS\times \Sigma_\mathrm{g}$ solutions for different branes, where $\Sigma_{\mathrm{g}}$ is a Riemann surface of genus $\mathrm{g}$, equipped with a constant curvature metric. 
In the second row, the references discussing the ``simplest''  supersymmetric $\AdS\times \spindle$ solutions, where $\spindle$ is the spindle.}
\label{minispindles}
\end{table}

The solution presented in \cite{Ferrero:2020laf}  opened up a new, unexpected, direction of exploration in the landscape of
AdS/CFT constructions.
This  comprises a supersymmetric  AdS$_3\times \spindle$ background of  minimal $D=5$ gauged supergravity, where $\spindle=\mathbb{WCP}^1_{[n_-,n_+]}$ is a weighted projective space, also known as a spindle. This  uplifts to an AdS$_3\times M_7$ solution of type~IIB supergravity and it has been argued to be dual to a class of $d=4$, ${\cal N}=1$ SCFTs compactified 
on the spindle with a novel type of twist, \emph{different} from the topological twist, that was later dubbed ``anti-twist''.  A similar construction, for 
AdS$_2\times \spindle$ solutions of  minimal $D=4$ gauged supergravity, was presented in~\cite{Ferrero:2020twa}. These have been later extended to spindle solutions of STU gauged supergravities in $D=5$ \cite{Hosseini:2021fge,Boido:2021szx} and $D=4$ \cite{Ferrero:2021ovq,Couzens:2021rlk}, respectively. 
A supersymmetric AdS$_5\times \spindle $ solution corresponding to M5-branes wrapped on the spindle was constructed in~\cite{Ferrero:2021wvk} and, differently from the previous constructions, it realises supersymmetry by means of a ``topologically topological twist''. Namely, the background $R$-symmetry gauge field is identified with a connection on the tangent bundle of the spindle, as for the topological twist, but the corresponding local curvatures are not equal.
It turns out that these local solutions comprising spindles contain, as interesting degenerate limits, solutions corresponding to branes wrapped on disks or Riemann surfaces with non-constant curvature \cite{Bah:2021hei,Couzens:2021tnv,Suh:2021ifj,Suh:2021aik,Suh:2021hef,Couzens:2021rlk}.

In this paper we will construct an AdS$_4\times \spindle$ solution, corresponding to  
 D4-branes wrapped on the spindle, thus filling the  outstanding entry in Table \ref{minispindles}. 
 We will show that our construction realises the topologically topological twist, as for the M5-brane solution in \cite{Ferrero:2021wvk}, with which it shares some similarities. 
 We will first present the solution in a $D=6$ gauged supergravity model and then we will discuss how to uplift this to a globally consistent solution in massive type IIA supergravity.
 We will elucidate   the global structure of the Killing spinors,  identifying the precise bundles of which they are sections and showing how they differ from the Killing spinors of the previous constructions for M2 \cite{Ferrero:2020twa} 
 and D3-branes  \cite{Ferrero:2020laf}.  Our solution completes the panorama of the ``basic'' branes wrapped on spindles.

While for the SCFTs compactified on Riemann surfaces with the standard topological twist various supersymmetric partition functions have been computed and studied in the large $N$ limit, for compactifications on spindles similar results are not yet available. For theories in $d=4$ and $d=6$ this lack of knowledge can be  bypassed employing the recipe of~\cite{Ferrero:2020laf} for extracting the trial central charge of the $(d-2)$-dimensional theories from the  anomaly polynomials of the parent theories. 
In $d=3$ an entropy function was obtained in \cite{Cassani:2021dwa}, 
 from the on-shell gravitational action of the suitably regularised black hole solutions, employing the method of \cite{Cabo-Bizet:2018ehj}.
 Extremizing this  reproduces the entropy associated to the AdS$_2\times \spindle$ solution of \cite{Ferrero:2020twa}. 
An extension of this entropy function was conjectured in  \cite{Ferrero:2021ovq} and shown to reproduce correctly the entropy  of 
 multi-charge spindle solutions. Taking inspiration from that, in this paper we will propose a conjectural off-shell free energy, whose extremization will, remarkably, reproduce
 the gravitational free energy associated to our solutions.

In the last part of the paper we will propose a universal class of off-shell free energies for various branes wrapped on spindles, analogous to the entropy functions, to which these reduce in $d=3$. Specifically, we conjecture that for a large class of SCFTs in dimensions $d=3,4,5,6$, possessing large $N$ gravity duals, when these are compactified on a spindle~$\spindle$, the exact superconformal $R$-symmetry of the SCFTs in dimension $d-2$ is determined extremizing the following off-shell free energies 
\begin{equation}
\boxed{ \Ispindle^\pm(\varphi_i,\epsilon;\ts_i,n_+,n_-,\sigma) = \frac{1}{\epsilon} \bigl( {\cal F}_d (\varphi_i + \ts_i \epsilon) \pm {\cal F}_d (\varphi_i - \ts_i \epsilon) \bigr)\, ,  }
\label{conjectureintrod}
\end{equation}
where the variables $\varphi_i,\epsilon$ and the magnetic fluxes $\ts_i$ satisfy the constraints 
\begin{equation}
\boxed{ \sum_{i=1}^\di  \varphi_i -   \frac{n_+ - \sigma n_-}{n_+n_-}\epsilon =2\, ,  \qquad \quad   \sum_{i=1}^\di  \ts_i= \frac{n_++\sigma n_-}{n_+n_-} \, .}
\end{equation}

The form \eqref{conjectureintrod} is suggested by the idea of gluing universal contributions called gravitational blocks, advocated in \cite{Hosseini:2019iad} for the entropy functions of SCFTs compactified on  different manifolds. The ``building blocks'' are the functions ${\cal F}_d$ above, which have different interpretations in the different dimensions $d$, being proportional to either the central charge or the sphere partition function of the SCFTs.
They are also related to the prepotentials of the various gauged supergravities in dimension $D=d+1$. Their precise form will be given later, see Table~\ref{tab:F-block}.
The sign $\sigma =\pm 1$ labels the different  twists that may occur on spindles. The sign $\sigma=+1$ corresponds to the topologically topological twist, which includes the standard topological twist as a special case, while the sign $\sigma=-1$ corresponds to the anti-twist,  realised by M2 \cite{Ferrero:2020twa} and D3-branes \cite{Ferrero:2020laf}. The sign $\pm$ depends on the gluing, in the language of \cite{Hosseini:2019iad}, and we shall comment below on its relation to the sign of~$\sigma$.
For example, in $d=3$, taking  $n_+=n_-=1$ in the above formulas leads to  the  entropy functions for the supersymmetric  black holes with  AdS$_2\times S^2$ near-horizon geometry  \cite{Hosseini:2019iad}.
In this case, for $\sigma=+1$ we must take $F^-$ and this reduces to the entropy function  \cite{Benini:2015eyy} of the supersymmetric AdS$_4$ black holes with a topological twist~\cite{Cacciatori:2009iz}. On the contrary, for $\sigma=-1$ we must take $F^+$ and this reduces to the entropy function \cite{Nian:2019pxj} for the supersymmetric rotating Kerr-Newmann AdS$_4$ black holes~\cite{Hristov:2019mqp}.

More generally, we will provide evidence that in $D=4,6$ the gluing sign $\pm$  coincides with $-\sigma$, while in $D=5,7$ they appear to be independent. 
 In $D=4$,  the fact that $-\sigma$ coincides with the sign $\pm$ may be understood as follows.  The AdS/CFT correspondence 
 implies that, in the large $N$ limit, the free energies $F^\pm$
should be identified with  the appropriately regularised gravitational on-shell action of the dual supergravity solutions. 
In \cite{BenettiGenolini:2019jdz} it has been proved, in the context of minimal gauged supergravity, that the on-shell action of any (Euclidean) supersymmetric solution 
takes the form of a sum over contributions from fixed points 
of the canonical Killing vector field, defined as a bilinear in the Killing spinors of the solution. 
The relative sign of these contributions is determined by the chirality of the Killing spinors at the fixed points, and in all the known 
supergravity solutions comprising spindle (including $S^2$ as a special case) we have that the chiralities at the north and south poles of the spindles are the same for the topologically topological twist and opposite for the anti-twist. A general proof  of this fact is given in  \cite{Ferrero:2021etw}.

Our proposal reproduces all the previously known results for $\AdS\times\spindle$ solutions, including the $\AdS\times S^2$ solutions as special cases\footnote{\label{recoversigma}As we shall discuss, for $\sigma=+1$,  by formally replacing $\frac{n_+ - n_-}{n_+n_-}\mapsto 0$ and $\frac{n_++ n_-}{n_+n_-} \mapsto \chi (\Sigma_{\mathrm{g}})$,
 our formulas cover also the AdS$\times \Sigma_{\mathrm{g}}$ solutions.}. 
Moreover, in  $d=5$, taking $\sigma=+1$, corresponding to the topologically topological twist, 
we will show that the extremization of the  function  $ \Ispindle^- (\varphi_i,\epsilon;\ts_i,n_+,n_-,+1)$ in \eqref{conjectureintrod} precisely reproduces the
gravitational $S^3$ free energy of our solution (see eq.~\eqref{spindle_free-energy}).  
We then conjecture that this should arise in the large $N$ limit of the localized partition function on $S^3\times \spindle$, with the topologically topological twist.
To add weight to our proposal, we will also  discuss  D4-branes wrapped on the four-dimensional orbifold $\spindle \times \Sigma_{\mathrm{g}}$. The effective field theory obtained from the twisted compactification of the $d=5$  SCFT is expected to be superconformal, at least in some ranges of the magnetic fluxes. We will discuss  the corresponding  supersymmetric $\AdS_2\times\spindle\times \Sigma_{\mathrm{g}}$ solutions of $D=6$ gauged supergravity and show that the entropy function constructed from the ``spindly'' gravitational blocks correctly reproduces the geometric entropy.

The rest of the paper is organised as follows.  In section~\ref{sec:6d-to-mIIA} we discuss the uplift of solutions of  a $D=6$, $U(1)^2$ gauged supergravity model to massive type IIA.
As a warm-up, we illustrate this obtaining the (global)  solutions of \cite{Brandhuber:1999np} and~\cite{Bah:2018lyv} from known solutions in $D=6$. 
In section~\ref{newsolutions_section} we construct new supersymmetric AdS$_4\times \spindle$ solutions and discuss global properties of these both in $D=6$ and $D=10$. 
In section~\ref{ftsection} we discuss aspects of the field theory duals of these solutions. In particular, we conjecture a field-theoretic large~$N$ off-shell free energy and show that extremizing this reproduces the holographic free energy 
associated to our solutions. In section~\ref{blockssection} we discuss how our proposal fits in a general scheme of off-shell free energies for field theories compactified on spindles (as well as on genus~$\mathrm{g}$ Riemann surfaces), comprising and extending entropy functions and trial central charges, previously discussed in the literature. 
In section~\ref{ads2solution} we begin investigating D4-branes wrapped on four-dimensional orbifolds, focussing on a class of supersymmetric  
AdS$_2\times \spindle\times \Sigma_\mathrm{g}$ solutions. We conclude with a discussion in section~\ref{discusssection}. Appendix~\ref{app:all} contains technical details useful for comparing known solutions, in different conventions. In 
appendix~\ref{app:osp} we demonstrate how the Killing spinors of the AdS$_4\times \spindle$ solutions encapsulate the  $OSp(2|4)$ superalgebra of the dual SCFTs.

\section{Uplift of $D=6$ solutions to massive type IIA}
\label{sec:6d-to-mIIA}

In this paper we discuss solutions of a  $D=6$ gauged supergravity with  gauge group  $U(1)^2$, comprising two gauge fields $A_1,A_2$, a two-form $B$ and two real scalar fields $\vec{\varphi}=(\varphi_1,\varphi_2)$. These can be uplifted locally to solutions of massive type IIA supergravity by means of the consistent truncation formulas presented in~\cite{Cvetic:1999xx}. However, we will see that globally the solutions uplifted through this ansatz are incompatible with quantization of the fluxes and need to be supplemented by an additional parameter that arises in $D=10$.

\subsection{The $D=6$ gauged supergravity}
\label{D6sugra}

The $D=6$ supergravity model of interest can also be obtained as a sub-sector of an extension of Romans $F(4)$ gauged supergravity~\cite{Romans:1985tw}, coupled to three vector multiplets~\cite{DAuria:2000afl}.  The bosonic part of the action reads\footnote{Here and in what follows we define, for any $p$-form $\omega$, $|\omega|^2 = \frac{1}{p!} \, \omega_{\mu_1\ldots\mu_p} \omega^{\mu_1\ldots\mu_p}$.}
\begin{equation} \label{6d-action} 
S_\text{6D} = \frac{1}{16\pi G_{(6)}} \int \dd^6x \, \sqrt{-g} \biggl( R - V - \frac12 |\dd\vec{\varphi}|^2 - \frac12 \sum_{i=1}^2 X_i^{-2} |F_i|^2 \biggr) \,,
\end{equation}
where $F_i=\dd A_i$ and the scalar fields $\vec{\varphi}$ are parameterised as
\begin{equation} \label{6d-scalar-para}
X_i = \e^{-\frac12 \vec{a}_i\cdot\vec{\varphi}}  \qquad  \text{with}  \qquad  \vec{a}_1 = \bigl(2^{1/2}, 2^{-1/2}\bigr) \,,  \qquad  \vec{a}_2 = \bigl(-2^{1/2}, 2^{-1/2}\bigr) \,.
\end{equation}
The scalar potential is
\begin{equation} \label{6d-potential}
V = m^2 X_0^2 - 4g^2 X_1 X_2 - 4mg \, X_0 (X_1 + X_2) \,,
\end{equation}
with $g$ the gauge coupling and $m$ the mass parameter, and where for later convenience we defined $X_0=(X_1 X_2)^{-3/2}$. Here we consistently set $B=0$ because in the first part of the paper we will restrict to configurations with $F_1\wedge F_2=0$. We will restore the two-form~$B$ in section~\ref{secads2}. It is worth mentioning that locally the ratio $m/g$ can be set to any non-zero value rescaling the scalar fields~$X_i$ and the field strengths~$F_i$. In particular, $m$ can be absorbed in the coupling constant transforming
\begin{equation}
X_i \mapsto \left(\frac{m}{l g}\right)^{1/4} X_i \,,  \qquad  \quad F_i \mapsto \left(\frac{m}{l g}\right)^{1/4} F_i \,,
\end{equation}
and defining the new gauge coupling as
$\tilde{g} = \big(\tfrac{m g^3}{l}\big)^{1/4}$,
with $l$ a positive constant. The action~\eqref{6d-action} keeps the same form,     with the scalar potential becoming
\begin{equation}
V = \tilde{g}^2 \bigl[ l^2 X_0^2 - 4 X_1 X_2 - 4l \, X_0 (X_1 + X_2)  \bigr]\,.
\end{equation}
However, for the time being we will keep both parameters $m$ and $g$. 

 A solution to the equations of motion of the model  is supersymmetric if and only if it satisfies also the following set of  Killing spinor equations~\cite{DAuria:2000afl}:
\begin{align}
\begin{split} \label{KSE_grav}
\mathcal{D}_\mu \epsilon^A + \frac18 \bigl[ g (X_1 + X_2) + m X_0 \bigr] \Gamma_\mu \epsilon^A & \\
+ \frac{\ii}{32} \bigl( X_1^{-1} F_1 + X_2^{-1} F_2 \bigr)_{\nu\lambda} \bigl( \Gamma_\mu^{\,\ \nu\lambda} - 6 \delta^\nu_\mu \, \Gamma^\lambda \bigr) (\sigma^3)^A_{\,\ B} \epsilon^B &= 0 \,,
\end{split} \\
\begin{split} \label{KSE_dila}
\frac14 \bigl( X_1^{-1} \partial_\mu X_1 + X_2^{-1} \partial_\mu X_2 \bigr) \Gamma^\mu \epsilon^A - \frac18 \bigl[ g (X_1 + X_2) - 3m X_0 \bigr] \epsilon^A & \\
- \frac{\ii}{32} \bigl( X_1^{-1} F_1 + X_2^{-1} F_2 \bigr)_{\mu\nu} \Gamma^{\mu\nu} (\sigma^3)^A_{\,\ B} \epsilon^B &= 0 \,,
\end{split} \\
\begin{split} \label{KSE_gauge}
\frac12 \bigl( X_1^{-1} \partial_\mu X_1 - X_2^{-1} \partial_\mu X_2 \bigr) \Gamma^\mu (\sigma^3)_{AB} \epsilon^B - g (X_1 - X_2) (\sigma^3)_{AB} \epsilon^B & \\
- \frac{\ii}{4} \bigl( X_1^{-1} F_1 - X_2^{-1} F_2 \bigr)_{\mu\nu} \Gamma^{\mu\nu} \epsilon_A &= 0 \,,
\end{split}
\end{align}
where
\begin{equation} \label{cov-D}
\mathcal{D}_\mu \epsilon^A \equiv  \partial_\mu \epsilon^A + \frac14 \, \omega_\mu^{\ ab} \Gamma_{ab} \epsilon^A - \frac{\ii}{2} \, g (A_1 + A_2)_\mu (\sigma^3)^A_{\,\ B} \epsilon^B \,.
\end{equation}
These follow from setting to zero the supersymmetry variations of the fermionic fields of the theory with three vector multiplets \cite{DAuria:2000afl},  that do not vanish automatically in the sub-truncation that we are considering. Here $(\sigma^3)^A_{\,\ B}$ is the usual third Pauli matrix and $\{\Gamma_a,\Gamma_b\}=2\eta_{ab}$. The $SU(2)$ indices $A,B$ are raised and lowered as $\epsilon^A=\varepsilon^{AB}\epsilon_B$ and $\epsilon_A=\epsilon^B\varepsilon_{BA}$, where $\varepsilon_{AB}=-\varepsilon_{BA}$ and its inverse matrix $\varepsilon^{AB}$ is defined such that $\varepsilon^{AB}\varepsilon_{AC}=\delta^B_C$. 
The supersymmetry parameter $\epsilon^A$ is an eight-component symplectic-Majorana spinor, hence it satisfies the condition
\begin{equation} \label{symp-Maj}
\varepsilon^{AB} \epsilon_B^* = \mc{B}_6 \epsilon_A \,,
\end{equation}
where $\mc{B}_6$ is related to the six-dimensional charge conjugation matrix $\mc{C}_6$ by $\mc{B}_6=\ii\,\mc{C}_6\Gamma^0$.

\subsection{Improved uplift to massive type IIA}
\label{sub:improved-uplift}

Any solution to the equations of motion of this theory can be embedded in massive type~IIA supergravity by means of the dimensional reduction ansatz presented in~\cite{Cvetic:1999xx}\footnote{The original ansatz was written for a four-scalar system, but the model with two scalars can be easily obtained setting to zero two of the four scalars in~\cite{Cvetic:1999xx}. The consistency of this sub-truncation was conjectured in \cite{Hosseini:2018usu}.}, provided the gauge coupling and mass parameter are related as $m=2g/3$. The metric in the string frame and the dilaton are given by
\begin{align}
\begin{split} \label{truncation_metric}
\dd s_\text{s.f.}^2 &= \mu_0^{-1/3} (X_1 X_2)^{-1/4} \bigl\{ \Delta^{1/2} \dd s_6^2 \\
 & + g^{-2} \Delta^{-1/2} \bigl[ X_0^{-1} \dd\mu_0^2 + X_1^{-1} \bigl(\dd\mu_1^2 + \mu_1^2 \sigma_1^2\bigr) 
 + X_2^{-1} \bigl(\dd\mu_2^2 + \mu_2^2 \sigma_2^2\bigr) \bigr] \bigr\} \,,
\end{split} \\
\label{truncation_dilaton}
\e^\Phi &= \mu_0^{-5/6} \Delta^{1/4} (X_1 X_2)^{-5/8} \,,
\end{align}
where $\dd s_6^2$ is the six-dimensional metric and we defined the one-forms $\sigma_i\equiv \dd\phi_i-gA_i$. The angular coordinates $\phi_1$, $\phi_2$ have canonical $2\pi$ periodicities, and the warp factor is
\begin{equation}
\Delta = \sum_{a=0}^2 X_a \mu_a^2 \,.
\end{equation}
The coordinates $\mu_a$, with $a=0,1,2$, satisfy the constraint $\sum\mu_a^2=1$, which can be solved for example defining 
\begin{equation}
\mu_0 = \sin\xi \,,  \qquad  \mu_1 = \cos\xi \sin\eta \,,  \qquad  \mu_2 = \cos\xi \cos\eta \,,
\end{equation}
and taking $\eta\in[0,\pi/2]$, $\xi\in(0,\pi/2]$, where the range of $\xi$ arises from the necessity of having $\mu_0>0$.  
At any point in the six-dimensional space-time, the metric inside the square brackets in~\eqref{truncation_metric} parameterises a four-dimensional hemisphere, that we will denote by $\hemi$.
This metric is in general squashed and it reduces to the metric on ``half the round four-sphere'' when $X_1=X_2=1$. The only non-vanishing fields of the RR sector are the ten-dimensional Romans mass
\begin{equation}
F_{(0)} = \frac{2g}{3}
\label{gromans}
\end{equation}
and the four-form flux. This is conveniently written in terms of its Hodge dual  as 
\begin{equation}
\star_{10} F_{(4)} = g U \, \vol{M_6} - \frac{1}{g^2} \sum_i X_i^{-2} \mu_i (\star_6 F_i) \wedge \dd\mu_i \wedge \sigma_i + \frac{1}{g} \sum_a X_a^{-1} \mu_a (\star_6 \dd X_a) \wedge \dd\mu_a \,,
\label{sixflux}
\end{equation}
where $\vol{M_6}$ is the volume form of the six-dimensional space and
\begin{equation}
U = 2 \sum_{a=0}^2 X_a^2 \mu_a^2- \left[\frac{4}{3}X_0 + 2(X_1+X_2) \right] \Delta  \,.
\end{equation}
The Hodge star operator $\star_{10}$ in~\eqref{sixflux} is computed using the string frame metric~\eqref{truncation_metric}, while $\star_{6}$ is defined using the six-dimensional metric~$\dd s_6^2$. 

Provided the equations of motion of the six-dimensional supergravity hold, the above field configuration solves the equations of motion of massive type IIA supergravity, whose action in the string frame reads\footnote{Here $B_{(2)}^n$ denotes the wedge product of $B_{(2)}$ with itself $n$ times, divided by $n!$.}
\begin{align}
S_\text{mIIA}  = \frac{1}{16\pi G_{(10)}} \bigg\{ \! &\int \! \dd^{10}x \, \sqrt{-g} \Bigl[ \e^{-2\Phi} \bigl( R + 4 |\dd\Phi|^2 - \tfrac12 |H_{(3)}|^2 \bigr) - \tfrac12 \bigl( F_{(0)}^2 + |F_{(2)}|^2 + |F_{(4)}|^2 \bigr) \Bigr] \nonumber\\
- \tfrac12  &\int \bigl( B_{(2)} \wedge \dd C_{(3)} \wedge \dd C_{(3)} + 2 F_{(0)} B_{(2)}^3 \wedge \dd C_{(3)} + 6 F_{(0)}^2 B_{(2)}^5 \bigr) \bigg\} \,,
\label{mIIA-action}
\end{align}
where the field strengths are defined from the NS two-form $B_{(2)}$ and the RR potentials $C_{(1)}$ and $C_{(3)}$ as
\begin{equation}
H_{(3)} = \dd B_{(2)} \,,  \quad  F_{(2)} = \dd C_{(1)} + F_{(0)} B_{(2)} \,,  \quad  F_{(4)} = \dd C_{(3)} - H_{(3)} \wedge C_{(1)} + \frac12 F_{(0)} B_{(2)} \wedge B_{(2)} \,.
\end{equation}

It will be important to notice that the equations of motion of massive type IIA are invariant if the fields are transformed as  
\begin{equation} \label{scaling-symm}
\begin{split}
\dd\hat{s}_\text{s.f.}^2 &= \lambda^2 \dd s_\text{s.f.}^2 \,,  \qquad \qquad \! \e^{\hat{\Phi}} = \lambda^2 \e^\Phi \,,  \qquad  \qquad \hat{B}_{(2)} = \lambda^2 B_{(2)}    \, ,\\
 \hat{F}_{(0)} &= \lambda^{-3} F_{(0)} \, , \qquad   \hat{C}_{(n-1)} = \lambda^{n-3} C_{(n-1)} \, , 
 \end{split}
\end{equation}
with $n=2,4$, where $\lambda$ is any strictly positive constant. However, this scaling  symmetry holds only at the classical level in supergravity and it is broken upon imposing the  Dirac quantization conditions on the fluxes. 
As we will discuss momentarily, this additional parameter  will be crucial for  ensuring that six-dimensional solutions yield globally 
regular solutions in 
$D=10$, in particular that the fluxes are correctly quantized.

Notice that  the reduction ansatz of \cite{Cvetic:1999xx} applies only after setting $m=2g/3$ in the six-dimensional theory and  it implies that the Romans mass of the ten-dimensional theory is 
fixed in terms of the gauge coupling constant $g$ as in \eqref{gromans}. It is natural to suspect that there may exist a more general truncation ansatz that relates the six-dimensional mass parameter
 $m$ to the ten-dimensional parameter $\lambda$, so that the Romans mass $F_{(0)}$ is an independent parameter.   This would be a mechanism analogous 
   to the ten-dimensional origin of dyonic four-dimensional supergravity 
   discussed in \cite{Guarino:2015jca}. We leave this interesting question for the future and proceed to discuss different globally regular 
   solutions of massive type IIA originating in $D=6$.   
   
   In summary, our strategy will be as follows. We construct the solutions in $D=6$ and after setting $m=2g/3$ we  uplift these to local solutions in $D=10$, using the formulas in \cite{Cvetic:1999xx}. Then we introduce the parameter $\lambda$ and proceed to quantize the fluxes, finding globally consistent solutions of massive type IIA supergravity.

\subsection{The AdS$_6$ solution and its uplift}
\label{BOsolution}

The equations of motion following from the action~\eqref{6d-action} admit the well-known  supersymmetric vacuum with constant scalars $X_1 = X_2 = \big(\tfrac{3m}{2g}\big)^{1/4}$, vanishing gauge fields $A_1 = A_2 = 0 $ and metric
\begin{align}
\dd s_6^2 &= \frac{9}{2 \left(6mg^3\right)^{1/2} }\, \dd s_{\AdS_6}^2 \,, 
\end{align}
where $\dd s_{\AdS_6}^2$ is the metric on $\AdS_6$ with unit radius.
We now set $m = 2g/3$ and  uplift this solution to massive type IIA using the formulas in~\cite{Cvetic:1999xx}. After introducing the parameter~$\lambda$ using the local 
scaling symmetry \eqref{scaling-symm}, we obtain 
\begin{equation}
\begin{split}
\label{10d_vacuum-metric}
\dd s_\text{s.f.}^2 &= \lambda^2 (\sin\xi)^{-1/3} L_{\AdS_6}^2 \biggl[ \dd s_{\AdS_6}^2 + \frac49 \bigl( \dd\xi^2 + \cos^2\!\xi \, \dd s_{S^3}^2 \bigr) \biggr] \,, \\
\e^\Phi &= \lambda^2 (\sin\xi)^{-5/6} \,, \\
F_{(0)} &= \frac{1}{\lambda^3L_{\AdS_6}} \,,  \qquad  F_{(4)} = -\lambda \frac{10 \cos^3\!\xi \sin^{1/3}\!\xi}{3g^3} \, \dd\xi \wedge \vol{S^3} \, ,
\end{split}
\end{equation}
where 
$\dd s_{S^3}^2$ denotes the metric on a unit radius round three-sphere
\begin{equation}
\dd s_{S^3}^2 = \dd\eta^2 + \sin^2\!\eta \, \dd\phi_1^2 + \cos^2\!\eta \, \dd\phi_2^2 \,,
\end{equation}
and $\vol{S^3}$ its associated volume form.  
The quantization conditions of the (non-zero) fluxes in massive type IIA read
\begin{equation} \label{genfluxquant}
(2\pi\ell_s) F_{(0)} = n_0 \in \ZZ  \qquad  \text{and}  \qquad  \frac{1}{(2\pi\ell_s)^3} \int_{\hemi} F_{(4)} = N \in \ZZ \,,
\end{equation}
where $\ell_s$ is the string length. In the  solution \eqref{10d_vacuum-metric} these imply that\footnote{For consistency, we need to identify the flux of $F_{(4)}$ with $-N$, where $N$ is the number of D4-branes.}
\begin{equation}
g^8 = \frac{1}{(2\pi\ell_s)^8} \frac{18\pi^6}{N^3 n_0} \,,  \qquad  \quad \lambda^8 = \frac{8\pi^2}{9N n_0^3} \, .
\label{fluxquantwithlambda}
\end{equation}

 It is clear from the second equation that setting  $\lambda=1$ leads to an inconsistent relation  between the integers $N$ and $n_0$. 
 This problem arises because without introducing $\lambda$ there is only one free dimensionless parameter~($g\ell_s$) and two conditions to impose. 
 Thus the scaling symmetry~\eqref{scaling-symm} plays a crucial role in making the uplifted solution globally consistent.  
After imposing \eqref{fluxquantwithlambda}
the uplifted solution~\eqref{10d_vacuum-metric} can be matched with the
solution of~\cite{Brandhuber:1999np} identifying
\begin{equation}
L_{\AdS_6}^2 = \frac94 \, \ell_s^2 \left( \frac{3(8-N_f)}{4\pi C} \right)^{1/2}  Q_4^{3/4} \,,  \qquad  \lambda^2 = \left( \frac{3(8-N_f)}{4\pi} \right)^{-5/6} C^{1/6} Q_4^{-1/4} \,,
\end{equation}
with $n_0=8-N_f$, where $N_f$ is the number of D8-branes, and $Q_4$ is related to the number of D4-branes  $N$ by
\begin{equation}
N = \frac{3Q_4}{8\pi}  \left( \frac{3(8-N_f)}{4\pi C^2} \right)^{1/3}  \,.
\end{equation}
Note that the constant $C$ is a trivial parameter which can be set to any non-zero value redefining $Q_4\mapsto C^{2/3} Q_4$.
Although both the six-dimensional vacuum and the ten-dimensional solution of~\cite{Brandhuber:1999np} are well-known, clarifying their relationship 
 will allow us to discuss the six-dimensional origin of more interesting solutions in the following.

As discussed in~\cite{Jafferis:2012iv}, the effective six-dimensional Newton constant that should be proportional to the large $N$ limit of the $S^5$ free energy  of the dual field theory is divergent, due to the singularity of the ten-dimensional solution on the boundary of the hemisphere~$\hemi$, \ie\ for $\xi\to0$. This problem was circumvented in~\cite{Jafferis:2012iv} by calculating the holographic entanglement entropy across a three-sphere and then extracting from this the free energy~$F_{S^5}$. We refer the reader to~\cite{Jafferis:2012iv} for the details and here, for completeness, we only quote the result
\begin{equation}
F_{S^5} 
= -\frac{3^5 \pi^6\lambda^4}{5(2\pi\ell_s g)^8}   = -\frac{9\sqrt{2}\pi}{5} \frac{N^{5/2}}{\sqrt{8-N_f}} \,.
\label{jafferisFS5}
\end{equation}

\subsection{The AdS$_4 \times \Sigma_\mathrm{g}$ solutions and their uplift}
\label{bahsolutionssection}

In this section we will discuss a class of supersymmetric solutions of the $D=6$, $U(1)^2$ supergravity comprising an $\AdS_4$ factor, that were constructed in~\cite{Karndumri:2015eta}
(see also~\cite{Hosseini:2018usu}). We will use these to illustrate our procedure for uplifting to solutions of massive type IIA supergravity, showing that the uplifted solutions coincide with the solutions constructed in~\cite{Bah:2018lyv}, by directly solving the supersymmetry conditions of~\cite{Passias:2018zlm} in ten dimensions.
The solutions have the form of a product $\AdS_4 \times \Sigma_\mathrm{g}$, where $\Sigma_\mathrm{g}$ is a genus $\mathrm{g}$ Riemann surface equipped with a constant curvature metric. As usual, we can distinguish three cases for the curvature $\kappa=\pm1,0$. When $\kappa=+1$ we have the round two-sphere with $\mathrm{g}=0$, while for $\kappa=-1$ locally we have the metric on the two-dimensional hyperbolic space~$H^2$, which can be quotiented to obtain a constant curvature Riemann surface with genus $\mathrm{g}>1$. In order to encompass both non-zero curvature cases\footnote{The case $\kappa =0$ is straightforward to include, but we will not discuss this further.} we denote by $\dd s_{\Sigma_\mathrm{g}}^2$ the metric on~$\Sigma_\mathrm{g}$ and define the one-form~$\omega_\mathrm{g}$ such that $\dd\omega_\mathrm{g}=\vol{\Sigma_\mathrm{g}}$. Explicitly, we can take
\begin{equation}
\begin{aligned}
\dd s_{\Sigma_\mathrm{g}}^2 &= \dd\theta^2 + \sin^2\!\theta \, \dd\phi^2 \,,  \qquad  & \omega_\mathrm{g} &= -\cos\theta \, \dd\phi  \qquad  & (\kappa &= +1) \,, \\
\dd s_{\Sigma_\mathrm{g}}^2 &= \dd\theta^2 + \sinh^2\!\theta \, \dd\phi^2 \,,  \qquad  & \omega_\mathrm{g} &= \cosh\theta \, \dd\phi  \qquad  & (\kappa &= -1) \,.
\end{aligned}
\end{equation}
In our conventions,
the metric, scalars and gauge potentials 
take the form 
\begin{equation}
\begin{split}
\dd s^2_6 &= L_{\AdS_4}^2 \dd s_{\AdS_4}^2 + \e^{2G} \, \dd s_{\Sigma_\mathrm{g}}^2 \,, \\
X_1 &= k_8^{1/8} k_2^{1/2} \,,  \qquad  X_2 = k_8^{1/8} k_2^{-1/2} \,, \\
A_1 &= \frac{\p1}{2g} \, \omega_\mathrm{g} \,,  \qquad  A_2 = \frac{\p2}{2g} \, \omega_\mathrm{g} \,,
\end{split}
\end{equation}
where
\begin{equation} \label{kar_6d-functions}
\begin{aligned}
L_{\AdS_4}^2 &= \frac{k_8^{3/4} }{m^2} \,,  \qquad  & \e^{2G} &= \left[ \frac{\p2^2 k_2^{3/2} - \p1^2 k_2^{-1/2}}{16mg^3 (k_2-1)} \right]^{1/2} \,, \\
k_8 &= \frac{4m^2 k_2}{g^2 (1+k_2)^2} \,,  \qquad  & k_2 &= -\frac{3(\p1-\p2) + \sqrt{9\p1^2 - 14\p1 \p2 + 9\p2^2}}{2\p2} \,,
\end{aligned}
\end{equation}
with $\p1$, $\p2$ two constant parameters obeying the supersymmetry constraint\footnote{A sign ambiguity in the above formulas~\cite{Karndumri:2015eta} has been fixed by noticing that the solution corresponding to the opposite sign is obtained by changing both signs of $\p1$ and $\p2$.} 
\begin{equation}
\p1 + \p2 = 2\kappa \,.
\label{BihCYcondition}
\end{equation}
The fluxes of the gauge fields through $\Sigma_\mathrm{g}$ are given by\footnote{Recall that, for $\kappa\neq0$, the integrated volume of the Riemann surface $\Sigma_\mathrm{g}$ is 
$\mathrm{Vol}(\Sigma_\mathrm{g}) = 4\pi \kappa (1-\mathrm{g})$.}
\begin{equation}
P_i = \frac{g}{2\pi} \int_{\Sigma_\mathrm{g}} F_i = \p{i} \kappa (1-\mathrm{g}) \ \in \  \ZZ \,,
\label{fluxessigmag}
\end{equation}
where the quantization condition above arises from the requirement that $g A_i$ be well-defined connection 
one-forms on $U(1)$  bundles over $\Sigma_\mathrm{g}$. 
The six-dimensional solution is therefore specified by the genus $\mathrm{g}$ and one integer, say $\p1(1-\mathrm{g})$. 

We now set $m=2g/3$ and, after uplifting to massive type IIA using the formulas in~\cite{Cvetic:1999xx}, we introduce the parameter $\lambda$. The ten-dimensional metric and dilaton are 
\begin{align}
\begin{split} \label{kar_10d-metric}
\dd s_\text{s.f.}^2 &=  \lambda^2 k_8^{3/4} g^{-2} \mu_0^{-1/3} \biggl\{ \tilde{\Delta}^{1/2} \biggl[ \frac94 \, \dd s_{\AdS_4}^2 + \e^{2G} k_8^{-3/4} g^2 \, \dd s_{\Sigma_\mathrm{g}}^2 \biggr] \\
& + \tilde{\Delta}^{-1/2} k_8^{-1} \bigl[ k_8^{1/2} \dd\mu_0^2 + k_2^{-1/2} \bigl(\dd\mu_1^2 + \mu_1^2 \sigma_1^2\bigr) + k_2^{1/2} \bigl(\dd\mu_2^2 + \mu_2^2 \sigma_2^2\bigr) \bigr] \biggr\} \,,
\end{split} \\
\e^\Phi &= \lambda^2 \mu_0^{-5/6} \tilde{\Delta}^{1/4} k_8^{-1/8} \,, 
\end{align}
while the Romans mass and four-form flux read
\begin{align}
\label{kar_10d-F0}
F_{(0)} &= \frac{2g}{3\lambda^3} \,, \\
\label{kar_10d-F4}
F_{(4)} &= \frac{\lambda\mu_0^{1/3}}{g^3  k_8^{1/2}\tilde{\Delta}} \biggl\{ \frac{\tilde{U}}{\tilde{\Delta}} \frac{\mu_1 \mu_2}{\mu_0} \, \dd\mu_1 \wedge \dd\mu_2 \wedge \sigma_1 \wedge \sigma_2 \\
& - g \Bigl[ F_1 \wedge \dd\phi_2 \wedge \Bigl( \mu_0 \mu_2 \dd\mu_2 - \mu_2^2 \frac{k_8^{1/2}}{k_2^{1/2}} \, \dd\mu_0 \!\Bigr) + F_2 \wedge \dd\phi_1 \wedge \bigl( \mu_0 \mu_1 \dd\mu_1 - \mu_1^2 k_8^{1/2} k_2^{1/2} \dd\mu_0 \bigr) \Bigr] \biggr\} \,. \nonumber 
\end{align}
We have $\sigma_i=\dd\phi_i-\frac{\p{i}}{2}\,\omega_\mathrm{g}$ and we defined
\begin{equation} \label{kar_10d-func}
\begin{aligned}
\tilde{\Delta} &= k_8^{-1/2} \mu_0^2 + k_2^{1/2} \mu_1^2 + k_2^{-1/2} \mu_2^2 \,, \\
\tilde{U} &= -\frac23 \bigl[ k_8^{-1/2} \mu_0^2 + 3k_8^{1/2} \bigl(1 - \mu_0^2\bigr) + 2\tilde{\Delta} \bigr] \,.
\end{aligned}
\end{equation}

We have checked that the above configuration satisfies the ten-dimensional equations of motion and that $\dd F_{(4)}=0$.
Imposing flux quantization \eqref{genfluxquant} we obtain the relations~\eqref{fluxquantwithlambda},
exactly as for the vacuum solution, showing that $\lambda=1$ would be again inconsistent. We have therefore obtained a globally regular (modulo the ever-present singularity at $\xi=0$) ten-dimensional solution with an $\AdS_4$ factor, parameterised by the genus $\mathrm{g}$ and the integers $n_0,N$, $\p1(1-\mathrm{g})$. 
This is precisely the solution presented in \cite{Bah:2018lyv}, as  we show in detail in appendix \ref{app:map-to-Bah}.

Writing the ten-dimensional metric \eqref{kar_10d-metric} in the form
\begin{equation}
\dd s_\text{s.f.}^2 = \e^{2A} \bigl( \dd s_{\AdS_4}^2 + \dd s_{M_6}^2 \bigr) \,,
\end{equation}
we have that   $\dd s_{M_6}^2$ is the metric  on the internal space $M_6$,  that is the total space of an $\hemi$ bundle over $\Sigma_\mathrm{g}$, namely 
\begin{equation}
\hemi \ \hookrightarrow \ M_6 \ \rightarrow \ \Sigma_\mathrm{g}\, . 
\end{equation}

Recall that $\hemi$ is a hemisphere of the four-sphere due to the fact that the warp factor is singular at  $\mu_0\to 0$, corresponding to the location of the O8-plane \cite{Brandhuber:1999np}. 
However,  we can also think of the internal geometry as an $S^4$ bundle over $\Sigma_\mathrm{g}$, before the inclusion of the O8-plane.
 Either way, there is a  $U(1)\times U(1)\subset SU(2)_R \times SU(2)_F$ symmetry acting on~$\hemi$ and the gauge fields $gA_i$ are connections on the associated circle bundles, twisting these over $\Sigma_\mathrm{g}$, with Chern numbers $P_i$.
The solution can be interpreted as follows \cite{Bah:2018lyv}: one starts with a geometry of the type $\mathbb{R}^{1,2}\times \mathbb{R}\times Y_6$, with
$Y_6$ a local Calabi-Yau three-fold of the form 
\begin{equation}
{\cal O}(-p_1)\oplus {\cal O}(-p_2) \ \hookrightarrow \ Y_6  \to \Sigma_\mathrm{g}\, , 
\end{equation}
where  
$-g A_i$ are Hermitian connections on the line bundles ${\cal O}(-p_i)$.
Then \eqref{BihCYcondition} implies that  the total space of $Y_6$ has vanishing first Chern
class and hence is a Calabi-Yau three-fold. 
 At the origin  of $\mathbb{R}$ there are an O8-plane and $N_f=8-n_0$ coincident D8-branes, and in addition one 
wraps $N$  D4-branes over
the zero section  $\Sigma_\mathrm{g}$ of  $Y_6$.

At low energies the effective theory on the D4/D8-system will be a $d = 3$, ${\cal N}=2$ 
field theory, 
obtained from the compactification on $\Sigma_\mathrm{g}$ of the $d=5$, ${\cal N}=1$ SCFT with gauge group $USp(2N)$ \cite{Brandhuber:1999np}, with the standard topological twist.
The supergravity solution above strongly suggests that in the large $N$ limit this is a SCFT and its $S^3$ free energy can be computed holographically, from the full ten-dimensional solution. Specifically, this can be obtained using the formula presented in~\cite{Guarino:2015jca} adapted to the string frame, and it reads~\cite{Bah:2018lyv} 
\begin{equation} \label{kar_free-energy}
\begin{split}
F_{S^3\times \Sigma_{\mathrm{g}}} = \frac{16\pi^3}{(2\pi\ell_s)^8} \int \e^{8A-2\Phi} \, \vol{M_6} =
 \frac{16 \pi \kappa  (1-\mathrm{g}) N^{5/2} \left(\z^2-\kappa^2\right)^{3/2} \left(\sqrt{\kappa^2 + 8\z^2}-\kappa\right)}{5 \sqrt{8-N_f} \left(\kappa \sqrt{\kappa^2 + 8\z^2}-\kappa^2 + 4\z^2\right)^{3/2}}\, ,
 \end{split}
\end{equation}
where the parameter $\z$ is related to our parameters as $\p1=\kappa+\z$, $\p2=\kappa-\z$.
This expression has been reproduced exactly by a direct field theory computation, using the large~$N$ expansion of the localized partition function on $S^3\times \Sigma_{\mathrm{g}}$~\cite{Crichigno:2020ouj,Crichigno:2018adf}.

\section{The AdS$_4 \times \spindle$ solutions}
\label{newsolutions_section}

\subsection{Local form of the solutions}
\label{newlocal_section}

Our starting point are the following local solutions to the equations of the $D=6$ gauged supergravity model of section \ref{D6sugra}
\begin{equation} \label{spindle_6d}
\begin{aligned}
\dd s^2 &= \bigl(y^2 h_1 h_2\bigr)^{1/4} \left( \dd s_{\AdS_4}^2 + \frac{y^2}{F} \, \dd y^2 + \frac{F}{h_1 h_2} \, \dd z^2 \right) \,, \\
A_i &= \left(\alpha_i - \frac{y^3}{h_i}\right) \dd z \,,  \qquad  X_i = \bigl(y^2 h_1 h_2\bigr)^{3/8} h_i^{-1} \,, \\
F(y) &= m^2 h_1 h_2 - y^4 \,,  \qquad  h_i(y) = \frac{2g}{3m} \, y^3 + \Li \,,
\end{aligned}
\end{equation}
where $\dd s_{\AdS_4}^2$ denotes the unit radius metric on $\AdS_4$ and $\LA$, $\LB$ are two real parameters. The real constants $\alpha_i$ are pure gauge and we have included them as
 they will play a crucial role for understanding the global properties of the solution. These backgrounds can be obtained
 by doing an  analytic continuation~\cite{Lu:2003iv} of a class of  six-dimensional BPS black holes \cite{Cvetic:1999un}.
A curvature singularity lies at $y=0$, hence  without loss of generality, in the subsequent analysis 
we will make sure that the globally regular solutions  will be  restricted to $y>0$. 

Before turning attention to the global structure of the solutions, we will demonstrate that they are supersymmetric by constructing 
the local form of the Killing spinors solving the equations \eqref{KSE_grav} - \eqref{KSE_gauge}.
We employ the following orthonormal frame
\begin{equation} \label{6bein}
e^{\hat{a}} = y^{1/4} (h_1 h_2)^{1/8} \, \hat{e}^{\hat{a}} \,,  \qquad  e^4 = \frac{y^{5/4} (h_1 h_2)^{1/8}}{F^{1/2}} \, \dd y \,,  \qquad  e^5 = \frac{y^{1/4} F^{1/2}}{(h_1 h_2)^{3/8}} \, \dd z \,,
\end{equation}
where $\hat{e}^{\hat{a}}$, $\hat{a}=0,\ldots,3$, is the vierbein on $\AdS_4$, whose coordinates are denoted as $x^{\hat{\mu}}$. Equation~\eqref{KSE_grav} then splits in the following  equations 
\begin{equation} 
\label{KSE2_grav_group}
\begin{split}
\left( \partial_{\hat{\mu}} + \frac14 \omega_{\hat{\mu}}^{\ \hat{a}\hat{b}} \Gamma_{\hat{a}\hat{b}} \right) \epsilon^A  - \frac{\ii}{2} \, \hat{e}^{\hat{a}}_{\hat{\mu}} \, \Gamma_{\hat{a}}^{\ 45} (\sigma^3)^A_{\,\ B} \epsilon^B =0 \,,\\
\partial_y \epsilon^A - \frac{1}{16 y} \left[ (2 + y \tilde h') \epsilon^A - \ii\,\frac{4y^2}{F^{1/2}} (4 - y \tilde h') \Gamma^5 (\sigma^3)^A_{\,\ B} \epsilon^B \right]=0  \,,\\
\partial_z \epsilon^A - \ii\,\frac{g}{2} \left( \alpha_1+\alpha_2 - \frac{2m}{g}\right) (\sigma^3)^A_{\,\ B} \epsilon^B =0 \,,
\end{split}
\end{equation}
where we defined $\tilde h \equiv \log ( h_1h_2)$.
Equations~\eqref{KSE_dila} and~\eqref{KSE_gauge} yield the same constraint
\begin{equation} \label{KSE2_grav}
F^{1/2} \Gamma^4 \epsilon^A + m (h_1 h_2)^{1/2} \epsilon^A + \ii\,y^2 \Gamma^{45} (\sigma^3)^A_{\,\ B} \epsilon^B = 0 \,.
\end{equation}
From the third equation in~\eqref{KSE2_grav_group} we see immediately that setting 
\begin{equation} 
\alpha_1+\alpha_2=\frac{2m}{g}
\end{equation}
 leads to Killing spinors independent of~$z$ and  we will adopt this  
choice in the reminder of this section. 
 We  consider the specific decomposition of the gamma matrices
\begin{equation} \label{gamma-deco}
\Gamma^{\hat{a}} = \gamma^{\hat{a}} \otimes I_2 \,,  \qquad \quad  \Gamma^{\hat{\imath}+3} = \gamma_5 \otimes \rho^{\hat{\imath}} \,,
\end{equation}
where $\gamma^{\hat{a}}$ are the  (Lorentzian) gamma matrices in $D=4$, $\gamma_5=\ii\,\gamma^0\gamma^1\gamma^2\gamma^3$ is the related chiral matrix and $\rho^{\hat{\imath}}$, $\hat{\imath}=1,2$, are the (Euclidean) gamma matrices in $D=2$. For $\rho^{\hat{\imath}}$ we choose the following representation
\begin{equation} \label{2d-gamma_rep}
\rho^{\hat{\imath}} = \sigma^{\hat{\imath}} \, ,   \qquad\quad   \rho_* = -\ii \, \rho^1 \rho^2 = \sigma^3 \,,
\end{equation}
and we take $\mc{B}_2=-\sigma^2$ so that $\mc{B}_6=\mc{B}_4\otimes\mc{B}_2$. The ansatz for the symplectic-Majorana Killing spinors $\epsilon^A$ is
\begin{equation} \label{spinor-deco}
\epsilon^A = \vartheta_+ \otimes \eta^A_+ + \vartheta_- \otimes \eta^A_- \,,
\end{equation}
where $\vartheta=\vartheta(x^{\hat{\mu}})$ is a Majorana Killing spinor on $\AdS_4$ and $\vartheta_\pm$ are its chiral components, \ie\ $\gamma_5\vartheta_\pm=\pm\vartheta_\pm$. Thus we have $\vartheta_\pm^*=\mc{B}_4\vartheta_\mp$ and $\hat{\nabla}_{\hat{\mu}}\vartheta_\pm=\frac12\gamma_{\hat{\mu}}\vartheta_\mp$. The spinors $\eta^A_\pm=\eta^A_\pm(y)$ are two-component Dirac spinors defined on the spindle\footnote{Notice that $\eta^A_\pm$ are \emph{not} chiral spinors, despite the notation suggests otherwise. 
The index $A=1,2$ is an internal index and we will see below that these four Dirac spinors are actually not all independent.}.

Adopting the decomposition~\eqref{gamma-deco}, the symplectic-Majorana condition~\eqref{symp-Maj} and the Killing spinor equations~\eqref{KSE2_grav_group}, \eqref{KSE2_grav} can be solved to give
\begin{equation}
\begin{aligned}
\eta^1_+ &= \xi \, y^{1/8} (h_1 h_2)^{-3/16} \begin{pmatrix} f_1^{1/2} \\ -f_2^{1/2} \end{pmatrix} \,,  \quad &
\eta^1_- &= -\xi \, y^{1/8} (h_1 h_2)^{-3/16} \begin{pmatrix} f_1^{1/2} \\ f_2^{1/2} \end{pmatrix} \,, \\
\eta^2_+ &= \ii\,\xi^* \, y^{1/8} (h_1 h_2)^{-3/16} \begin{pmatrix} f_2^{1/2} \\ -f_1^{1/2} \end{pmatrix} \,,  \quad &
\eta^2_- &= \ii\,\xi^* \, y^{1/8} (h_1 h_2)^{-3/16} \begin{pmatrix} f_2^{1/2} \\ f_1^{1/2} \end{pmatrix} \,,
\end{aligned}
\end{equation}
where $\xi$ is a complex constant and we have defined
\begin{equation}
f_1(y) \equiv m (h_1 h_2)^{1/2} + y^2 \,,  \qquad  f_2(y) \equiv m (h_1 h_2)^{1/2} - y^2 \,,
\end{equation}
which satisfy $F(y)=f_1(y)f_2(y)$. 
Notice  that the four two-dimensional spinors above can be expressed in terms of just one of them, say $\eta^1_+$, by means of the relations
\begin{equation} \label{spinor_relations}
\eta^1_- = -\sigma^3 \eta^1_+ \,,  \qquad  \eta^2_+ = -\ii\,\sigma^1 (\eta^1_+)^* \,,  \qquad  \eta^2_- = \sigma^2 (\eta^1_+)^* \,.
\end{equation}
Notice also that all the spinors  never vanish,  as it can be seen from their norm, given by 
\begin{equation}
(\eta^1_+)^\dagger \, \eta^1_+  = 2|\xi|^2 m \, y^{1/4} (h_1 h_2)^{1/8} \, . 
\end{equation}
 
 We can now count the number of supersymmetries preserved  by our $\AdS_4\times\spindle$ solution. $\vartheta$ is a Majorana spinor, hence it has
four real degrees of freedom, while the spinors $\eta^A_\pm$ are fully determined by the complex constant $\xi$. Therefore, there are eight real 
independent Killing spinors, that is half the number of supersymmetries of the  six-dimensional $\mc{N}=(1,1)$ theory, hence the solution is  $1/2$-BPS.
The eight Killing spinors correspond to four Poincar\'e supercharges $Q$ and four superconformal supercharges $S$
in the   $d=3$, ${\cal N}=2$ SCFTs. The precise identification of the spinors with 
the supercharges is  discussed in appendix  \ref{app:osp}. In particular, we show that $\partial_z$ is  part of the superconformal 
$R$-symmetry, namely the $U(1)$ generator in the $OSp(2|4)$ superalgebra.

\subsection{Global analysis I: metric and magnetic fluxes}

From now on we set $m=2g/3$
without loss of generality. 
In order to have a well-defined metric on the spindle $\spindle$,
\begin{equation} \label{spindle_metric}
\dd s_\spindle^2 = \frac{y^2}{F} \, \dd y^2 + \frac{F}{h_1 h_2} \, \dd z^2 \, , 
\end{equation}
and positive scalars $X_i$ we need to take  $F>0$, $h_1>0$, $h_2>0$ in a closed interval not containing the curvature singularity in $y=0$, 
thus without loss of generality we restrict to $y>0$.
Taking a look at the explicit form of $F$ and its first derivative
\begin{equation}
\begin{aligned}
F(y) &= \frac{4g^2}{9} \biggl[ y^6 - \frac{9}{4g^2} \, y^4 + (\LA+\LB) y^3 + \LA \LB \biggr] \,, \\
F'(y) &= \frac{8g^2}{3} y^2 \Bigl[ y^3 - \frac{3}{2g^2} \, y + \frac{\LA+\LB}{2} \Bigr] \,,
\end{aligned}
\end{equation}
from the expression of $F'$ we see that there exist at most three turning points, hence at most four distinct real roots of $F$. Since the coefficient of $y^6$ in $F(y)=0$ is positive and we restricted to $y>0$, we need at least three positive roots and, to this end, Descartes' rule of signs implies the necessary conditions 
\begin{equation}
\LA+\LB>0 \qquad  \mathrm{and} \qquad  \LA\LB<0\, . 
\end{equation}
 Without loss of generality we take $\LA>0$, $\LB<0$. Recalling that
\begin{equation}
F = \frac{4g^2}{9} h_1 h_2 - y^4 \,,
\end{equation}
we have $h_1 h_2>0$ when $F>0$. Moreover, for positive $y$ and $\LA$, $h_1>0$ and, accordingly, $h_2>0$ as well.

The conditions for $\spindle$ to be a spindle are obtained studying $\dd s_\spindle^2$ in the neighbourhood of the zeros of $F$. Denoting $[\A,\B]$ the range of the coordinate~$y$, as the latter approaches one of the end-points of this interval, say $y_\alpha$, the metric becomes
\begin{equation} \label{spindle_metric-near-poles}
\dd s_\spindle^2 \simeq \dd\varrho^2 + \varrho^2 \, \frac{g^2 F'(y_\alpha)^2}{9y_\alpha^6} \, \dd z^2 \,,
\end{equation}
where we defined $\varrho^2=|y-y_\alpha|$. As a consequence, $\dd s_\spindle^2$ is a smooth orbifold metric on the spindle
if the following conditions hold
\begin{equation} \label{TAB-def}
\frac{g F'(\A)}{3\A^3} \, \Delta z = \frac{2\pi}{n_-} \,, \quad  \qquad  \frac{g F'(\B)}{3\B^3} \, \Delta z = -\frac{2\pi}{n_+} \,,
\end{equation}
where the minus sign in the second relation is due to the fact that $F'(\B)<0$.
Here $n_\pm$ are two co-prime integers and  $\Delta z$ is the periodicity of the $z$ coordinate. 
The Euler characteristic of metric~\eqref{spindle_metric} can be computed noticing that
\begin{equation}
\sqrt{g_\spindle} \, R_\spindle = \frac{\dd}{\dd y} \left( \frac{F \, \partial_y(h_1 h_2) - (h_1 h_2) \, \partial_y F}{y\,(h_1 h_2)^{3/2}} \right) \,.
\end{equation}
We then find
\begin{equation} \label{quant_euler}
\chi(\spindle) = \frac{1}{4\pi} \int_\spindle R_\spindle \, \vol{\spindle} = \frac{n_+ + n_-}{n_+ n_-} \,,
\end{equation}
where we employed $F(y_\alpha)=0$ and the following identity:
\begin{equation} \label{TAB-2}
\frac{g F'(y_\alpha)}{3y_\alpha^3} = \frac{4g^3}{9} \biggl[ \frac{3}{2g^2} - \frac{(\LA+\LB) y_\alpha^3 + 2\LA\LB}{y_\alpha^4} \biggr] \,.
\end{equation}

We now proceed to the quantization conditions for the magnetic fluxes of our $\AdS_4\times\spindle$ solution across the spindle. The integrated fluxes of~$F_i$ are given by
\begin{equation} \label{quant_fluxes}
P_i = \frac{g}{2\pi} \int_\spindle F_i = - g \, \Li \, \frac{\B^3 - \A^3}{ h_i(\A)h_i(\B)} \, \frac{\Delta z}{2\pi} = \frac{\p{i}}{n_+ n_-} \,,  \qquad  \p{i}\in\ZZ \,,
\end{equation}
where the quantization of the~$\p{i}$ arises from the requirement that $g A_i$ be well-defined connection one-forms on $U(1)$ bundles over~$\spindle$ (\emph{c.f.} appendix~A of~\cite{Ferrero:2020twa}). In particular, the total flux reads
\begin{equation} \label{Ps_sum}
\begin{split}
P_1 + P_2 &= \frac{4g^3}{9} \left[ \frac{(\LA+\LB) \B^3 + 2\LA\LB}{\B^4} - \frac{(\LA+\LB) \A^3 + 2\LA\LB}{\A^4} \right] \frac{\Delta z}{2\pi} \\
&= \frac{n_+ + n_-}{n_+ n_-} = \chi(\spindle) \,,
\end{split}
\end{equation}
where we made used  the relation~\eqref{TAB-2} and $F(y_\alpha)=0$. 
 From~\eqref{Ps_sum} it follows that
\begin{equation}
\p1 + \p2 = n_+ + n_- \,,
\end{equation}
and therefore the two integers $p_i$ can be conveniently parameterised as
\begin{equation}
\p1 = \frac{n_+ + n_-}{2} (1 + \z) \,, \quad  \qquad  \p2 = \frac{n_+ + n_-}{2} (1 - \z) \,, 
\end{equation}
where $\z$ is an appropriate rational number\footnote{We must take $\z < -1$ to ensure that 
$\p1 < 0$ ,    $\p2 > 0$, as follows from~\eqref{quant_fluxes} and the signs of $\Li$. Moreover $\z$ must be chosen such  that $p_1$ and $p_2$ are integers}. The situation  is analogous to that obtained for M5-branes wrapped on the spindle  \cite{Ferrero:2021wvk} as here we also see that the ``topologically topological twist''  is realised by the solution, in contrast to the ``anti-twist'' that was first encountered for D3-branes and M2-branes wrapped on spindles.   In~\cite{Ferrero:2021etw} it is shown that very generally  on the spindle only these two types of twists are possible. As explained in this reference, the occurrence of the twist case in our solution is 
correlated with the behavior of the Killing spinors at the north and south poles of the spindle, which we will 
discuss  in section \ref{globalspinorssec}.

\subsection{Solution of the regularity conditions}

In this subsection we elaborate on the conditions worked out in the global analysis above. Specifically, 
we will aim to derive  expressions for $\A$, $\B$ and $\Delta z$ in terms of the spindle parameters $n_\pm$ and the flux parameter~$\p1$ (or $\z$).
Imposing $F(y_\alpha)=0$ we  obtain the sum and product of $\LA$ and $\LB$ in terms of the roots $\A$ and $\B$ as
\begin{equation}
\label{q_sum_prod}
\begin{split}
\LA + \LB &= -(\B^3 + \A^3) + \frac{9}{4g^2} \frac{\B^4 - \A^4}{\B^3 - \A^3} \,, \\
\LA \LB &= \B^3 \A^3 \biggl( 1 - \frac{9}{4g^2} \frac{\B - \A}{\B^3 - \A^3} \biggr) \,,
\end{split}
\end{equation}
while conditions~\eqref{TAB-def} yield
\begin{equation} \label{reg-cond_1}
\frac{n_-}{\A} \biggl( \A^3 - \frac{3}{2g^2} \, \A + \frac{\LA+\LB}{2} \biggr) = -\frac{n_+}{\B} \biggl( \B^3 - \frac{3}{2g^2} \, \B + \frac{\LA+\LB}{2} \biggr) \,.
\end{equation}
Plugging the first of the equations~\eqref{q_sum_prod} into~\eqref{reg-cond_1} and changing variables as
\begin{equation}
\A \equiv \Cd (1 - \x) \,,  \qquad  \B \equiv \Cd (1 + \x) \,,
\end{equation}
we obtain the equation
\begin{equation}
2\Cd^2 g^2 \bigl(x^2 + 3\bigr)^2 (x - \mu) - 3x \bigl(x^2 + 5\bigr) + 9\mu \bigl(x^2 + 1\bigr) = 0 \,,
\end{equation}
where we defined $\mu_\pm\equiv n_+\pm n_-$ and $\mu\equiv \mu_-/\mu_+$. From $0<\A<\B$ it immediately follows that $\Cd>0$ and $0<\x<1$. This quadratic equation for $\Cd$ can be easily solved, 
\begin{equation} \label{C2solution}
\Cd = \frac{1}{g(x^2+3)} \sqrt{\frac{9\mu (x^2+1) - 3x (x^2+5)}{2 (\mu - x)}} \,,
\end{equation}
giving the following   expressions for the two roots in terms of the parameter $\mu$ and the new variable $x$ 
\begin{equation} \label{yABsolutions}
\begin{split}
\A &= \frac{1-x}{g(x^2+3)} \sqrt{\frac{9\mu (x^2+1) - 3x (x^2+5)}{2 (\mu - x)}} \,, \\
\B &= \frac{1+x}{g(x^2+3)} \sqrt{\frac{9\mu (x^2+1) - 3x (x^2+5)}{2 (\mu - x)}} \,.
\end{split}
\end{equation}
Notice that in order for $\A,\B$ to be real, there are two alternative set of conditions, namely
\begin{align}
\label{branch-2}
& 3\mu \bigl(x^2 + 1\bigr) - x \bigl(x^2 + 5\bigr) > 0  \qquad   \text{and}  \qquad  \mu - x > 0 \,, \\[0.5em]
\label{branch-1}
& 3\mu \bigl(x^2 + 1\bigr) - x \bigl(x^2 + 5\bigr) < 0  \qquad   \text{and}  \qquad   \mu - x < 0 \,.
\end{align}

Below we will turn attention to the variable $x$ and we will check which of these two conditions hold.
From~\eqref{quant_fluxes} and~\eqref{TAB-def} we obtain the following useful relations
\begin{equation} \label{Qs_prod}
\frac{\p1 \p2}{(n_+ n_-)^2} = \frac{16g^6}{81} \frac{\LA\LB (\B^3 -\A^3)^2}{\B^4 \A^4} \left(\frac{\Delta z}{2\pi}\right)^2 \,,
\end{equation}
\begin{equation} \label{red-cond_2}
\frac{\A}{n_-} + \frac{\B}{n_+} = -\frac{8g^3}{9} \biggl[ (\B^3 - \A^3) - \frac{3}{2g^2} (\B-\A) \biggr] \frac{\Delta z}{2\pi} \, .
\end{equation}
Extracting $\Delta z$ from~\eqref{Qs_prod} and inserting it into~\eqref{red-cond_2}, along with the expression for the roots  $\A,\B$ from \eqref{yABsolutions},
we obtain 
\begin{equation} \label{old_quartic}
\mu_+ \sqrt{-x^4 + \bigl(9\mu^2 - 5\bigr) x^2 - 12\mu x + 9\mu^2} - 4\eta \sqrt{-2\p1 \p2} \, x = 0 \,,
\end{equation}
where 
$\eta = \sign(\mu - x)$.
Recalling that $x>0$, in order for this equation to have a real solution we need $\eta=+1$, hence any solution will have to satisfy the conditions~\eqref{branch-2}. In particular, being $x$ positive by construction, we need $\mu>0$ as well, hence $\mu_->0$ and $n_-<n_+$. Bearing these existence conditions in mind and recalling that $\p{1,2}=\mu_+(1\pm\z)/2$, from~\eqref{old_quartic} we find that $x$ must be a root of  the following quartic  equation
\begin{equation} \label{eq_quartic}
P(x) \equiv x^4 + \bigl(8\z^2 - 3 - 9\mu^2\bigr) x^2 + 12\mu x - 9\mu^2 = 0 \,.
\end{equation}
From~\eqref{red-cond_2} we can then obtain the periodicity of~$z$, namely
\begin{equation} \label{delta-z}
\Delta z = \chi \, \frac{3\pi \left(x^2+3\right) (\mu - x)}{8g x^2} \,,
\end{equation}
and  from  \eqref{q_sum_prod}
we can express $\LA$ and $\LB$ in terms of $\x$ as
\begin{equation}
\LAB = \Cd \frac{3 (1 - x^2)}{g^2 (x^2 + 3)^2 (\mu - x)} \left[ 3\mu (1 + x^2) - 2x \mp x (x^2 + 3) \z \right] .
\end{equation}

All the relevant quantities are now given in terms of a solution $x$ of equation~\eqref{eq_quartic}, with $0<x<1$ and meeting the constraints~\eqref{branch-2}.
We are  left with the analysis of the existence and uniqueness of such solution. Focussing on the second condition of~\eqref{branch-2}, namely $x<\mu$, we know that, by definition, $\mu<1$, thus this constraint reduces the range of existence of~$x$ to $(0,\mu)$. 
Recalling that $\z^2>1$ and $\mu^2<1$, we have
\begin{equation}
\begin{aligned}
& P(0) = -9\mu^2 < 0 \,, \\
& P(\mu) = 8\bigl(\z^2 - \mu^2) \mu^2 > 0 \, .
\end{aligned}
\end{equation}
 Since $P(x)$ is continuous, there must exist at least one zero in the interval $(0,\mu)$. 
 Let us now prove by contradiction that the first condition of~\eqref{branch-2} holds.  Assuming that
\begin{equation}
3\mu \bigl(x^2 + 1\bigr) - x \bigl(x^2 + 5\bigr) \le0 \, ,
\end{equation}
multiplying this by the positive quantity $x$,  and using equation~\eqref{eq_quartic} to get rid of the quartic term, we obtain
\begin{equation}
3\mu \bigl[ x \bigl(x^2 + 5\bigr) - 3\mu \bigl(x^2 + 1\bigr) \bigr] + 8\bigl(\z^2 - 1\bigr) x^2 \le0 \,,
\end{equation}
hence
\begin{equation}
3\mu \bigl(x^2 + 1\bigr) - x \bigl(x^2 + 5\bigr) \geq \frac{8\bigl(\z^2 - 1\bigr) x^2}{3\mu} > 0 \,,
\end{equation}
in contradiction with the starting hypothesis, which must be false. In this way, we showed that there exists at least one real solution to equation~\eqref{eq_quartic}, lying inside the range $0<x<1$.

We shall now prove that this root is unique inside the interval $(0,\mu)$. The first derivative of $P(x)$ reads
\begin{equation}
P'(x) = 4x^3 + 2\bigl(8\z^2 - 3 - 9\mu^2\bigr) x + 12 \mu \,.
\end{equation}
When $8\z^2-3-9\mu^2\geq0$, $P'(x)>0$ for $x\in(0,\mu)$, hence $P(x)$ is strictly increasing and can thus have only one root inside the considered range. In the other case, we can focus on the zeros of the polynomial, multiply the condition $P'(x)>0$ by $x$ and remember that, at these points, $P(x)=0$, ending up with
\begin{equation}
-\bigl(8\z^2 - 3 - 9\mu^2\bigr) x^2 + 18 \mu (\mu - x) > 0 \,.
\end{equation}
This inequality is true since $8\z^2 - 3 - 9\mu^2<0$ for hypothesis and $\mu-x>0$. Hence, in every zero of $P(x)$ in the range $(0,\mu)$ the polynomial must be increasing, but being $P(x)$ continuous the root must be unique.

\subsection{Global analysis II: gauge fields and Killing spinors}
\label{globalspinorssec}

We shall now discuss global properties of the Killing spinors and the gauge fields $A_i$, following closely the exposition in~\cite{Ferrero:2020twa}. 
Recall that the Killing spinors that we wrote in section \ref{newlocal_section} were obtained in the frame  \eqref{6bein} and with the gauge fields in 
the gauge given by the expressions  \eqref{spindle_6d}, subject to 
\begin{equation}
\alpha_1+\alpha_2=\frac{4}{3}\, ,
\end{equation}
which was motivated by the fact in this gauge they are independent of~$z$. However, both the frame and the gauge fields are singular at the north and south poles of the spindle, where the azimuthal coordinate $z$ is ill-defined. 
In order to shed light on the global properties of the spinors and of the gauge fields, we shall therefore cover the spindle by two patches $U_\pm$ as usual, and check that spinors and gauge fields can be correctly glued across the equator, where they overlap, identifying the  correct bundles of which they are sections and on which they are  connections, respectively.

We begin introducing the angular  coordinate $\varphi$ defined as 
\begin{equation}
\varphi = \frac{2\pi}{\Delta z} \, z \, , 
\end{equation}
with canonical $2\pi$ periodicity, so that the two gauge potentials in~\eqref{spindle_6d} read
\begin{equation}
A_i = \left(\alpha_i - \frac{y^3}{h_i}\right) \frac{\Delta z}{2\pi} \, \dd\varphi \,,
\end{equation}
and we define the $R$-symmetry gauge field $A_R\equiv g(A_1+A_2)$.
The open sets $U_\pm$ cover the  two hemispheres containing the south and north poles, respectively.
Specifically, we have  $\B\in U_+ \simeq  \RR/\ZZ_{n_+}$ and  $\A \in U_- \simeq \RR^2 / \ZZ_{n_-}$. 
In these two patches independently we can perform the gauge transformations
\begin{equation} \label{reg_gauge-trans}
U_\pm:  \quad  A_i^\pm = A_i + \Lambda_i^\pm \, \dd\varphi \, , 
\end{equation}
so that the transformed gauge fields $A_i^\pm$ are non-singular in their respective patches provided that 
\begin{equation}
\begin{aligned}
\Lambda_i^- &= \frac{1}{4g} \left( s_i \chi \z \, \frac{1 + x}{x} + \frac{2}{n_-} \right) - \biggl(\alpha_i - \frac23\biggr) \frac{\Delta z}{2\pi} \,, \\
\Lambda_i^+ &= \frac{1}{4g} \left( s_i \chi \z \, \frac{1 - x}{x} - \frac{2}{n_+} \right) - \biggl(\alpha_i - \frac23\biggr) \frac{\Delta z}{2\pi} \,,
\end{aligned}
\end{equation}
where $s_1=+1$, $s_2=-1$. 
We then  have that  $A_i^- |_{y=\A} =0$ and $A_i^+ |_{y=\B}=0$, implying  that both gauge fields are non-singular at the poles, as required. The corresponding gauge transformations for the $R$-symmetry gauge field $A_R$ are given by 
\begin{equation} \label{reg_gauge-para-tot}
\Lambda_R^- = g (\Lambda_1^- + \Lambda_2^-) = \frac{1}{n_-} \,,  \qquad  \Lambda_R^+ = g (\Lambda_1^+ + \Lambda_2^+) = -\frac{1}{n_+} \,,
\end{equation} 
so that on  the overlap $U_-\cap U_+$ the gauge fields  transform as
\begin{equation}
\begin{split}
A_i^- &= A_i^+ + \left(  \Lambda_i^-  - \Lambda_i^+\right) \dd \varphi  = A_i^+ + \frac{p_i}{gn_+ n_-} \, \dd\varphi \,, \\
A_R^- &= A_R^+ + \left(  \Lambda_R^-  - \Lambda_R^+\right) \dd \varphi  = A_R^+ + \frac{n_+ + n_-}{n_+ n_-} \, \dd\varphi \, ,
\end{split}
\end{equation}
meaning that $gA_i$ are connections on $O(p_i)$ bundles and $A_R$ is a connection on the $O(n_++n_-)$ bundle (which is the tangent bundle)  on the spindle, respectively. 
From the covariant derivative~\eqref{cov-D} we see that the $R$-symmetry charge of $\epsilon^1$ is $1/2$, while that of $\epsilon^2$ is $-1/2$. This implies that a gauge transformation $A_R\mapsto A_R+\Lambda_R\dd\varphi$ acts on the Killing spinors as $\epsilon^1\mapsto\e^{\ii\varphi\Lambda_R/2}\epsilon^1$ and $\epsilon^2\mapsto\e^{-\ii\varphi\Lambda_R/2}\epsilon^2$, with the spinors~$\eta^1_\pm$ and $\eta^2_\pm$ behaving accordingly. 

We now move to the analysis of the global structure of the Killing spinors. 
Since the frame spanning the spindle $\{e^4,e^5\}$ in~\eqref{6bein}  is again singular at the poles, we consider two distinct local frames in each of the two patches. 
We define $\varrho_\pm$ the geodesic distance between $y$ and each root contained in $U_\pm$, $\B$ and $\A$ respectively. We can thus write (\cf~\eqref{spindle_metric-near-poles})
\begin{equation}
\begin{aligned}
& U_-:  \quad  && e^4 \sim \dd\varrho_- \,,  \quad  && e^5 \sim \varrho_- \frac{\dd\varphi}{n_-} \,, \\
& U_+:  \quad  && e^4 \sim -\dd\varrho_+ \,,  \quad  && e^5 \sim \varrho_+ \frac{\dd\varphi}{n_+} \,, \\
\end{aligned}
\end{equation}
where the sign of $e^4$ in $U_+$ is due to the fact that approaching $\B$ the coordinate $y$ is increasing, while $\varrho_+$ is decreasing. In the patch $U_-$ we introduce the complex coordinate $z_-=\varrho_-\e^{\ii\varphi/n_-}=x_-+\ii\,y_-$, which is non-singular in $\A$ and, thus, defines a smooth one-form $\dd z_-$ on the orbifold. This one-form, in turn, determines a non-singular frame, that can be obtained rotating the initial frame as follows:
\begin{equation} \label{reg_frame-rotation}
\begin{pmatrix} e^4 \\ e^5 \end{pmatrix} \mapsto
\begin{pmatrix}
\cos\frac{\varphi}{n_-} & -\sin\frac{\varphi}{n_-} \\
\sin\frac{\varphi}{n_-} & \cos\frac{\varphi}{n_-}
\end{pmatrix}
\begin{pmatrix} e^4 \\ e^5 \end{pmatrix} \sim
\begin{pmatrix} \dd x_- \\ \dd y_- \end{pmatrix} \,.
\end{equation}
This is an $SO(2)\cong U(1)$ rotation of the frame in the patch $U_-$, which induces a
transformation of the spinors given by the action of the exponential of the spinor representation of the infinitesimal version of the $SO(2)$ frame rotation. Explicitly, this is a
$U(1)$ rotation of the components of the spinors by means of the matrix
\begin{equation} \label{reg_spinor-rotation}
\exp \left( -\frac{\varphi}{2n_-} \sigma^1 \sigma^2 \right) =
\begin{pmatrix}
\e^{-\ii\varphi/2n_-} & 0 \\
0                   & \e^{\ii\varphi/2n_-}
\end{pmatrix} \,.
\end{equation}
Performing in $U_-$ the frame rotation~\eqref{reg_frame-rotation} and the gauge transformation~\eqref{reg_gauge-trans} with gauge parameter $\Lambda_R^-$~\eqref{reg_gauge-para-tot}, the spinors undergo 
an $R$-symmetry rotation plus a $U(1)$ rotation. The total action on, \eg, $\eta^1_+$ is
\begin{equation}
U_-:  \quad  \eta^1_+ \simeq
\begin{pmatrix} f_1^{1/2} \\ -f_2^{1/2} \end{pmatrix} \mapsto \e^{\ii\varphi/2n_-}
\begin{pmatrix}
\e^{-\ii\varphi/2n_-} & 0 \\
0                   & \e^{\ii\varphi/2n_-}
\end{pmatrix}
\begin{pmatrix} f_1^{1/2} \\ -f_2^{1/2} \end{pmatrix} =
\begin{pmatrix} f_1^{1/2} \\ -\e^{\ii\varphi/n_-} f_2^{1/2} \end{pmatrix} \,.
\label{sectionminus}
\end{equation}
The $\varphi$ coordinate is not well-defined in $\A$, however since $f_2(\A)=0$, the transformed spinor is smooth and well-defined at this pole of the spindle and, thus, in the whole patch~$U_-$. Of course the same is true  for the spinors $\eta^1_-$ and $\eta^2_\pm$.
A similar analysis can be performed in the patch $U_+$ defining the non-singular coordinate $z_+=-\varrho_+\e^{-\ii\varphi/n_+}$, related to the initial frame by the rotation
\begin{equation}
\begin{pmatrix} e^4 \\ e^5 \end{pmatrix} \mapsto
\begin{pmatrix}
\cos\frac{\varphi}{n_+}  & \sin\frac{\varphi}{n_+} \\
-\sin\frac{\varphi}{n_+} & \cos\frac{\varphi}{n_+}
\end{pmatrix}
\begin{pmatrix} e^4 \\ e^5 \end{pmatrix} \,.
\end{equation}
This transformation is analogous to the frame rotation in $U_-$, but is performed in the opposite direction, and the same happens to the spinors. The corresponding spinor rotation and gauge transformation combine so that the spinor $\eta^1_+$ transforms as
\begin{equation}
U_+:  \quad  \eta^1_+ \simeq
\begin{pmatrix} f_1^{1/2} \\ -f_2^{1/2} \end{pmatrix} \mapsto \e^{-\ii\varphi/2n_+}
\begin{pmatrix}
\e^{\ii\varphi/2n_+} & 0 \\
0                   & \e^{-\ii\varphi/2n_+}
\end{pmatrix}
\begin{pmatrix} f_1^{1/2} \\ -f_2^{1/2} \end{pmatrix} =
\begin{pmatrix} f_1^{1/2} \\ -\e^{-\ii\varphi/n_+} f_2^{1/2} \end{pmatrix} \,,
\label{sectionplus}
\end{equation}
giving, again, a well-defined spinor in $U_+$.

In conclusion, we have shown that the Killing spinors on the spindle 
are smooth and well-defined, in the appropriate orbifold sense, in line with all the previous constructions of supersymmetric solutions involving spindles.  
The spinor transition function in going from the path $U_+$  to 
the patch $U_-$  reads
\begin{equation}
\begin{pmatrix}
\e^{-\ii\varphi/2n_-} & 0 \\
0                  & \e^{\ii\varphi/2n_-}
\end{pmatrix} \cdot
\begin{pmatrix}
\e^{-\ii\varphi/2n_+} & 0 \\
0                   & \e^{\ii\varphi/2n_+}
\end{pmatrix} =
\begin{pmatrix}
\e^{-\ii\varphi\frac{n_+ + n_-}{2n_+ n_-}} & 0 \\
0                   & \e^{\ii\varphi\frac{n_+ + n_-}{2n_+ n_-}}
\end{pmatrix} \,,
\end{equation}
where the sign of the rotation in $U_+$ is reversed because we started with a non-singular spinor in the patch $U_+$.
This identifies the positive and negative chirality spin bundles $S^{(\pm)}$ on the spindle as the bundles $O(\mp \tfrac{1}{2}(n_++n_-))$, respectively. Recalling that our spinors are also charged under $A_R$ and therefore they are sections of the bundles
$O(\pm(n_++n_-))^{1/2}=O(\pm\tfrac{1}{2}(n_++n_-))$, we conclude that, for example 
\begin{equation}
\begin{split}
\tfrac{1}{2}(1+\rho_*)\eta^1_+  & \quad \mathrm{is~a~section~of}  \quad  O(- \tfrac{1}{2}(n_++n_-)) \otimes O(\tfrac{1}{2}(n_++n_-)) = O(0) \, , \\
\tfrac{1}{2}(1-\rho_*)\eta^1_+  & \quad \mathrm{is~a~section~of}  \quad  O( \tfrac{1}{2}(n_++n_-)) \otimes O(\tfrac{1}{2}(n_++n_-) ) =  O(n_++n_-)\, , 
\end{split}
\end{equation}
as it was indeed obvious from the explicit transition functions obtained from passing from the expression in  \eqref{sectionminus} to that in \eqref{sectionplus}.

Notice that all the Killing spinors have definite chirality at the north and south poles of the spindle, and this is the same at both poles.  For example, at the poles the spinor~$\eta^1_+ $ reads
\begin{equation}
\eta^1_+ = \xi \sqrt{2}(m y_\alpha)^{3/8}\begin{pmatrix} 1 \\ 0 \end{pmatrix} \qquad \quad \mathrm{for}\qquad \alpha=N,S \, , 
 \end{equation}
which have both positive chirality. The other spinors behave similarly. As discussed in~\cite{Ferrero:2021etw}, this behaviour is indeed consistent with having a global topological twist.

\subsection{Uplift to massive type IIA and holographic free energy}

By means of the reduction ansatz of~\cite{Cvetic:1999xx} we can uplift our six-dimensional
$\AdS_4\times\spindle$ background~\eqref{spindle_6d}
to massive type IIA and  subsequently introduce $\lambda$ as in~\eqref{scaling-symm}. The metric and dilaton read
\begin{align}
\begin{split}
\label{mIIAmetric}
\dd s_\text{s.f.}^2 &= \lambda^2 \mu_0^{-1/3} \Bigl\{ y^{-1} \Delta_h^{1/2} \bigl( \dd s_{\AdS_4}^2 + \dd s_\spindle^2 \bigr) \\
 & + g^{-2} \Delta_h^{-1/2} \bigl[ y^3 \, \dd\mu_0^2 + h_1 \bigl(\dd\mu_1^2 + \mu_1^2 \sigma_1^2\bigr) + h_2 \bigl(\dd\mu_2^2 + \mu_2^2 \sigma_2^2\bigr) \bigr] \Bigr\} \,,
\end{split} \\[0.25em]
\e^\Phi &= \lambda^2 \mu_0^{-5/6} y^{-3/2} \Delta_h^{1/4} \,,
\end{align}
the Romans mass is 
\begin{equation}
F_{(0)} = \frac{2g}{3\lambda^3} \,,
\end{equation}
while the four-form flux takes the form
\begin{equation}
\begin{split}
F_{(4)} &= \frac{\lambda \mu_0^{1/3} h_1 h_2}{g^3 \Delta_h} \biggl\{ \frac{U_h}{\Delta_h} \frac{\mu_1 \mu_2}{\mu_0} \, \dd\mu_1 \wedge \dd\mu_2 \wedge \sigma_1 \wedge \sigma_2 \\
& - g \sum_{i\neq j} F_i \wedge \dd\phi_j \wedge \bigl( \mu_0 \mu_j \, \dd\mu_j - y^3 h_j^{-1} \mu_j^2 \, \dd\mu_0 \bigr) \\
& + \frac{y^3}{\Delta_h} \sum_{i\neq j}  \frac{h_j (h_i' - 3y^{-1} h_i)}{h_i} \mu_i^2 \, \dd y \wedge \sigma_i \wedge \sigma_j \wedge \bigl( \mu_0 \mu_j \, \dd\mu_j - y^3 h_j^{-1} \mu_j^2 \, \dd\mu_0 \bigr) \biggr\} \,.
\end{split}
\end{equation}
For convenience, we defined the functions
\begin{equation}
\begin{aligned}
\Delta_h &= h_1 h_2\,\mu_0^2 + y^3 h_2\,\mu_1^2 + y^3 h_1\,\mu_2^2 \,, \\
U_h &= 2 \bigl[ (y^3 - h_1)(y^3 - h_2) \mu_0^2 - y^6 \bigr] - \frac43 \Delta_h \,.
\end{aligned}
\end{equation}

The quantization of the fluxes proceeds as in~\eqref{genfluxquant} and it yields again the relations~\eqref{fluxquantwithlambda}, which fix the parameters $g$ and $\lambda$ in terms of the integers 
$N$ and $n_0$.
Therefore our ten-dimensional solution is characterised by five integers: the  pair $n_0$ and $N$, determining the dual five-dimensional theory,
$n_\pm$ defining the spindle, and $\z$ related to the magnetic charges threading this.

The ten-dimensional geometry has a form analogous to that of the solutions in section~\ref{bahsolutionssection}. 
The internal six-dimensional space $M_6$ has a fibration structure
\begin{equation}
\hemi \ \hookrightarrow \ M_6 \ \rightarrow \ \spindle\, ,
\end{equation}
with the twisting of the bundle specified by the connection one-forms $gA_i$ with Chern numbers 
\begin{equation}
P_i = \frac{g}{2\pi} \int_\spindle \dd A_i = \frac{\p{i}}{n_+ n_-} \,,  \qquad  \p{i}\in\ZZ \,,
\end{equation}
subject to 
\begin{equation}
\begin{split}
P_1 + P_2 &= \frac{n_+ + n_-}{n_+ n_-} = \chi(\spindle) \,.
\label{banana}
\end{split}
\end{equation}
The solution can then be interpreted as follows: one starts with a geometry of the type $\mathbb{R}^{1,2}\times \mathbb{R}\times Y_6$, where $Y_6$ is the total space of the vector 
bundle
\begin{equation}
{\cal O}(-p_1)\oplus {\cal O}(-p_2) \ \hookrightarrow \ Y_6  \to \spindle\, , 
\end{equation}
and the condition \eqref{banana} guarantees that the first Chern class of this bundle vanishes, thus $Y_6$ is a local Calabi-Yau three-fold.
At the origin  of $\mathbb{R}$ there are an O8-plane and $N_f=8-n_0$ coincident D8-branes, and in addition one 
wraps $N$  D4-branes over
the zero section  $\spindle$ of  $Y_6$. 

At low energies the effective theory on the D4/D8-system will be a $d = 3$, ${\cal N}=2$ 
field theory,
obtained from the compactification on $\spindle$ of the
 $d=5$, ${\cal N} = 1$ SCFT with gauge group $USp(2N)$  \cite{Brandhuber:1999np}, with the ``topologically topological twist".
The supergravity solution above strongly suggests 
that in the large $N$ limit this is a SCFT and its $S^3$ free energy can be computed holographically as before~\cite{Guarino:2015jca}. Specifically, we have
\begin{equation} \label{spindle_free-energy}
\begin{split}
F_{S^3 \times \spindle} &= \frac{16\pi^3}{(2\pi\ell_s)^8} \frac{3\pi^2 \lambda^4}{10g^4} \, (\B^3 - \A^3) \, \Delta z \\
&= \chi \, \frac{\sqrt{3}\pi N^{5/2}}{5\sqrt{8-N_f}} \, \frac{[3\mu (x^2+1) - x (x^2+5)]^{3/2}}{x (x^2+3) (\mu - x)^{1/2}} \, .
\end{split}
\end{equation}
Notice that the dependence of $x$ on the parameter  $\z$  and  $\mu$  could be made explicit by writing out the solution to the quartic  \eqref{eq_quartic}, however this is extremely cumbersome and we will refrain from doing so.  
Alternatively, one could think of $x$ and $\mu$ as the two independent parameters, with $\z$ given in terms of these two by  solving  \eqref{eq_quartic}, which is a  simple quadratic equation. In any case, in the next section we will reproduce the
 expression \eqref{spindle_free-energy}
analytically, starting from a conjectural large $N$ free energy of the dual field theories.

Noticing that $\mu$ is a free ``small'' parameter\footnote{Although $\mu\in \mathbb{Q}$, we can treat it formally as a real variable taking values in the interval $(0,1)$.}, it is useful to expand in series of $\mu$ near to $\mu\to 0 $ (holding $\chi$ fixed), 
which formally corresponds to reducing to a spindle with equal 
conical deficits, and in particular it includes the two-sphere for $n_+=n_-=1$. 
The root of the quartic equation~\eqref{eq_quartic}  meeting the required constraints then has the following expansion
\begin{equation}
x = \frac{3}{2+t} \, \mu + \frac{27 (1+t) (3+t)}{2t (2+t)^4} \, \mu^3 + \mc{O}(\mu^5)\, , 
\end{equation}
where we have defined $t\equiv\sqrt{8\z^2+1}$, with $t>3$. 
Inserting this in  the free energy \eqref{spindle_free-energy} we obtain the expansion
\begin{equation}
\begin{split}
F_{S^3\times \spindle} &= \chi \,  \frac{\pi N^{5/2}}{5\sqrt{8-N_f}}\biggl[ \frac{(t-3)^{3/2}}{(t-1)^{1/2}} + \frac{6 (t-3)^{3/2}}{(2+t) (t-1)^{3/2}} \, \mu^2 \biggr]  + \mc{O}(\mu^4) \, , 
\end{split}
\label{freeperturb}
\end{equation}
which, after setting $\chi=2$, at leading order in $\mu$ agrees with the free energy~\eqref{kar_free-energy} 
for $\mathrm{g}=0$ and $\kappa=1$~\cite{Bah:2018lyv}. This suggests that it may be possible to recover the  $\AdS_4\times S^2$  solution in~\cite{Bah:2018lyv} by performing a suitable scaling limit of our solutions, but we will not attempt to do so here.
For future reference, let us also record the expansion for $\Delta z$, that reads
\begin{equation}
\begin{split}
\Delta z &= \chi \, \frac{\pi}{8g} \biggl[ \frac{(t-1) (2+t)}{\mu} - \frac{3 (9+34t+25t^2+4t^3)}{2 t (2+t)^2} \, \mu \biggr] + \mc{O}(\mu^3) \,  . 
\end{split}
\end{equation}

\section{Field theory}
\label{ftsection}

We conjecture that the solutions we have constructed in section~\ref{newsolutions_section} are holographically dual to three-dimensional SCFTs obtained by compactifying on a spindle~$\spindle$ the five-dimensional SCFTs dual to the solution of~\cite{Brandhuber:1999np}. In the reminder of this section we will provide evidence for this by proposing an \emph{off-shell free energy} whose extremization reproduces exactly the holographic free energy~\eqref{spindle_free-energy}. This is an extension of the entropy functions that have been shown to provide an efficient method for reproducing the entropy of supersymmetric AdS black holes in various dimensions. 
\emph{A priori}, this function should be derived from first principles, by computing (minus the logarithm of) the localized partition function of the $d=5$ SCFT, placed on the background of $S^3\times\spindle$, and then taking the large~$N$ limit. 
This strategy has been implemented in \cite{Crichigno:2020ouj,Crichigno:2018adf} for the background $S^3\times\Sigma_\mathrm{g}$
and, indeed, it led to reproducing the holographic free energy~\eqref{kar_free-energy} previously obtained in~\cite{Bah:2018lyv}. 
Instead, we will follow a short-cut inspired by the ``gravitational blocks'' advocated in~\cite{Hosseini:2019iad}. 
We will infer from the supergravity description the main ingredients involved in the field-theoretic construction
and we will propose a large $N$ off-shell free energy on $S^3\times\spindle$ obtained by suitably gluing the $S^5$ free energy of the $d=5$ theories. 
We will then show that extremizing it will reproduce exactly the holographic free energy~\eqref{spindle_free-energy}.

\subsection{$d=5$ SCFTs dual to the AdS$_6$ solution}
\label{5dtheorysection}

Let us begin by recalling the salient features of the five-dimensional theory that is holographically dual to the $\AdS_6\times\hemi$ background of massive type IIA, arising in the near-horizon limit of $N$ D4-branes and $N_f$ D8-branes, that we reviewed in 
section \ref{BOsolution}. This is an ${\cal N}=1$ gauge theory with gauge group $USp(2N)$, coupled to $N_f$ massless hypermultiplets in the fundamental representation and one hypermultiplet in the antisymmetric representation of $USp(2N)$ \cite{Seiberg:1996bd}.
At low energies, this theory flows to an interacting SCFT, with global symmetry $SU(2)_R\times SU(2)_F \times E_{N_f+1}$, where the first two factors are  realised as symmetries of the $\AdS_6\times\hemi$ solution~\cite{Passias:2012vp}. 
Placing this theory on a rigid $S^5$ background, one can compute the exact localized partition function $Z_{S^5}$ and consider the associated free energy, namely 
 \begin{equation}
F_{S^5} \equiv -\log Z_{S^5}\, , 
\end{equation}
as a good measure of the degrees of freedom of the theory. This was computed in \cite{Jafferis:2012iv}, that also showed that in the large $N$ limit it becomes 
\begin{equation}
F_{S^5} =  - \frac{9\sqrt{2} \pi }{5}\frac{N^{5/2}}{\sqrt{8-N_f}} \, 
\end{equation}
and is reproduced by a holographic calculation in the solution of  \cite{Brandhuber:1999np}. 
The $S^5$ free energy may also be ``refined'',  promoting it to an off-shell free energy, regarded as a function of the fugacities for the  $U(1)\times U(1)$  Cartan subgroup of  $SU(2)_R\times SU(2)_F$, which we will denote as $\Delta_i$, with $i=1,2$. \emph{A priori}, the $R$-symmetry can mix with 
any flavour symmetry and the $\Delta_i$ parameterise this mixing. For the present theory this is actually not necessary, as the $R$-symmetry is non-Abelian, nevertheless this will be useful in the sequel. 
We can then write 
\begin{equation}
F_{S^5}(\Delta_i) = -    \frac{9\sqrt{2} \pi }{5}\frac{N^{5/2}}{\sqrt{8-N_f}} (\Delta_1 \Delta_2)^{3/2}
\equiv         \frac{27}{4}  {\cal F} (\Delta_i)\, , 
\label{defcalF}
\end{equation}
with the fugacities obeying, in a canonical normalization, the $R$-symmetry constraint
\begin{equation}
\Delta_1 + \Delta_2= 2\, . 
\label{const12}
\end{equation}
For later convenience we have defined the auxiliary function  ${\cal F} (\Delta_i)$. Extremizing $F_{S^5}(\Delta_i)$ gives $\Delta_1^*=\Delta_2^*=1$ and inserting these values back one reproduces  the initial free energy 
 \begin{equation}
F_{S^5} \equiv F_{S^5} (\Delta_i^*) = - \frac{9\sqrt{2} \pi }{5}\frac{N^{5/2}}{\sqrt{8-N_f}} \, . 
\end{equation}
Let us now move to discussing  compactifications of this theory to $d=3$ dimensions and the corresponding off-shell free energies.

\subsection{$d=3$ SCFTs dual to the AdS$_4 \times \Sigma_\mathrm{g}$ solutions}

\label{Bahtheorysection}

We can obtain three-dimensional ${\cal N}=2$ theories by compactifying the above $d=5$ SCFT on a Riemann surface $\Sigma_\mathrm{g}$ of arbitrary genus $\mathrm{g}$, performing the standard topological twist~\cite{Bah:2018lyv}. 
Specifically, we place the theory on $\Sigma_\mathrm{g}$ and couple it to two background gauge fields~$A_i$ for the $U(1)\times U(1)$ Cartan subgroup of $SU(2)_R\times SU(2)_F$, with appropriately quantized magnetic fluxes\footnote{Here and in the following we shall rename the background gauge fields as $gA_i \mapsto A_i$, which is more natural from the field theory point of view. The magnetic fluxes $\ts_i$ correspond precisely to the fluxes~$P_i$ defined in~\eqref{fluxessigmag}. However, we denote these with different symbols to emphasise the fact that the $P_i$ were defined as integrals of supergravity fields, living in 
$D=6$, while the $\ts_i$ are defined as integrals of background gauge fields, living in $d=5$.} 
\begin{equation} 
\ts_i  = \frac{1}{2\pi} \int_{\Sigma_\mathrm{g}} F_i = \p{i} \kappa (1-\mathrm{g})  \ \in \ \ZZ \, . 
\end{equation}
The topological twist implies that the $R$-symmetry gauge field
 $A_R=A_1+A_2$ is identified with a connection on the tangent bundle, thus 
\begin{equation}
 \frac{1}{2\pi} \int_{\Sigma_\mathrm{g}} \dd A_R = \chi (\Sigma_\mathrm{g}) = 2 (1-\mathrm{g})   =  \ts_1 + \ts_2 \, ,
\end{equation}
and the Killing spinors become just constant. It is then convenient to parameterise the magnetic fluxes as
\begin{equation} \label{zparametersigmag}
\ts_1 = (1-\mathrm{g}) (1 + \kappa\,\z) \,,  \qquad  \ts_2 = (1-\mathrm{g}) (1- \kappa\,\z) \,.
\end{equation}
The exact $R$-symmetry of the $d=3$ theory will be determined by extremizing the off-shell $S^3$ free energy \cite{Jafferis:2010un}, viewed as a function of the fugacities $\Delta_i$. Equivalently, this quantity may be thought of as the off-shell free energy  of the $d=5$ theory on $S^3\times \Sigma_\mathrm{g}$. The latter quantity was computed in \cite{Crichigno:2020ouj,Crichigno:2018adf} using  localization, and  in the large $N$ limit it was shown to reproduce the holographic free energy \eqref{kar_free-energy}.
Below we will show that it can also be reproduced by a formula obtained by gluing two gravitational blocks.
We begin defining the following  conjectural large $N$ off-shell free energy\footnote{For $\mathrm{g}=0$, the form \eqref{Foffshellsigmag} has been recently proved in 
 \cite{Hosseini:2021mnn}  for $S^3_b \times S^2_\epsilon$.}
 \begin{equation}
\Ispindle (\Delta_i,\epsilon; \ts_i) \equiv  \frac{1}{\epsilon}\left( {\cal F} (\Delta_i^+ ) -   {\cal F} (\Delta_i^-)  \right)\, , 
\label{Foffshellsigmag}
\end{equation} 
where  ${\cal F} (\Delta_i)$ is defined in \eqref{defcalF} and
 \begin{equation}
 \begin{split}
 \Delta_i^+  \,  \equiv  \Delta_i +   \ts_i \epsilon \, , \qquad \Delta_i^- \,  \equiv   \Delta_i  -      \ts_i\epsilon \,, 
\end{split}
\end{equation}
with $\Delta_i$ satisfying \eqref{const12}.

Notice that in addition to the fugacities $\Delta_i$ of the parent $d=5$ theory and the magnetic fluxes $\ts_i$, \eqref{Foffshellsigmag} depends \emph{a priori} also on the parameter $\epsilon$, although we shall see below that the extremization equations automatically set $\epsilon=0$. At least in the case $\mathrm{g}=0$, $\epsilon$ 
 may be interpreted as the fugacity associated to the $U(1)_J\subset SU(2)_J$ rotational symmetry of the two-sphere. 
  Extremizing \eqref{Foffshellsigmag} with respect to $\epsilon$ and $\Delta_i$, subject to \eqref{const12}, we easily find the critical values 
\begin{equation}
  \Delta_1^* = 1 + \frac{\kappa+ \sqrt{8\z^2 + \kappa^2}}{4\z}\, , \qquad \quad \epsilon^* = 0 \, .
  \label{cane}
  \end{equation}
For the two-sphere, the fact that $\epsilon^*$ vanishes means that 
 the $R$-symmetry of the compactified theory does not have a component along $U(1)_J$, as expected.  
In any case, inserting the critical values in \eqref{Foffshellsigmag} we get 
 \begin{equation}
\Ispindle  (\Delta_i^*,0;\z) = \frac{16 \pi  (1-\mathrm{g}) \kappa N^{5/2} \left(\z^2-\kappa^2\right)^{3/2} \left(\sqrt{\kappa^2 + 8\z^2}-\kappa\right)}{5 \sqrt{8-N_f} \left(\kappa \sqrt{\kappa^2 + 8\z^2}-\kappa^2 + 4\z^2\right)^{3/2}}\, , 
\label{gatto}
 \end{equation}
which agrees with \eqref{kar_free-energy}.
Since $\epsilon^* = 0$, we could have started
 setting  $\epsilon = 0$ in \eqref{Foffshellsigmag}, thus 
 reducing to the known $\epsilon$-independent off-shell free energy
\begin{equation}
 \lim_{\epsilon\to0} \Ispindle(\Delta_i,\epsilon; \ts_i)
= 2 \sum_{j=1}^2 \ts_j  \frac{\partial {\cal F}(\Delta_i) }{\partial \Delta_j} 
= - \frac{4\sqrt{2}\pi }{5}\frac{N^{5/2}}{\sqrt{8-N_f}} (\Delta_1 \Delta_2)^{1/2}(\ts_2\Delta_1+ \ts_1\Delta_2) \,,
\label{criceto}
\end{equation}
corresponding to the standard topological twist~\cite{Crichigno:2020ouj,Crichigno:2018adf,Hosseini:2018uzp,Hosseini:2018usu}.
In particular, extremizing~\eqref{criceto} reproduces~\eqref{gatto} with $\Delta_1^*$ given in~\eqref{cane}.

We notice that $\Delta_1^*$, $\Delta_2^*$ match with the scaling dimensions of two particular 1/2-BPS operators, corresponding to  D2-branes wrapped on calibrated surfaces, embedded in the internal six-dimensional geometries, and sitting at the center of $\AdS_4$, described in~\cite{Bah:2018lyv}.  For $\mathrm{g}>1$ we can set $\z=0$ so that $\Delta_1^*=\Delta_2^*=1$ and
the free energy~\eqref{gatto} reduces to the universal relation~\cite{Bobev:2017uzs} 
\begin{equation}
\Ispindle (\Delta_i^*,0; \ts_1=\ts_2=1-\mathrm{g}) = \frac{16 \pi  (\mathrm{g}-1) N^{5/2}  }{5\sqrt{2} \sqrt{8-N_f}} = - \frac{8}{9} (\mathrm{g} - 1) F_{S^5} \, .
\end{equation}

\subsection{$d=3$ SCFTs dual to  the AdS$_4 \times \spindle$ solutions}
\label{D4onspindlesection}

We now consider the compactification of the $d=5$, ${\cal N}=1$ SCFT on a spindle, that we expect to be dual to the solutions we constructed in section~\ref{newsolutions_section}. In particular, 
we perform a global  topological twist, which means that we place the theory
 on $\spindle$ and couple it to two background gauge fields $A_i$ for the 
$U(1)\times U(1)$  Cartan subgroup of  $SU(2)_R\times SU(2)_F$, with appropriately quantized magnetic fluxes
\begin{equation} 
\ts_i  = \frac{1}{2\pi} \int_\spindle F_i  = \frac{\p{i}}{n_+ n_-} \,,  \qquad  \p{i}\in\ZZ \, . 
\end{equation}
As for the standard topological twist, the 
$R$-symmetry gauge field
 $A_R=A_1+A_2$ becomes  a connection on the tangent bundle, thus 
\begin{equation}
 \frac{1}{2\pi} \int_{\spindle} \dd A_R = \chi (\spindle) = \frac{n_++n_-}{n_+n_-}=  \ts_1 + \ts_2 \, ,
\end{equation}
but crucially, this \emph{does not} imply that the metric on the spindle has constant curvature, nor that the spinors are chiral and constant. 
Specifically, the rigid Killing spinors are expected to behave precisely as the spinors $\eta_\pm^A$ arising in the supergravity solution.  
As before in the paper, we will continue to  parameterise the magnetic fluxes as
 \begin{equation}
\ts_1 = \frac{\chi}{2} (1 + \z) \,,  \qquad  \ts_2 = \frac{\chi}{2} (1 - \z) \, . 
\label{defzspindle}
\end{equation}

Taking  inspiration from the entropy function proposed in  \cite{Ferrero:2021ovq}, 
we conjecture that the large $N$ off-shell free energy for these theories is given by
 \begin{equation}
 \Ispindle  (\Delta_i,\epsilon;\ts_i,n_+,n_-) \equiv   \frac{1}{\epsilon}\left( {\cal F} (\Delta_i^+ ) -   {\cal F} (\Delta_i^-)  \right)\, , 
\label{offshellFconjecture}
\end{equation}
 where  ${\cal F} (\Delta_i)$ is defined in \eqref{defcalF} and
 \begin{equation}
 \begin{split}
 \Delta_i^+  \equiv  \Delta_i  +     \epsilon \left(\ts_i+\frac{1}{2} \frac{n_+ - n_-}{n_+n_-}\right) \, , \qquad \Delta_i^-   \equiv  \Delta_i -   \epsilon \left(\ts_i -\frac{1}{2} \frac{n_+ - n_-}{n_+n_-}\right) \, .
\end{split}
\end{equation}
The $\Delta_i$ are the fugacities parameterising the $R$-symmetry within the  $U(1)\times U(1)\subset SU(2)_R\times SU(2)_F$ global 
symmetries  of the $d=5$ theory, and are therefore still subject to the constraint   \eqref{const12}, 
while $\epsilon$ is an equivariant parameter for  the spindle, that is a fugacity for the 
$U(1)_J$ rotational symmetry. In general, we expect that this will parameterise a non-trivial mixing of the $R$-symmetry of the parent theory, with the  $U(1)_J$ of the spindle. 
The off-shell free energy~\eqref{offshellFconjecture} 
bares a close resemblance to the off-shell central charge for D3-branes wrapped on spindle in~\cite{Hosseini:2021fge} and we shall elaborate on this in the next section.

Employing the parametrisation \eqref{defzspindle} we see that 
upon redefining $\chi\epsilon=\hat \epsilon$, the off-shell free energy  \eqref{offshellFconjecture} becomes $\Ispindle =\chi \cdot 
f(\Delta_i,\hat \epsilon;\mu,\z)$, implying that the free energy at the critical point must be a function of $\mu$ and $\z$ only, with an overall factor of $\chi$.
Notice that setting $n_+=n_-=1$ the present setup reduces to the case of twisted compactification on~$S^2$, that is the case $\mathrm{g}=0$ discussed in the previous subsection.

In order to implement the constraint 
\eqref{const12}  it is useful to introduce 
 a Lagrange multiplier and consider the extremization of the following function
\begin{equation}
{\cal S} (\Delta_i,\epsilon,\Lambda;\ts_i,n_+,n_-)  =  \Ispindle  (\Delta_i,\epsilon;\ts_i,n_+,n_-) + \Lambda (\Delta_1 + \Delta_2 - 2) \, ,
\end{equation}
that is analogous to the entropy functions studied in the literature.  The corresponding extremality equations read
 \begin{equation}
 \begin{split}
\Lambda+   \frac{1}{\epsilon} \left[ \frac{\partial\mc{F}(\Delta_i^+)}{\partial\Delta_j^+} - \frac{\partial\mc{F}(\Delta_i^-)}{\partial\Delta_j^-} \right]  &= 0 \,, \\
 \frac{\Ispindle}{\epsilon}- \frac{1}{\epsilon} \sum_{j=1}^2 \left[\left(\ts_j + \frac{n_+ - n_-}{2n_+n_-}\right) \frac{\partial\mc{F}(\Delta_i^+)}{\partial\Delta_j^+}  + \left(\ts_j -  \frac{n_+ - n_-}{2n_+n_-}\right)\frac{\partial\mc{F}(\Delta_i^-)}{\partial\Delta_j^-}  \right] &  = 0 \,, \\
 \Delta_1 + \Delta_2 - 2   & = 0 \, .
\end{split}
\end{equation}
These are four equations for the variables  $\Delta_1$, $\Delta_2$, $\epsilon$, $\Lambda$ and, in order to solve these, it is convenient to process them further. 
Noticing that~\eqref{offshellFconjecture} is homogeneous of degree two in~$\Delta_i$ and~$\epsilon$, by Euler's theorem we have
  \begin{equation}
\sum_{j=1}^2\Delta_j  \frac{\partial \Ispindle}{\partial \Delta_j} + \epsilon  
\frac{\partial \Ispindle}{\partial \epsilon} = 2  \Ispindle\, . 
\end{equation}
The extremization equations for ${\cal S}$   written as
  \begin{equation}
  \begin{aligned}
 \frac{\partial \Ispindle}{\partial \Delta_i}  +\Lambda  = 0\, ,  \quad \qquad \frac{\partial \Ispindle}{\partial \epsilon}  = 0\, , 
\end{aligned}
\end{equation}
then immediately imply $\Lambda = -\Ispindle$.
We can therefore eliminate $\Lambda$ from the system and
 write the remaining two  independent equations as
\begin{equation}
\begin{split}
 \frac{\partial\mc{F}(\Delta_i^+)}{\partial\Delta_1^+} - \frac{\partial\mc{F}(\Delta_i^-)}{\partial\Delta_1^-}  & =\frac{\partial\mc{F}(\Delta_i^+)}{\partial\Delta_2^+} - \frac{\partial\mc{F}(\Delta_i^-)}{\partial\Delta_2^-}  \,, \\
\label{multi-lag_Delta_alternativeplus}
\left(1+\frac{2\epsilon}{n_+} \right) \Ispindle &= 
2 \sum_{j=1}^2 \ts_j  \frac{\partial\mc{F}(\Delta_i^+)}{\partial\Delta_j^+} \, ,
\end{split}
\end{equation}
where one has to use also the constraint  $\Delta_1+\Delta_2=2$. Notice that taking $\epsilon\to 0$ the second equation reduces to the first equality in~\eqref{criceto}.

After some work,  we determined the critical values
\begin{equation}
\Delta_1^* = 1 + \frac{2\z x}{(x^2 + 3)(\mu - x)} \,,  \qquad  \quad \epsilon^* = \frac{1}{\chi} \, \frac{4x^2}{(x^2 + 3) (\mu - x)}\, , 
\label{criticalvalues}
\end{equation}
in terms of the parameter $x$, that is the unique root in the interval $(0,1)$ of the quartic~\eqref{eq_quartic} that we introduced in the discussion of the gravitational solution.
Inserting  these values back into \eqref{offshellFconjecture} we obtain
\begin{equation}
\Ispindle (\Delta_i^*,\epsilon^*;\z,\chi,\mu)  = \chi \, \frac{\sqrt{3}\pi N^{5/2}}{5\sqrt{8-N_f}} \, \frac{[3\mu (x^2+1) - x (x^2+5)]^{3/2}}{x (x^2+3) (\mu - x)^{1/2}} \, , 
\label{freeenergyreproduced}
\end{equation}
which, remarkably, agrees exactly with the gravitational free energy  \eqref{spindle_free-energy}!

To arrive at the solution \eqref{criticalvalues}, we first solved the extremality equations perturbatively in $\mu$ around $\mu=0$, obtaining agreement with the expansion \eqref{freeperturb}, up to high powers of~$\mu$. We then noticed that the result for $\epsilon^*$ could be rewritten as 
\begin{equation}
\epsilon^* = \frac{3}{4g} \frac{2\pi}{\Delta z} \,,
\end{equation}
which is  a universal relation holding in all previous spindle solutions 
\cite{Ferrero:2020laf,Hosseini:2021fge,Boido:2021szx,Ferrero:2021wvk,Ferrero:2020twa,Ferrero:2021ovq,Couzens:2021rlk}. 
Using this, we then obtained the result for $\Delta_1^*$ in \eqref{criticalvalues}.

As discussed in section~\ref{newsolutions_section}, in the limit  $n_+=n_-=1$ \eqref{freeenergyreproduced} reproduces the free energy~\eqref{gatto} for $\mathrm{g}=0$, corresponding to the compactification of the $d=5$ 
theory on the two-sphere, with the standard topological twist. 
Moreover, expanding $\Delta_1^*$ in \eqref{criticalvalues} in series of $\mu$ around $\mu=0$ we find that 
\begin{equation}
\Delta_1^*=1 + \frac{1+ \sqrt{8\z^2 + 1}}{4\z} + {\cal O}(\mu^2)\, , 
\end{equation}
again in agreement with the two-sphere value given in~\eqref{cane}. 
It would be interesting to reproduce $\Delta_1^*$, $\Delta_2^*$ by computing the scaling dimensions of some supersymmetric probe D2-branes wrapped on calibrated two-cycles in the ten-dimensional geometry~\eqref{mIIAmetric}.

\section{Gravitational blocks for branes on spindles}
\label{blockssection}

The off-shell free energy that we discussed in the previous section may be regarded as a particular instance of a general class of \emph{off-shell free energies} $\Ispindle^\pm$, for twisted compactifications of $d$-dimensional theories on the spindle $\spindle$.
Below we will state our conjecture and we will then illustrate how it encapsulates and generalises various extremal functions discussed in the literature. 
From the constructions of M2, D3 and M5-branes wrapped on a spindle and the results we discussed so far in this paper, 
it has emerged that supersymmetry on a spindle can be preserved in two different ways, 
  that can be referred to  as \emph{twist} and \emph{anti-twist}. 
 These are   characterised by   two types of  background $R$-symmetry gauge field $A_R$, with fluxes given by
\begin{equation}
\frac{1}{2\pi} \int_{\spindle} \dd A_R = \frac{n_+ + \sigma n_-}{n_+ n_-} \,,
\end{equation}
where $\sigma=+1$ for the twist and $\sigma=-1$ for the anti-twist.  The sign $\sigma=+1$ corresponds to the choice we made in section~\ref{D4onspindlesection}, and that was also realised for M5-branes wrapped on the spindle in \cite{Ferrero:2021wvk}. 
The sign $\sigma=-1$ has been realised by the supergravity solutions for D3-branes~\cite{Ferrero:2020laf,Hosseini:2021fge,Boido:2021szx} and M2-branes~\cite{Ferrero:2020twa,Ferrero:2021ovq,Couzens:2021rlk}.
 Below  we will treat both cases simultaneously, with the understanding that not all cases may have a counterpart as gravity solutions.  
   In  \cite{Ferrero:2021etw} it is proved  that  these are the only two possible twists preserving supersymmetry on the spindle.

A large class of SCFTs in different dimensions are expected to be characterised by supersymmetric partition functions $Z_{\cal M}$, where ${\cal M}$ are rigid geometries comprising a metric on an appropriate curved space, the background gauge fields for the $R$-symmetry  and possibly other flavour symmetries. 
The associated free energy is defined by 
\begin{equation}
F_{{\cal M}}  \equiv -\log Z_{{\cal M}} 
\end{equation}
 and it is regarded as a function of the fugacities $\Delta_i$ associated to the Cartan subgroup of the continuous global  symmetry group of the theory.
 For theories with an Abelian $R$-symmetry, that we are interested in, the fugacities obey a
 constraint that we can always normalise  to be 
\begin{equation}
\sum_{i=1}^\di \Delta_i = 2\, ,
\label{deltaconstraintgeneral}
\end{equation}
where  $\di$  is the rank of the global symmetry group, of which the Abelian  $R$-symmetry is part. In general, these are complicated matrix models, which however simplify drastically in special limits, such as the large $N$ limit or Cardy-like limits, reducing to  simple local functions of the fugacities $\Delta_i$.  For example, in four dimensions the partition function corresponding to ${\cal M}=S^1\times S^3$ is the (refined) superconformal index and in either limits its logarithm is related to the trial central charge $a_4 (\Delta_i)$ of the theory. In all SCFTs possessing an Abelian $R$-symmetry, it has been either proved or conjectured that the exact superconformal $R$-symmetry is determined by extremizing these quantities.

  We conjecture  that for a general class of $d$-dimensional SCFTs compactified on the spindle, with either the twist or the anti-twist, 
the exact $R$-symmetry is determined by extremizing the following off-shell free energies
\begin{equation}
\Ispindle^\pm  (\Delta_i,\epsilon;\ts_i,n_+,n_-,\sigma) =  \frac{1}{\epsilon}\left( {\cal F}_d (\Delta_i^+ ) \pm   {\cal F}_d (\Delta_i^-)  \right)\, , 
\label{superspindleconjecture}
\end{equation}
where the variables $\Delta_i^+$, $\Delta_i^-$ are defined as 
 \begin{equation}
 \begin{split}
 \Delta_i^+  \equiv   \Delta_i  +     \epsilon \left(\ts_i+\frac{r_i}{2} \frac{n_+ - \sigma n_-}{n_+n_-}\right)\, , \qquad \Delta_i^-   \equiv  \Delta_i -   \epsilon \left(\ts_i -\frac{r_i}{2} \frac{n_+ - \sigma n_-}{n_+n_-}\right) \,, 
 \end{split}
\end{equation}
and the magnetic fluxes through the spindle, $\ts_i$, satisfy the constraint 
\begin{equation}
   \sum_{i=1}^\di  \ts_i= \frac{n_++\sigma n_-}{n_+n_-}\, . 
\label{spindleconst12}
\end{equation}
The building blocks are the functions ${\cal F}_d (\Delta_i)$ summarised in Table~\ref{tab:F-block}. 
They  are proportional to: 
the $S^3$ off-shell free energy of the ABJM theory, the trial central charge of the  ${\cal N}=4$ SYM theory, the $S^5$ off-shell free energy of the $d=5$, ${\cal N}=1$ SCFT and the trial central charge of the $d=6$, $(2,0)$  SCFT, respectively. In the first two cases, it is straightforward to replace these with the corresponding quantities for more general $d=3$, ${\cal N}=2$ theories and $d=4$, ${\cal N}=1$ theories, but in this paper we will not pursue this.

\begin{table}[h]
\centering
\begin{tabular}{ | c || C{3.25cm} | C{3.25cm} | C{3.25cm} | C{3.25cm} | }
\hline
& $d=3$ & $d=4$ & $d=5$ & $d=6$ \\
\hline\hline
&&&& \\[\dimexpr-\normalbaselineskip+0.25em]
$\mc{F}_d$ & $\kk_3 (\Delta_1\Delta_2\Delta_3\Delta_4)^{1/2}$ & $\kk_4 (\Delta_1\Delta_2\Delta_3)$ & $\kk_5 (\Delta_1\Delta_2)^{3/2}$ & $\kk_6 (\Delta_1\Delta_2)^2$ \\
\hline
&&&& \\[\dimexpr-\normalbaselineskip+0.2em]
\multirow{2}{*}{$\kk_d$} & $-\frac{\sqrt2 \pi }{3} N^{3/2}$ & $ -\frac{3}{2}N^2 $ & $-\frac{2^{5/2} \pi }{15} \frac{N^{5/2}}{\sqrt{8-N_f}}$ & $-\frac{9}{256}N^3$ \\
& $-F_{S^3}$ & $ -6a_{4}$ & $\frac{4}{27} F_{S^5}$ & $-\frac{63}{4096} a_{6}$ \\[0.3em]
\hline
$\pm$ & $-\sigma$ & $-$ & $-\sigma$ & $-$ \\
\hline
\end{tabular}
\caption{Various  $\mc{F}_d(\Delta_i)$ in  different spacetime dimension~$d$. The constants~$\kk_d$ are  given in terms of the free energy on~$S^d$ ($d=3,5$) or the central charge $a_d$ ($d=4,6$).
In $d=6$ and $d=5$ the rank of the global symmetry group is $\di=2$, in $d=4$ it is $\di=3$,
 while in $d=3$ it is $\di=4$. 
In the last row we summarised the relations between the gluing sign $\pm$ and the sign $\sigma$, characterising the  type of twist.}
\label{tab:F-block}
\end{table}

We expect  that the form \eqref{superspindleconjecture} arises, in the large $N$ limit,  from a fixed point formula, with the ``blocks''  ${\cal F}_d (\Delta_i)$ evaluated at the north and south poles of the spindle. 
The superscripts in $\Ispindle^\pm$ refer to the relative choice of sign $\pm$ in \eqref{superspindleconjecture},  corresponding to the 
  type of gluing of the contributions of the two hemispheres of the spindle \cite{Hosseini:2019iad}.  
    We will give circumstantial  evidence that in $D=4,6$ the type of gluing is correlated with the type of twist, specifically that the gluing sign is $-\sigma$. On the other hand, in $D=5,7$ the results of 
  the explicit supergravity 
  solutions are all reproduced by the minus gluing sign.  
In the examples that we discuss below, we will explain which choice of twist and gluing  is relevant, but a more systematic  understanding of these choices is clearly desirable.

The variable $\epsilon$ is a fugacity associated to the $U(1)_J$ rotational symmetry of the spindle and the significance of the fact that at the critical point of \eqref{superspindleconjecture} this takes a non-zero value is that the $R$-symmetry of the parent $d$-dimensional theory mixes with $U(1)_J$ to give the exact superconformal $R$-symmetry of the $(d-2)$-dimensional theory arising in the IR, when this flows to  an SCFT. 
The constants $r_i$ are arbitrary, but subject to the constraint   
\begin{equation}
 \sum_{i=1}^\di r_i = 2\, ,
 \label{constraintgeneral}
\end{equation}
and parameterise the ambiguities of defining flavour symmetries
\cite{Hosseini:2021fge}. In the  previous section we picked the most symmetric choice, corresponding to $r_i=\frac{\di}{2}=1$. However, it is simple to show that the functions~\eqref{superspindleconjecture} evaluated at the critical point are independent of the choice of $r_i$. 
Introducing a new  set of variables defined as 
\begin{equation}
\varphi_i \equiv  \Delta_i + \frac{r_i}{2} \frac{n_+ - \sigma n_-}{n_+n_-}\epsilon\, , 
\end{equation}
the off-shell free energies simply read
\begin{equation}
\Ispindle^\pm(\varphi_i,\epsilon;\ts_i) = \frac{1}{\epsilon} \bigl( \mc{F}_d (\varphi_i + \ts_i \epsilon) \pm \mc{F}_d (\varphi_i - \ts_i \epsilon) \bigr)\, , 
\label{superspindleconjecturebetter}
\end{equation}
where, from now on, we will omit $n_+$, $n_-$, $\sigma$ from the arguments of the function, in order not to clutter the subsequent formulas. 
The variables  $\varphi_i,\epsilon$  satisfy the constraint
\begin{equation}
\sum_{i=1}^\di  \varphi_i -   \frac{n_+ - \sigma n_-}{n_+n_-}\epsilon =2  \, ,
\label{constraintnice}
\end{equation}
inherited from~\eqref{deltaconstraintgeneral} and~\eqref{constraintgeneral}. Since \eqref{superspindleconjecturebetter} does not depend on the constants $r_i$, it follows that 
the critical values $\varphi_i^*$, $\epsilon^*$ and $\Ispindle^\pm(\varphi_i^*,\epsilon^*;\ts_i)$ do not depend on the $r_i$ either.

Our proposal unifies previous proposals concerning entropy functions   \cite{Hosseini:2019iad,Hosseini:2020mut,Ferrero:2021ovq,Hosseini:2020wag} and central charges \cite{Hosseini:2021fge},
extending these to compactifications of $d$-dimensional theories on spindles with both twist and anti-twist. 
Below we shall illustrate this, recovering known results and discussing some generalisations. 
The constrained extremization problem  can be carried  out introducing a Lagrange multiplier.  Defining 
\begin{equation}
\mc{S}^\pm(\varphi_i,\epsilon,\Lambda;\ts_i,n_+,n_-) = \Ispindle^\pm(\varphi_i,\epsilon;\ts_i,n_+,n_-) + \Lambda \left( \sum_{j=1}^\di \varphi_j - \frac{n_+ - \sigma n_-}{n_+ n_-}\epsilon - 2 \right) \,,
\end{equation}
and using the fact that \eqref{superspindleconjecture} is homogeneous of degree\footnote{$h=1$ for $d=3$, $h=2$ for $d=4,5$ and $h=3$ for $d=6$.} $h$ in $\varphi_i$ and $\epsilon$, by means of Euler's theorem we have 
that $\Lambda = -\frac{h}{2} \Ispindle^\pm$, and we can therefore  eliminate $\Lambda$ from the system. The resulting extremization equations can be written as 
\begin{equation}
\begin{split}
\frac{h \epsilon}{2} \Ispindle^\pm &= \frac{\partial\mc{F}_d(\varphi_i^+)}{\partial\varphi_j^+} \pm \frac{\partial\mc{F}_d(\varphi_i^-)}{\partial\varphi_j^-} \,,  \qquad\qquad  (j=1,\ldots,\di) \\
\left(1 + \frac{\sigma h \epsilon}{n_+} \right) \Ispindle^\pm &= 
2 \sum_{j=1}^\di \ts_j  \frac{\partial\mc{F}_d(\varphi_i^+)}{\partial\varphi_j^+} \, ,
\end{split}
\end{equation}
where $\varphi_i^\pm \equiv \varphi_i \pm \ts_i \epsilon$. Notice that the last equation can  be replaced by
\begin{align}
\left(1-\frac{h \epsilon}{n_-} \right) \Ispindle^\pm &= 
\mp 2 \sum_{j=1}^\di \ts_j  \frac{\partial\mc{F}_d(\varphi_i^-)}{\partial\varphi_j^-} \, .
\end{align}


\subsection{M2-branes}

Supergravity solutions describing M2-branes wrapped on the spindle were first constructed in~\cite{Ferrero:2020twa} in minimal $D=4$ gauged supergravity and generalised to $U(1)^2$ gauged supergravity in~\cite{Ferrero:2021ovq,Couzens:2021rlk}. They realise the anti-twist, $\sigma=-1$. 
The corresponding dual field theory is the ABJM model compactified on the spindle, with two background gauge fields with magnetic fluxes
\begin{equation}
  \ts_1+   \ts_2= \frac{n_+ + \sigma  n_-}{2n_+n_-}\, ,
\end{equation}
 for $\sigma = -1$. 
However, it is straightforward to carry out the extremization leaving the twist unspecified. Picking the plus gluing sign in~\eqref{superspindleconjecturebetter}  gives
\begin{equation}
\Ispindle^+ (\varphi_i,\epsilon;\ts_i)   = - \frac{2\sqrt{2}\pi N^{3/2}}{3}\, \left(\ts_1 \ts_2 \epsilon+\frac{\varphi _1 \varphi _2}{\epsilon}\right)\, , 
\label{IM2varphi}
\end{equation}
with $\varphi_i,\epsilon$ satisfying the constraint
 \begin{equation}
  \varphi_1 + \varphi_2 -   \frac{n_+ - \sigma n_-}{2n_+n_-}\epsilon =1  \, .
   \label{M2varphiconstN4}
  \end{equation}
For $\sigma=-1$   this is exactly the entropy function proposed in \cite{Ferrero:2021ovq}, in the case of vanishing electric charges and angular momentum.  
Performing the extremization of  \eqref{IM2varphi}, subject to \eqref{M2varphiconstN4}, we get
 \begin{equation}
 \begin{split}
\varphi _1^* &=  \frac{-n_+ + \sigma n_- + \sqrt{16 n_-^2 n_+^2 \ts_1 \ts_2+(n_+-n_- \sigma )^2}}{2 \sqrt{16 n_-^2 n_+^2 \ts_1 \ts_2+(n_+-n_- \sigma )^2}}\, ,\\
\epsilon^* & =  -\frac{2 n_- n_+}{\sqrt{16 n_-^2 n_+^2 \ts_1 \ts_2+(n_+- \sigma n_-)^2}}\, ,
\end{split}
\end{equation}
and inserting these values back in \eqref{IM2varphi} we find
\begin{equation}
\Ispindle^+  (\varphi_i^*,\epsilon^*;\ts_i) = \frac{\sqrt{2}\pi N^{3/2}}{3} \, \frac{ -n_+ +  \sigma n_-  + \sqrt{16 n_+^2 n_-^2 \ts_1 \ts_2+(n_+ - \sigma n_-)^2} }{2 n_+ n_- }\, ,
\end{equation}
which coincides with the entropy in \cite{Ferrero:2021ovq} upon setting $\sigma=-1$. 
In minimal $D=4$ gauged supergravity  the entropy function \eqref{IM2varphi}, for $\varphi_1=\varphi_2$ and $\ts_1=\ts_2$, was derived in \cite{Cassani:2021dwa}, 
and its plus gluing sign is consistent with the fact that it indeed arises as the sum of contributions from the north and south poles of the spindle, which are the fixed points of the canonical Killing vector field \cite{BenettiGenolini:2019jdz}.

The precise map between the variables here and those used in \cite{Ferrero:2021ovq} is as follows 
\begin{equation}
 P_i^\mathrm{there} = 2 \ts_i , \qquad Q^\mathrm{there}=0 \, ,\qquad \varphi_i^\mathrm{there}= \pm \frac{\pi \ii}{2} \varphi_i^\mathrm{here}   \, , \qquad \omega^\mathrm{there} = \pm 2\pi \ii  \epsilon  \, , 
\end{equation}
and using this, the constraint   \eqref{M2varphiconstN4} becomes 
\begin{equation}
2 (\varphi_1^\mathrm{there} + \varphi_2^\mathrm{there}) - \frac{\chi}{4} \omega^\mathrm{there} = \pm \ii \pi \,.
\label{M2varphiconstOLD}
\end{equation}
Note that with vanishing angular momentum and electric charges the entropy function is purely real (or purely imaginary) and therefore the critical points are purely real.

For the other type of gluing we obtain (for either choice of $\sigma$) the function 
\begin{equation}
\Ispindle^-(\varphi_i,\epsilon;\ts_i) = - \frac{2\sqrt{2}\pi N^{3/2}}{3} \left(\ts_2 \varphi _1+\ts_1 \varphi _2\right) \,,
\end{equation}
whose extremization, however, leads to a degenerate result.

The entropy function for the general four-charge model, for either type of gluing and either type of twist, reads 
 \begin{equation}
\begin{split}
\label{topolino}
\Ispindle^\pm  (\varphi_i,\epsilon;\ts_i)  & =- \frac{\sqrt{2}\pi N^{3/2}}{3\epsilon}\bigg(\sqrt{\left( \varphi _1+\ts_1 \epsilon \right) \left(\varphi _2+\ts_2 \epsilon \right) \left(\varphi _3+\ts_3 \epsilon \right) \left(\varphi _4+\ts_4 \epsilon\right)}\\
& \qquad \qquad  \qquad \pm\sqrt{\left(\varphi _1-\ts_1 \epsilon \right) \left(\varphi _2-\ts_2 \epsilon \right) \left(\varphi_3-\ts_3 \epsilon\right) \left(\varphi_4 - \ts_4 \epsilon \right)}\bigg)\, . 
\end{split}
\end{equation}
The magnetic fluxes obey
\begin{equation}
  \ts_1+   \ts_2+  \ts_3+ \ts_4= \frac{n_++\sigma n_-}{n_+n_-}\, ,
\end{equation}
and the variables $\varphi_i,\epsilon$ satisfy the constraint
\begin{equation}
\varphi_1 + \varphi_2 + \varphi_3 + \varphi_4 - \frac{n_+ - \sigma n_-}{n_+ n_-} \epsilon = 2 \,.
\label{M2varphiconst}
\end{equation}
For $n_+=n_-=1$ these reduce to the entropy functions proposed  in \cite{Hosseini:2019iad}. 
In particular,  for $\sigma=+1$, $F^-$ reduces to the entropy function of the supersymmetric AdS$_4$ black holes with a topological twist~\cite{Cacciatori:2009iz}, whereas for  $\sigma=-1$, $F^+$  reduces to the entropy function for the supersymmetric rotating  Kerr-Newmann  AdS$_4$ black holes~\cite{Hristov:2019mqp}. More generally, it is natural to expect  that the correct gluing for either type of twists is $-\sigma$, as reported in Table~\ref{tab:F-block}. It would be nice to corroborate this proposal showing that extremizing the entropy functions $F^{-\sigma}$ in \eqref {topolino} reproduces the entropy of the AdS$_2\times \spindle$ solutions of STU gauged supergravity.

\subsection{D3-branes}

Supergravity solutions describing D3-branes wrapped on the spindle were first constructed in~\cite{Ferrero:2020laf} in minimal $D=5$  gauged supergravity and generalised to $U(1)^3$ gauged supergravity in~\cite{Hosseini:2021fge,Boido:2021szx}. They realise the anti-twist, $\sigma=-1.$
The corresponding dual field theory is ${\cal N}=4$ SYM compactified on the spindle, with three background gauge fields with magnetic fluxes satisfying
\begin{equation}
  \ts_1+   \ts_2+  \ts_3= \frac{n_++\sigma n_-}{n_+n_-}\, ,
\end{equation}
for $\sigma=-1$.
However, it is straightforward to carry out the extremization leaving the twist unspecified.
 In this case we find that, for \emph{both} types of twist, the correct gluing
 sign that reproduces all known results 
  is the lower sign in \eqref{superspindleconjecturebetter}, which gives
\begin{equation}
\Ispindle^-  (\varphi_i,\epsilon;\ts_i) =  - 3 N^2 \left(\ts_1\varphi _2 \varphi _3+ \varphi _1\ts_2 \varphi _3+\varphi _1  \varphi _2\ts_3 +\ts_1 \ts_2 \ts_3 \epsilon ^2\right)\, , 
\label{ED3varphi}
\end{equation}
with $\varphi_i$, $\epsilon$ satisfying the constraint
\begin{equation}
  \varphi_1 + \varphi_2+ \varphi_3 -   \frac{n_+ - \sigma n_-}{n_+n_-}\epsilon =2  \, .
  \label{D3varphiconst}
\end{equation}
Extremizing  \eqref{ED3varphi} subject to the constraint \eqref{D3varphiconst} we find 
\begin{equation}
 \begin{split}
\varphi _1^* & =  \frac{\ts_1 \left(\ts_1-\ts_2-\ts_3\right)}{2  \big(\frac{\sigma}{n_+n_-} - (\ts_1\ts_2+\ts_1\ts_3+\ts_2\ts_3)\big)}\, ,\\
\varphi _2^* & =  \frac{\ts_2 \left(\ts_2-\ts_3-\ts_1\right)}{2  \big(\frac{\sigma}{n_+n_-} - (\ts_1\ts_2+\ts_1\ts_3+\ts_2\ts_3)\big)}\, ,\\
\epsilon^* & =  \frac{\frac{ n_+-\sigma n_-}{n_+n_-} }{2  \big(\frac{\sigma}{n_+n_-}  -(\ts_1\ts_2+\ts_1\ts_3+\ts_2\ts_3)\big)}\, , 
\end{split}
\end{equation}
and inserting these back in \eqref{ED3varphi} we get
\begin{equation}
\Ispindle^-  (\varphi_i^*,\epsilon^*;\ts_i)  =   \frac{3 N^2 \ts_1 \ts_2\ts_3}{ \frac{\sigma}{n_+n_-}- (\ts_1\ts_2+\ts_1\ts_3+\ts_2\ts_3) }\, . 
\label{ED3nice}
\end{equation}
Setting $\sigma=-1$ this reduces to the $a_2$ central charge obtained in \cite{Hosseini:2021fge,Boido:2021szx}.
On the other hand, setting $\sigma=+1$ we reproduce also the central charge for D3-branes wrapped on the spindle with the twist, for which supergravity solutions were recently presented in \cite{Ferrero:2021etw}.
Further setting $n_+=n_-=1$, \eqref{ED3nice} reduces to the result for D3-branes wrapped on $S^2$ with the standard topological twist~\cite{Benini:2013cda}. 
Equivalently, we can extremize over the variables $\Delta_i$, $\epsilon$, with $\Delta_1+\Delta_2+\Delta_3=2$ obtaining the same result as in~\eqref{ED3nice} for the critical central charge. However, the variables $\Delta_i$ are affected by the ambiguity related to the choice of the constants $r_i$.

Notice that for the other type of gluing  we obtain (for either choice of $\sigma$) the function 
\begin{equation}
\Ispindle^+  (\varphi_i,\epsilon;\ts_i)  =  - 3 N^2 \left[  \left(\ts_1 \varphi _2\ts_3 + \varphi _1\ts_2 \ts_3 +\ts_1 \ts_2 \varphi _3\right)\epsilon +\frac{ \varphi _1 \varphi _2 \varphi _3}{\epsilon }\right]\, . 
\end{equation}
It may be possible that this corresponds to a different type of twisted compactification of D3-branes on the spindle with a corresponding class of supergravity constructions.

\subsection{D4-branes}

Let us return to the D4-branes and write down the main ingredients involved in considering 
 simultaneously the case of twist and anti-twist.
 The magnetic fluxes satisfy
\begin{equation}
  \ts_1+   \ts_2=  \frac{n_++\sigma n_-}{n_+n_-}\, . 
\end{equation}
In terms of the variables $\varphi_i,\epsilon$ satisfying the constraint
\begin{equation}
  \varphi_1 + \varphi_2 -   \frac{n_+ - \sigma  n_-}{n_+n_-}\epsilon =2  \, ,
  \label{D4varphiconst}
\end{equation}
the off-shell free energies read
\begin{equation}
\Ispindle^\pm  (\varphi_i,\epsilon;\ts_i) = -\frac{4\sqrt{2} \pi}{15\epsilon}\frac{N^{5/2}}{\sqrt{8-N_f}} \Bigl[ \bigl((\varphi _1 + \ts_1 \epsilon) (\varphi_2 + \ts_2 \epsilon)\bigr)^{3/2} \pm \bigl((\varphi _1 - \ts_1 \epsilon) (\varphi _2 - \ts_2 \epsilon)\bigr)^{3/2} \Bigr] \,.
\label{ID4}
\end{equation}

As for the M2-branes, we expect that the correct gluing for either type of twists is $-\sigma$, as reported in Table~\ref{tab:F-block}. 
Extremizing $\Ispindle^-(\varphi_i,\epsilon;\ts_i)$, subject to the constraint~\eqref{D4varphiconst}, we find (leaving $\sigma$ unspecified)
\begin{equation}
\begin{aligned}
\varphi_1^* &= 1 + \frac{2x [(n_+ - \sigma n_-)x + n_+ n_- (\ts_1 - \ts_2)]}{(x^2 + 3)[(n_+ - \sigma n_-) - (n_+ + \sigma n_-)x]} \,, \\
\epsilon^* & = \frac{4n_+ n_- x^2}{(x^2 + 3) [(n_+ - \sigma n_-) - (n_+ + \sigma n_-)x]} \,,
\end{aligned}
\end{equation}
where $x$ is the only solution in the interval $(0,1)$ of the quartic equation
\begin{equation}
\begin{split}
& (n_+ + \sigma n_-)^2 x^4 + 4\bigl[ 2n_+^2 n_-^2 (\ts_1-\ts_2)^2 - 3(n_+^2 + n_-^2 - \sigma n_+ n_-) \bigr] x^2 \\
&\qquad + 12(n_+^2 - n_-^2) x - 9(n_+ - \sigma n_-)^2 = 0 \,.
\end{split}
\end{equation}
Inserting the values of~$\varphi_i^*$ and~$\epsilon^*$ back  in~\eqref{ID4} we get
\begin{equation}
\Ispindle^-(\varphi_i^*,\epsilon^*;\ts_i) = \frac{\sqrt3 \pi}{5} \frac{N^{5/2}}{\sqrt{8-N_f}} \frac{[3(n_+ - \sigma n_-) (x^2 + 1) - (n_+ + \sigma n_-) x (x^2 + 5)]^{3/2}}{n_+ n_- x (x^2 + 3) [(n_+ - \sigma n_-) - (n_+ + \sigma n_-)x]^{1/2}} \,,
\end{equation}
which  reduces to the results of the previous section for $\sigma=+1$.

We expect the choice of  minus gluing sign to be correlated to the fact that in our solutions the Killing spinors on the spindle have the same chirality at the north and south poles. It would be interesting to investigate the extremization of  $F^{+}$ and to find out whether for $\sigma=-1$ it has a physical critical point, corresponding to $\AdS_4\times\spindle$ solutions with the anti-twist, yet to be constructed.

\subsection{M5-branes}

Supergravity solutions describing M5-branes wrapped on the spindle were  constructed in~\cite{Ferrero:2021wvk} in a $D=7$, $U(1)^2$ gauged supergravity model. They realise the twist, $\sigma=+1$.
The corresponding dual field theory is the $d=6$, $(2,0)$ SCFT compactified on the spindle, with two background gauge fields with magnetic fluxes 
\begin{equation}
  \ts_1+   \ts_2  = \frac{n_++\sigma n_-}{n_+n_-}\, ,
\end{equation}
for $\sigma=+1$. 
However, it is  straightforward to carry out the extremization  for either type of twist. 
In this case we find that the correct gluing sign that reproduces the results of~\cite{Ferrero:2021wvk} 
is the lower sign in \eqref{superspindleconjecturebetter},  which gives
\begin{equation}
\Ispindle^-  (\varphi_i,\epsilon;\ts_i)  = - \frac{9}{64}  N^3 \left(\ts_2 \varphi _1+\ts_1 \varphi _2\right) \left(\ts_1 \ts_2 \epsilon ^2+\varphi _1 \varphi _2\right)\, , 
 \label{IM5varphi}
\end{equation}
with $\varphi_i,\epsilon$ satisfying the constraint
\begin{equation}
  \varphi_1 + \varphi_2 -   \frac{n_+ - \sigma  n_-}{n_+n_-}\epsilon =2  \, .
  \label{M5varphiconst}
\end{equation}
Performing the extremization of \eqref{IM5varphi}, subject to the constraint \eqref{M5varphiconst}, we find 
\begin{equation}
\begin{aligned}
\varphi_1^* &= 1 + \frac{(\mathtt{s} + \ts_1 + \ts_2) [2n_+^2 n_-^2 (\ts_1^2 - \ts_2^2) + 3(n_+ - \sigma n_-)^2 + n_+^2 n_-^2 (\ts_1 - \ts_2) \mathtt{s}]}{12n_+ n_- (\sigma - n_+ n_- \ts_1 \ts_2) [\mathtt{s} + 2(\ts_1 + \ts_2)]} \,, \\
\epsilon^* &= \frac{(n_+ - \sigma n_-) (\mathtt{s} + \ts_1 + \ts_2)}{2(\sigma - n_+ n_- \ts_1 \ts_2) [\mathtt{s} + 2(\ts_1 + \ts_2)]} \,,
\end{aligned}
\label{critvaluesM5}
\end{equation}
where
\begin{equation}
\mathtt{s} \equiv \sqrt{7(\ts_1^2 + \ts_2^2) + 2\ts_1 \ts_2 - 6 \, \frac{n_+^2 + n_-^2}{n_+^2 n_-^2}} \, ,
\end{equation}
and inserting these values back in \eqref{IM5varphi} we get
\begin{equation}
\Ispindle^-  (\varphi_i^*,\epsilon^*;\ts_i) = \frac{3N^3}{8} \frac{\ts_1^2 \ts_2^2 \left(\mathtt{s} + \ts_1 + \ts_2\right)}{(\frac{\sigma}{n_+n_-} -  \ts_1 \ts_2) [\mathtt{s} + 2(\ts_1 + \ts_2)]^2} \,.
\end{equation}
Setting $\sigma=+1$ this reduces to the $a_4$ central charge obtained in~\cite{Ferrero:2021wvk}  integrating  the M5-brane anomaly polynomial on the spindle.
However, in~\cite{Ferrero:2021wvk} the extremization was carried out over the variables $\Delta_i$, subject to $\Delta_1+\Delta_2=2$, and it was found that the critical values are given by $\Delta_1^*=\Delta_2^*=1$, while the critical $\epsilon^*$ coincides with the one given in \eqref{critvaluesM5}, up to a convention-dependent factor,
specifically, $\epsilon^*_\mathrm{here}=-\frac{1}{2} \epsilon^*_\mathrm{there}$.
One can check that the variables $\Delta_i$ utilised in~\cite{Ferrero:2021wvk} correspond (for $\sigma=+1$) to the choice of constants $r_1=2-r_2$ given by 
\begin{equation}
r_1= 1 + \frac{n_+^2 n_-^2 (\ts_1 - \ts_2) [\mathtt{s} + 2(\ts_1 + \ts_2)]}{3(n_+ -  \sigma n_-)^2} \,.
\end{equation}
It would be interesting  to find out whether the critical points for  $\sigma=-1$ correspond  to AdS$_5\times \spindle$ solutions with the anti-twist, yet to be constructed.

We note that, as for the case of AdS$_4\times \Sigma_{\mathrm{g}}$ solutions,  the extremization of the off-shell free energy~\eqref{IM5varphi}  reproduces also the 
central charge\footnote{For $\mathrm{g}=0$, \eqref{IM5varphi}  can also be viewed as the off-shell free energy on $S^3\times S^2_\epsilon$ of ${\cal N}=2$ SYM in $d=5$. Using this, the form  
\eqref{IM5varphi}  has been  recently proved in  \cite{Hosseini:2021mnn} for  $S_b^3\times S^2_\epsilon$.}
 of $d=4$ SCFTs dual to the $\AdS_5\times\Sigma_\mathrm{g}$ solutions~\cite{Bah:2012dg}. As anticipated in footnote \ref{recoversigma}, 
in this case the variables~$\varphi_i$ and the magnetic fluxes are subject to the constraints
\begin{equation}
\varphi_1 + \varphi_2 = 2 \,, \quad \qquad \ts_1 + \ts_2  = 2(1 - \mathrm{g}) \,,
\end{equation}
and as usual the latter may be   parameterised as
\begin{equation}
\ts_1 = (1 - \mathrm{g}) \left(1 + \z\, \kappa \right) \,,  \qquad  \ts_2 = (1 - \mathrm{g}) \left(1 - \z\, \kappa \right) \,.
\end{equation}
The free energy~\eqref{IM5varphi} is extremized by
\begin{equation}
\varphi_1^* = 1 + \frac{\kappa + \sqrt{\kappa^2 + 3\z^2}}{3\z} \,,  \qquad  \epsilon^* = 0 \,,
\end{equation}
to which corresponds the critical value \cite{Bah:2012dg} 
\begin{equation}
\Ispindle^-(\varphi_i^*,0;\ts_i) = (1 - \mathrm{g}) N^3 \, \frac{\kappa^2 - 9\z^2 + \kappa (\kappa^2 + 3\z^2)^{3/2}}{48\z^2} \, .
\label{bababa}
\end{equation}
Since $\epsilon^*=0$, we could have started setting $\epsilon=0$ in \eqref{IM5varphi}, thus reducing to the known $\epsilon$-independent off-shell free energy
\begin{equation}
\lim_{\epsilon\to0} \Ispindle^-  (\varphi_i,\epsilon;\ts_i)
 =2 \sum_{j=1}^2 \ts_j  \frac{\partial {\cal F}_6(\varphi_i) }{\partial \varphi_j}  
 = - \frac{9N^3}{64} \varphi _1 \varphi _2\left(\ts_2 \varphi _1+\ts_1 \varphi _2\right)  \,, 
\label{geko}
\end{equation}
corresponding to the standard topological twist.
In particular, extremizing~\eqref{geko} reproduces~\eqref{bababa}.
For $\mathrm{g}>1$ we can set $\z=0$ so that $\varphi_1^*=\varphi_2^*=1$ and the free energy~\eqref{geko} reduces to the universal relation~\cite{Bobev:2017uzs}
\begin{equation}
\Ispindle^- (\varphi_i^*,0; \ts_1=\ts_2=1-\mathrm{g})   =   \frac{9N^3}{32} (\mathrm{g}-1)=  \frac{63}{512} (\mathrm{g} - 1) a_6\, .
\end{equation}

Notice that for the other type of gluing  we obtain (for either choice of $\sigma$) the function 
\begin{equation}
\Ispindle^+  (\varphi_i,\epsilon;\ts_i) = -\frac{9}{128}N^3 \left(\ts_1^2  \ts_2^2 \epsilon^3+ \left(\ts_1^2\varphi _2^2+\ts_2^2 \varphi _1^2 +4 \ts_1 \ts_2 \varphi _1 \varphi _2 \right)\epsilon+\frac{\varphi _1^2\varphi _2^2}{\epsilon}\right)\, . 
\end{equation}
It may be possible that this corresponds to a different type of twisted compactification of M5-branes on the spindle with a corresponding class of supergravity constructions.

\section{D4-branes wrapped on $\spindle\times\Sigma_\mathrm{g}$}
\label{ads2solution}

In this section we will begin investigating constructions corresponding to D4-branes wrapped on orbifolds of dimension higher than two, focussing on a class of  explicit solutions that can be easily obtained uplifting to massive type IIA supergravity the AdS$_2$ solutions of four-dimensional minimal gauged supergravity of \cite{Ferrero:2020twa}.  Generically, for four-dimensional orbifolds $\mathbbl{M}_4$,  the field theories are $d=1$ SCQMs
 obtained from twisted compactifications of the $d=5$, ${\cal N}=1$ $USp(2N)$ gauge theory. 
The off-shell free energies are in this case  entropy functions, whose extremization determines the entropy of supersymmetric AdS$_6$ black holes with AdS$_2\times \mathbbl{M}_4$ near-horizon geometry. 
As we have discussed,  we expect that these will take the form of a sum of gravitational blocks over the set of fixed points of the canonical Killing vector,
which includes also the non-orbifold geometries as special cases. In particular, the entropy function for the product of two constant-curvature 
Riemann surfaces $\Sigma_{\mathrm{g}_1}\times  \Sigma_{\mathrm{g}_2}$ with the standard topological twist may be recovered considering\footnote{The overall factor of $-\tfrac{1}{4}$ may be fixed by splitting 
 the compactification on $\Sigma_{\mathrm{g}_1}\times  \Sigma_{\mathrm{g}_2}$ in two steps, first reducing from $d=5$ to $d=3$ and then from $d=3$ to $d=1$. This factor is then consistent with the 
 rules summarised in 
 Table \ref{tab:F-block}.} 
\begin{equation}
\label{Ssigma12}
\begin{split}
S & = -\frac{1}{4\epsilon_1\epsilon_2}\Big[  \mc{F}_5 (\varphi_i + \ts_i \epsilon_1 +   \tr_i \epsilon_2) - \mc{F}_5 (\varphi_i - \ts_i \epsilon_1 +   \tr_i \epsilon_2)  \\
& \qquad \qquad -  \mc{F}_5 (\varphi_i + \ts_i \epsilon_1 -   \tr_i \epsilon_2) + \mc{F}_5 (\varphi_i - \ts_i \epsilon_1 -   \tr_i \epsilon_2) \Big]\, , 
\end{split}
\end{equation}
subject to the constraints
\begin{equation}
     \ts_1 + \ts_2 = 2 (1-\mathrm{g}_1)\, , \qquad   \tr_1 + \tr_2 =  2 (1-\mathrm{g}_2) \, , \qquad \varphi_1+ \varphi_2  =2  \, .
\end{equation}
 Extremizing \eqref{Ssigma12} with respect to $\epsilon_1$ and $\epsilon_2$ sets  $\epsilon_1=\epsilon_2=0$ so that the entropy function reduces to 
\cite{ Hosseini:2018uzp}
 \begin{equation}
\begin{split}
\lim_{\epsilon_1,\epsilon_2 \to 0} S(\varphi_i,\epsilon_{1},\epsilon_2; \ts_i,\tr_i) &
=  -\sum_{j,k=1}^2 \ts_j  \tr_k \, \frac{\partial^2 {\cal F}_5(\varphi_i) }{\partial \varphi_j\partial \varphi_k} \\
&= \frac{4\sqrt{2} \pi }{15} \frac{N^{5/2}}{\sqrt{8-N_f}} \sum_{j,k=1}^2 \ts_j \tr_k \frac{\partial^2(\varphi_1\varphi_2)^{3/2}}{\partial \varphi_j\partial \varphi_k} \, . 
\end{split}
\end{equation}
Extremizing this with respect to $\varphi_i$ reproduces the entropy of an associated  class of AdS$_2\times \Sigma_{\mathrm{g}_1}\times  \Sigma_{\mathrm{g}_2}$ supersymmetric solutions 
\cite{Hosseini:2018usu}. Below we will discuss the case of $\mathbbl{M}_4=\spindle \times \Sigma_\mathrm{g}$, leaving $\mathbbl{M}_4=\spindle_1 \times \spindle_2$ and more general orbifolds for future work.
Analogous solutions, corresponding to M5-branes wrapped on $\spindle \times \Sigma_\mathrm{g}$, were presented in \cite{Boido:2021szx}.

\subsection{AdS$_2\times\spindle\times\Sigma_{\mathrm{g}}$ solutions}
\label{secads2}

A class of supersymmetric $\AdS_2\times\spindle\times\riemann$ backgrounds may be easily obtained by lifting the $\AdS_2\times\spindle$ solutions of four-dimensional minimal gauged supergravity constructed in~\cite{Ferrero:2020twa} to $D=6$ matter-coupled gauged supergravity, using the consistent truncation presented in \cite{Hosseini:2020wag}. In our conventions, the bosonic part of the action 
reads
\begin{equation} \label{U13sugra}
\begin{split}
S_\text{6D} &= \frac{1}{16\pi G_{(6)}} \int \dd^6x \, \sqrt{-g} \biggl[ R - V - \frac12 |\dd\vec{\varphi}|^2 - \frac12 \sum_{i=1}^2 X_i^{-2} |F_i|^2 - \frac18 (X_1 X_2)^2 |H|^2 \\
& - \frac{m^2}{4} (X_1 X_2)^{-1} |B|^2 - \frac{1}{16} \frac{\varepsilon^{\mu\nu\rho\sigma\tau\lambda}}{\sqrt{-g}} B_{\mu\nu} \Bigl( F_{1\,\rho\sigma} F_{2\,\tau\lambda} + \frac{m^2}{12} B_{\rho\sigma} B_{\tau\lambda} \Bigr) \biggr] \,,
\end{split}
\end{equation}
where $F_i=\dd A_i$, $H=\dd B$ and the scalar fields~$X_i$ with the scalar potential~$V$ are given by~\eqref{6d-scalar-para} and~\eqref{6d-potential}, respectively. This model is the complete version of the truncation presented in section~\ref{D6sugra}, with non-vanishing two-form~$B$.
Below for simplicity we will restrict our attention to the static solution, which in the notation of~\cite{Ferrero:2020twa} corresponds to setting $\mathtt{j}=0$. 
The six-dimensional solution  then reads
\begin{equation} \label{asd2-sp-ri}
\begin{aligned}
\dd s_6^2 &= \e^{-2C} L_{\AdS_4}^2 \left[ \frac{y^2}{4} \, \dd s_{\AdS_2}^2 + \frac{y^2}{q(y)} \, \dd y^2 + \frac{q(y)}{4y^2} \, \dd z^2 \right] + \e^{2C} \dd s_{\riemann}^2 \,, \\
X_1 &= k_8^{1/8} k_2^{1/2} \,,  \qquad\qquad  X_2 = k_8^{1/8} k_2^{-1/2} \,, \\
B &= \mathtt{a} \, \frac{9k_8^{1/2}}{8g^2} \, \vol{\AdS_2} \,, \\
F_1 &= \frac{\mathtt{a}}{y^2} \frac{3k_8^{1/2} k_2^{1/2}}{4g} \, \dd y \wedge \dd z + \frac{\kappa + \z}{2g} \, \vol{\riemann} \,, \\
F_2 &= \frac{\mathtt{a}}{y^2} \frac{3k_8^{1/2} k_2^{-1/2}}{4g} \, \dd y \wedge \dd z + \frac{\kappa - \z}{2g} \, \vol{\riemann} \, , 
\end{aligned}
\end{equation}
where the function $q(y)$ and the constant  $\mathtt{a}$ are given by\footnote{In order to be consistent with our notation we have exchanged $n_+$ and $n_-$ with respect to~\cite{Ferrero:2020twa}.}
\begin{equation}
q(y) = y^4 - (2y - \mathtt{a})^2 \,, \qquad \quad  \mathtt{a} = \frac{n_+^2-n_-^2}{n_+^2+n_-^2} \, . 
\end{equation}
The constants $k_2$ and $k_8$ are those appearing in~\eqref{kar_6d-functions}, namely
\begin{equation}
k_2 = \frac{3\z + \sqrt{\kappa^2 + 8\z^2}}{\z - \kappa} \,,  \qquad  k_8 = \frac{16 k_2}{9 (1+k_2)^2} \,,
\end{equation}
while $C$ reads
\begin{equation}
\e^{-2C} = m^2 k_8^{1/4} k_4  \qquad  \text{with}  \qquad  k_4 = \frac{18}{-3\kappa + \sqrt{\kappa^2 + 8\z^2}} \,.
\end{equation}
As usual, the parameter~$g$ is fixed in terms of~$m$, while $L_{\AdS_4}$ is related to~$m$ by eq.~(4.24) of~\cite{Hosseini:2020wag},
\begin{equation}
g = \frac{3m}{2} \,,  \qquad  \quad L_{\AdS_4} = \frac{k_8^{1/4} k_4^{-1/2}}{m^2} \,.
\end{equation}
Notice that formally taking  $\mathtt{a}=0$ the solution~\eqref{asd2-sp-ri} reduces precisely to the $\AdS_4\times\riemann$ 
solution  discussed  in section~\ref{bahsolutionssection}.

Let us now consider the quantization of the fluxes. For the fluxes through the Riemann surface we have
\begin{equation}
\begin{aligned}
\tr_1 &= \frac{g}{2\pi} \int_{\riemann} F_1 = \left(1 + \z\, \kappa \right) (1-\mathrm{g})  \in\ZZ\,, \\
\tr_2 &= \frac{g}{2\pi} \int_{\riemann} F_2 = \left(1 - \z \, \kappa \right) (1-\mathrm{g})  \in\ZZ\,,
\end{aligned}
\end{equation}
with $\tr_1+\tr_2=2(1-\mathrm{g})$. 
Recalling  that~\cite{Ferrero:2020twa} 
\begin{equation}
 \left(\frac{1}{y_1} - \frac{1}{y_2}\right) \frac{\Delta z}{2\pi} = \frac{n_+^2 + n_-^2}{n_+ n_- (n_+ + n_-)} \, , 
\end{equation}
where $y_1,y_2$ are the two relevant roots of $q(y)$,
the fluxes through  the spindle are given by
\begin{equation}
\begin{aligned}
\ts_1 &= \frac{g}{2\pi} \int_\spindle F_1 = \frac{3}{4} k_8^{1/2} k_2^{1/2} \frac{n_+ - n_-}{n_+ n_-} = \frac{p_1}{n_+ n_-} \,, \\
\ts_2 &= \frac{g}{2\pi} \int_\spindle F_2 = \frac{3}{4} k_8^{1/2} k_2^{-1/2} \frac{n_+ - n_-}{n_+ n_-} = \frac{p_2}{n_+ n_-} \,,
\end{aligned}
\end{equation}
where $p_i\in\ZZ$. We then note that 
\begin{equation}
\frac{3}{2} k_8^{1/2} k_2^{1/2} = \Delta_1^* \,,  \qquad  \frac{3}{2} k_8^{1/2} k_2^{-1/2} = 2-\Delta_1^* \equiv \Delta_2^* \,,
\end{equation}
where $\Delta_1^*$ was given in~\eqref{cane}, thus  we have
\begin{equation} \label{keyrelation}
\ts_i = \Delta_i^* \frac{n_+ - n_-}{2n_+ n_-} \equiv  n \, \Delta_i^*  \,, 
\end{equation}
that will be important in the field theory extremization.
The constraint $\Delta_1^*+\Delta_2^*=2$ implies
\begin{equation} \label{antitwistwehave}
\ts_1 + \ts_2  = \frac{n_+ - n_-}{n_+ n_-} \, ,
\end{equation}
showing that there is an anti-twist over the spindle. Note that from the  relations~\eqref{keyrelation} we obtain
\begin{equation}
\bigg(1 + \frac{\kappa+ \sqrt{8\z^2 + \kappa^2}}{4\z}  \bigg) (n_+ - n_-) = 2 p_1 \,,
\end{equation}
which is a non-trivial Diophantine equation. One example of solution is given by the set of values $\kappa=-1$, $\z=6$, $n_+-n_-=6$, $p_1=5$ and we have checked that other combinations exist. Moreover, when $\kappa=-1$ it is possible to smoothly take the limit $\z\to0$, in which case the equation reduces to $n_+ - n_- = 2p_1$.

We can now uplift the solution~\eqref{asd2-sp-ri} to massive type~IIA using the recipe described in section~\ref{sub:improved-uplift}. For simplicity, we will only present 
the relevant ingredients for the computation of the entropy, namely the metric and the dilaton, that read
\begin{align}
\begin{split}
\dd s_\text{s.f.}^2 &= \lambda^2 \mu_0^{-1/3} \biggl\{ \tilde{\Delta}^{1/2} \biggl[ \e^{-2C} L_{\AdS_4}^2 \biggl( \frac{y^2}{4} \, \dd s_{\AdS_2}^2 + \frac{y^2}{q(y)} \, \dd y^2 + \frac{q(y)}{4y^2} \, \dd z^2 \biggr) + \e^{2C} \dd s_{\Sigma_\mathrm{g}}^2 \biggr] \\
& + g^{-2} \tilde{\Delta}^{-1/2} k_8^{-1/4} \bigl[ k_8^{1/2} \dd\mu_0^2 + k_2^{-1/2} \bigl(\dd\mu_1^2 + \mu_1^2 \sigma_1^2\bigr) + k_2^{1/2} \bigl(\dd\mu_2^2 + \mu_2^2 \sigma_2^2\bigr) \bigr] \biggr\} \,,
\end{split} \\
\e^\Phi &= \lambda^2 \mu_0^{-5/6} \tilde{\Delta}^{1/4} k_8^{-1/8} \,,
\end{align}
with $\sigma_i=\dd\phi_i-gA_i$ and $\tilde{\Delta}$ given in~\eqref{kar_10d-func}.
Expressing the metric in string frame as
\begin{equation}
\dd s_\text{s.f.}^2 = \e^{2A} \bigl( \dd s_{\AdS_2}^2 + \dd s_{M_8}^2 \bigr) \,,
\end{equation}
the effective two-dimensional Newton constant is given by
\begin{equation}
\frac{1}{G_{(2)}} = 
\frac{32\pi^2}{(2\pi \ell_s)^8} \int \e^{8A-2\Phi} \, \vol{M_8} \,, 
\end{equation}
and therefore the entropy reads
\begin{equation}
S = \frac{1}{4G_{(2)}} = \frac{1}{(2\pi\ell)^8} \frac{9 (3 \pi \lambda)^4 k_8^{1/2}}{20 g^8  k_4} \, 4\pi\kappa (1-\mathrm{g}) \, A_h \,,
\end{equation}
where $A_h$ is the area of the horizon of the four-dimensional black hole with $L_{\AdS_4}=1$, specifically~\cite{Ferrero:2020twa} 
\begin{equation}
A_h = \frac12 (y_2-y_1) \Delta z = \pi \, \frac{-(n_+ + n_-) + \sqrt{2} \sqrt{n_+^2 + n_-^2}} {n_+ n_-} \,.
\end{equation}

Both the Romans mass $F_{(0)}$ and the part of the four-form flux $F_{(4)}$ along the four-hemisphere~$\hemi$ remain unaltered with respect to~\eqref{kar_10d-F0} and~\eqref{kar_10d-F4}, therefore the quantization of the fluxes in ten dimensions
is unchanged, giving the   relations~\eqref{fluxquantwithlambda}. The final expression of the entropy is
\begin{equation}
S = \frac{8\kappa (1-\mathrm{g}) N^{5/2} (\z^2-\kappa^2)^{3/2} (\sqrt{\kappa^2 + 8\z^2}-\kappa)}{5\sqrt{8-N_f} (\kappa \sqrt{\kappa^2 + 8\z^2}-\kappa^2 + 4\z^2)^{3/2}} \, A_h \,.
\end{equation}

\subsection{Entropy function}

As before, we can deduce the main ingredients of the field theory construction from the supergravity solution. In particular,  the magnetic fluxes $\tr_i$ show that
 there is a  (standard) topological twist on  $\Sigma_\mathrm{g}$, while from  the form of the $\ts_i$ we see that there is an anti-twist on $\spindle$. Then applying 
 our conjecture twice, we obtain the following entropy function
\begin{equation}
\label{Ssigmaspindle}
\begin{split}
S(\varphi_i,\epsilon_1,\epsilon_2;\ts_i,\mathfrak{s}_i)  & = -\frac{1}{4\epsilon_1\epsilon_2}\Big[  \mc{F}_5 (\varphi_i + \ts_i \epsilon_1 +   \tr_i \epsilon_2) + \mc{F}_5 (\varphi_i - \ts_i \epsilon_1 +   \tr_i \epsilon_2)  \\
& \qquad \qquad -  \mc{F}_5 (\varphi_i + \ts_i \epsilon_1 -   \tr_i \epsilon_2) - \mc{F}_5 (\varphi_i - \ts_i \epsilon_1 -   \tr_i \epsilon_2) \Big]\, , 
\end{split}
\end{equation}
subject to the constraints
\begin{equation}
\label{pinco}
     \ts_1 + \ts_2 =  \frac{n_+ - n_-}{n_+ n_-}\, , \qquad   \tr_1 + \tr_2 =  2 (1-\mathrm{g}) \, , \qquad \varphi_1+ \varphi_2 -   \frac{n_+ +  n_-}{n_+ n_-}\epsilon_1 =2  \, .
\end{equation}
Extremizing this with respect to $\epsilon_2$ sets $\epsilon_2=0$ and after renaming $\epsilon_1\mapsto\epsilon$ we obtain
\begin{equation}
      \begin{split}
S(\varphi_i,\epsilon;\ts_i,\mathfrak{s}_i)   =  \frac{c}{\epsilon} \left[\sqrt{(\varphi_1+\ts_1\epsilon)(\varphi_2+\ts_2\epsilon)}(\mathfrak{s}_1(\varphi_2+\ts_2\epsilon)+\mathfrak{s}_2 (\varphi_1+\ts_1\epsilon))+
(\epsilon\to -\epsilon)\right]\, , 
\label{entropyfunk}
\end{split}
    \end{equation} 
where 
   \begin{equation}
c \equiv  \frac{\sqrt{2}\pi }{5}\frac{N^{5/2}}{\sqrt{8-N_f}}  \,, 
     \end{equation} 
which has to be extremized subject to the constraints \eqref{pinco}.

So far we have not used the additional input   \eqref{keyrelation}
 given by the supergravity solution discussed above. This strongly suggests that the extremization of \eqref{entropyfunk}, \emph{without} imposing 
\eqref{keyrelation}, will give the entropy of a more general supergravity solution\footnote{The recent paper \cite{Giri:2021xta} discusses 
 $\AdS_2\times\spindle\times\riemann$ supergravity solutions and it would be interesting to understand the relationship to our work.}. 
However, below we will proceed enforcing (\ref{keyrelation}).
Doing so, it is convenient to introduce the rescaled variables
\begin{equation}
\varphi_1 \equiv \Delta_1^* \lambda_1 \, ,\qquad  \varphi_2 \equiv \Delta_2^* \lambda_2 \, , 
\end{equation}
subject to the constraint
\begin{equation}
 \Delta_1^* \lambda_1+  \Delta_2^* \lambda_2 -   \frac{n_+ +  n_-}{n_+ n_-}\epsilon  =2  \, ,
\end{equation}
in terms of which the entropy function reads
\begin{equation}
\begin{split}
S(\lambda_i,\epsilon;\z) = \frac{c}{\epsilon} \Bigl[\sqrt{\Delta_1^*\Delta_2^*(\lambda_1+  n \epsilon)(\lambda_2+ n \epsilon)}(\mathfrak{s}_1\Delta_1^*(\lambda_2+n\epsilon)+\mathfrak{s}_2 \Delta_2^*(\lambda_1+n\epsilon))+
(\epsilon\to -\epsilon)\Bigr] \,. 
\label{entropyfunky}
\end{split}
\end{equation}

Extremizing this we find the critical values
\begin{equation} 
\epsilon^* = \eta \frac{2n_+n_-}{\sqrt{2n_+^2+2n_-^2}} \,, \qquad  \lambda_1^*=\lambda_2^* = 1+\eta \frac{n_++n_-}{\sqrt{2n_+^2+2n_-^2}}\, , 
\end{equation}
where the sign ambiguity $\eta =\pm 1$ arises by solving the equations over the complex numbers\footnote{Generically, in the presence of rotation, we are forced 
to work with the complex numbers \cite{Cabo-Bizet:2018ehj,Cassani:2021dwa} and we therefore continue to do so also in the static case.}. 
 Inserting these values  back into the entropy function, we compute
\begin{equation}
\begin{split}
S(\lambda^*_i,\epsilon^*;\z)  &  = 2  \frac{\sqrt{2}\pi }{5}\frac{N^{5/2}}{\sqrt{8-N_f}}   \sqrt{\Delta_1^*\Delta_2^*}(\mathfrak{s}_1 \Delta_2^*+\mathfrak{s}_2 \Delta_1^*)
\frac{-n_+- n_- -\eta \sqrt{2n_-^2+2n_+^2}}{n_+ n_-}\\
&= \frac{8\pi \kappa (1-\mathrm{g}) N^{5/2} (\z^2-\kappa^2)^{3/2} (\sqrt{\kappa^2 + 8\z^2}-\kappa)}{5\sqrt{8-N_f} (\kappa \sqrt{\kappa^2 + 8\z^2}-\kappa^2 + 4\z^2)^{3/2}}   \frac{-n_+ - n_- -\eta \sqrt{2n_+^2 + 2n_-^2}} {n_+ n_-} \, , 
\end{split}
\end{equation} 
where we need to pick $\eta=-1$ in order to get a positive entropy that agrees with the one computed from the ten-dimensional supergravity solution.

It should be straightforward to incorporate electric charge and rotation (along the spindle), by promoting the extremization of \eqref{entropyfunky} to the Legendre transform. The entropy obtained in this way should match that of the supergravity solution obtained by uplifting to massive type IIA the rotating AdS$_2\times \spindle$ solution of \cite{Ferrero:2020twa}\footnote{In the rotating solution, setting $n_+=n_-=1$ will reduce to a rotating AdS$_2\times S^2_\epsilon\times\Sigma_{\mathrm{g}}$ solution in $D=6$, whose entropy function has been recently discussed in \cite{Hosseini:2021mnn}.}.

\section{Discussion}
\label{discusssection}

The work of \cite{Ferrero:2020laf} opened the way to a novel class of examples of the AdS/CFT correspondence, by constructing a supergravity solution with an AdS$_3$  factor,
comprising the two-dimensional orbifold $\spindle$, known as the spindle. This was interpreted as the near-horizon limit of D3-branes wrapped on the spindle with the corresponding field theory duals being a class of $d=4$, ${\cal N}=1$ SCFTs compactified on the spindle, with a new type of supersymmetry-preserving twist, that was dubbed anti-twist. 
Following this, over the past year various extensions, including analogous constructions for M2 and M5-branes, have appeared, realising either the anti-twist \cite{Ferrero:2020laf},
 or a global generalization \cite{Ferrero:2021wvk} of the standard topological twist. These constructions have left out the notable class of solutions describing D4-branes wrapped on the spindle, that we constructed in this paper. 
We found these AdS$_4\times \spindle$ solutions in $D=6$ gauged supergravity and then uplifted them to massive type IIA supergravity.
The type of twist realised in our solutions is the one previously found in \cite{Ferrero:2021wvk} for M5-branes, which was referred to as a global topological twist. 
Differently from the standard topological twist, the $R$-symmetry gauge field does not cancel the spin connection,  despite the fact that its integrated flux is  equal to the Euler characteristic of the spindle.
As a consequence the Killing spinors are then non-trivial sections (in particular, they are not simply constant) of the same bundles that occur in standard topological twist. 

There are several  aspects of our solution that may be interesting to investigate in the future. For example, it may be instructive to cast it in the form of the classification of supersymmetric AdS$_4$ solutions of massive type IIA supergravity \cite{Passias:2018zlm}, or to study supersymmetric probe D-branes in our background in order to extract further information about the dual field theories.
Another question that arises from our work is whether there exists a more general consistent truncation of massive type IIA supergravity to $D=6$, analogous to the dyonic consistent truncation to $D=4$ supergravity found in \cite{Guarino:2015jca}. It is also intriguing to investigate whether there exist supergravity solutions corresponding to 
 D4-branes wrapped on the spindle with the anti-twist. Finally, as an important step towards improving control on the dual field theories,  it would be worthwhile computing an appropriately regularised on-shell action, that should prove the validity of our conjectural off-shell free energy \eqref{offshellFconjecture}. 
To do this, one may need to know the full ``black two-brane'' solution,  interpolating between  AdS$_6$ asymptotically and AdS$_4\times \spindle$ in the near-horizon, similarly to \cite{Cassani:2021dwa}.
However, it may be possible to employ the strategy of \cite{BenettiGenolini:2019jdz} to prove that the supergravity (Euclidean) on-shell action localises  at the poles of the spindle, which are the fixed points for the canonical Killing vector field associated to any (Euclidean) supersymmetric solution of the theory \cite{Alday:2015jsa}.
   
There is also a number of assorted interesting questions in the field theory side. The most direct one is  to study the five-dimensional SCFTs in the background of $S^3\times \spindle$, computing the localized partition function from first principles, and then showing that in the large $N$ limit the associated free energy reduces to \eqref{offshellFconjecture}. This is indeed an open problem for SCFTs compactified on spindles in different dimension.

It is  compelling that the five-dimensional off-shell free energy that we conjectured here and the entropy function that was conjectured in \cite{Ferrero:2021ovq} fit into a broader conjecture for the free energies (\ie\ minus the logarithm of partition functions) 
of SCFTs in various dimensions, namely \eqref{superspindleconjecturebetter}. These extend the idea of gravitational blocks put forward in \cite{Hosseini:2019iad} in two directions. Firstly, from  compactifications on smooth manifolds, to the realm of compactifications on orbifolds. It is remarkable that compactifications on 
spindles can be incorporated by a simple modification of  the constraint obeyed by the fugacities (see eq.~\eqref{constraintnice}). This depends on the type of twist performed and it includes the  standard topological twist and the no-twist 
as special cases.  Secondly, we have pointed out that the form \eqref{superspindleconjecturebetter} should hold also for observables of higher-dimensional theories, beyond the entropy, associated to AdS$_2$ solutions dual $d=1$ SCFTs.

 Using the AdS/CFT correspondence, the free energies \eqref{superspindleconjecturebetter} are generically expected to arise as gravitational (Euclidean) on-shell actions, and their structure is suggested by the fact  that for supersymmetric solutions, 
 this form should arise from summing  contributions at  fixed points
of the canonical Killing vector field,   defined as a bilinear in the Killing spinors of the solution  \cite{BenettiGenolini:2019jdz,Genolini:2021urf}. An immediate issue that would be nice to clarify is a better justification of 
 the $\pm$ signs in \eqref{superspindleconjecturebetter} and their relationship with the type of twist. For example, while for field theories in odd dimensions we have proposed that the gluing sign should be $-\sigma$, for field theories in even dimensions, 
the explicit examples indicate that one should always pick the minus sign. It would interesting to find out whether the functions  that have not been considered so far, may have critical points that correspond to new gravitational objects.

 It is quite clear that the structure of~\eqref{superspindleconjecturebetter} will extend to compactifications of SCFTs on higher-dimensional\footnote{The  recent paper \cite{Hosseini:2021mnn} discusses related setups, for compactifications of $d=5$ and $d=6$ SCFTs on various smooth manifolds, including $S^2_\epsilon \times \Sigma_{\mathrm{g}}$ and $S^2_{\epsilon_1} \times S^2_{\epsilon_2}$.} \emph{orbifolds} and in general we expect that the functions to extremize will take the form of a sum of contributions from fixed points of the canonical Killing vector on the compactification orbifold $\mathbbl{M}$. Assuming this is toric, for concreteness, there will be as many equivariant parameters $\epsilon_I$ as the rank of the torus acting on $\mathbbl{M}$. One of the simplest examples of this type of construction is given by twisted compactifications of the $d=5$ SCFT on  $\mathbbl{M}=\spindle \times \Sigma_\mathrm{g}$, that we discussed in section~\ref{ads2solution}. We have shown that the corresponding entropy function
 can be obtained  iterating our conjecture for the off-shell free energies twice and extremizing this reproduces the entropy of a corresponding  class of AdS$_2\times \spindle \times \Sigma_\mathrm{g}$  supergravity solutions.
 More generally, we expect that there should exist  solutions of the type AdS$_2\times \spindle_1\times \spindle_2$ in $D=6$ supergravity and of the type 
 AdS$_3\times \spindle_1\times \spindle_2$ in $D=7$ supergravity, for which the corresponding entropy function and trial central charge can be obtained following the rules we proposed in this paper. 
 Interestingly, the former case would be identified with the near-horizon limit of a novel class of supersymmetric black holes in AdS$_6$, that may be accelerating. Work along these lines is underway and we hope to report in the near future.

As we continue to navigate the landscape of supergravity solutions corresponding to branes wrapping orbifolds, it will be revealing to employ the approach developed in \cite{Couzens:2018wnk,Hosseini:2019ddy,Gauntlett:2019roi,Gauntlett:2019pqg}, adapting it to situations with orbifold singularities. While this is already well developed for solutions arising from M2 and D3-branes, it is tantalising to think that an analogous geometric approach may be concocted for studying AdS$_5$ solutions of $D=11$ supergravity or other backgrounds with an AdS factor.

\subsection*{Acknowledgments}

The research of FF is supported by the project ``Nonlinear Differential Equations'' of Universit\`a degli Studi di Torino. DM would like to thank J. Gauntlett and J. Sparks for insightful comments and enjoyable collaborations on related topics.


\appendix

\section{More details on the  AdS$_4 \times \Sigma_\mathrm{g}$ solutions}
\label{app:all}

\subsection{Relation with the Lagrangian of \cite{Karndumri:2015eta}}
\label{app:map-for-Lag-mg}

Here we make contact between the $D=6$ Lagrangian \eqref{6d-action},   that we use in the paper, and the Lagrangian used in~\cite{Karndumri:2015eta}, given explicitly in~\cite{Suh:2018szn}.
In order to minimise confusion we have relabelled some of the quantities in~\cite{Suh:2018szn} as follows
\begin{equation}
\varphi_{1,2} \mapsto \hat{\varphi}_{1,2} \,,  \qquad  F^{3,6} \mapsto \frac12 \mc{F}^{3,6} \,,  \qquad  m \mapsto m_1 \,,
\end{equation}
where the $1/2$ factor is due to the unusual definition of the field strengths as, \eg, $\mc{F}^3_{\mu\nu} = \frac12 (\partial_\mu\mc{A}^3_\nu - \partial_\nu\mc{A}^3_\mu)$. 
Consistency between the equations of motion and the supersymmetry equations requires $\hat{\varphi}_1=0$, and in this case the $D=6$ Lagrangian in~\cite{Suh:2018szn} reads
\begin{equation}
\begin{split}
e^{-1} \mc{L} &= \frac14 R - \hat{V} - |\dd\sigma|^2 - \frac14 |\dd\hat{\varphi}_2|^2 - \frac{1}{16} \e^{-2\sigma} \cosh(2\hat{\varphi}_2) \, |\mc{F}^3|^2 \\
& - \frac{1}{16} \e^{-2\sigma} \cosh(2\hat{\varphi}_2) |\mc{F}^6|^2 + \frac{1}{16} \e^{-2\sigma} \sinh(2\hat{\varphi}_2) \mc{F}^3_{\mu\nu} \mc{F}^{6\mu\nu} \,,
\end{split}
\end{equation}
where the scalar potential is
\begin{equation}
\hat{V} = -g_1^2 \e^{2\sigma} - 4g_1 m_1 \e^{-2\sigma} \cosh\hat{\varphi}_2 + m_1^2 \e^{-6\sigma} \,.
\end{equation}
Defining the quantities
\begin{equation}
\begin{split}
\vec{\varphi} = -\sqrt2 \, \bigl(\hat{\varphi}_2, 2\sigma\bigr)  \qquad  &\iff  \qquad  X_1 = \e^{\sigma+\hat{\varphi}_2} \,,  \quad  X_2 = \e^{\sigma-\hat{\varphi}_2} \,, \\
F_1 = \frac12 (\mc{F}^3 + \mc{F}^6) \,,  &\quad\qquad  F_2 = \frac12 (\mc{F}^3 - \mc{F}^6) \,, \\
g = g_1 \,,  &\quad  \qquad m = 2m_1 \,,
\end{split}
\end{equation}
we  obtain the Lagrangian given in~\eqref{6d-action}, divided by 4.

\subsection{Equivalence with the solutions of \cite{Bah:2018lyv}}
\label{app:map-to-Bah}

Below we  show that the ten-dimensional background~\eqref{kar_10d-metric} - \eqref{kar_10d-F4}, obtained from the uplift of the $D=6$ solution of \cite{Karndumri:2015eta},  
is equivalent to the solution to massive type IIA supergravity constructed in~\cite{Bah:2018lyv}. Let us start with the metric~(4.14) of~\cite{Bah:2018lyv} 
\begin{equation}
\label{bahmetrictend}
\dd s_\text{s.f.}^2 = \frac{L_{\AdS_4}^2 }{( H y)^{1/2}\Fm} \biggl( \dd s_{\AdS_4}^2 + \e^{2\nu} \dd s_{\Sigma_\mathrm{g}}^2 + \frac{H}{4} \, \dd s_{\hemi}^2 \biggr) \,, 
\end{equation}
where we restored the dimension-full $\AdS_4$ radius $L_{\AdS_4}$
that was set to one in~\cite{Bah:2018lyv} and we renamed the constant $F_0$  in \cite{Bah:2018lyv} as $\Fm$. 
The metric on the squashed hemisphere reads
\begin{equation}
\dd s_{\hemi}^2 = \frac83 \, \dd\mu_0^2 + 2 \bigl( \dd\hat{\mu}_1^2 + \hat{\mu}_1^2 \eta_+^2 + \dd\hat{\mu}_2^2 + \hat{\mu}_2^2 \eta_-^2 \bigr) \,,
\end{equation}
where the  coordinates $\mu_0,\hat \mu_1,\hat\mu_2$ satisfy the constraint $\mu_0^2+a_+\hat{\mu}_1^2+a_-\hat{\mu}_2^2=1$, and the one-forms $\eta_\pm$ are
$\eta_\pm=\dd\phi_\pm-m_\pm\,\omega_\mathrm{g}$.
The functions $H$ and $y$ are given by 
\begin{equation}
H = \frac{2}{3\mu_0^2 + 4 \bigl( a_+^2 \hat{\mu}_1^2 + a_-^2 \hat{\mu}_2^2 \bigr)} \,,  \quad  \qquad  y^3 \Fm^2 = \frac32 \mu_0^2 \,,
\end{equation}
while $\e^{2\nu}$ is a constant and reads
\begin{equation}
\e^{2\nu} = \frac{-\kappa + \sqrt{\kappa^2 + 8\z^2}}{4} \, . 
\end{equation}
The constants $a_\pm$, $m_\pm$ can be expressed in terms of~$\kappa$
and a constant parameter~$\z$ as
\begin{equation}
a_\pm = \frac{1 \pm \epsilon}{2} \,,   \qquad  \epsilon = \frac{\kappa \pm \sqrt{\kappa^2 + 8\z^2}}{4\z} \,,  \qquad  m_\pm = \frac{\kappa \pm \z}{2} \,  .
\end{equation}
In order to compare the two solutions we take
\begin{equation} \label{match_k-vs-a}
k_2 = \frac{a_+}{a_-} \,,  \qquad\quad  k_8 = \frac{16}{9} a_+ a_- \,,
\end{equation}
which are consistent with the constraint $a_++a_-=1$ required by the definition of $a_\pm$, and relate our coordinates $\mu_1$, $\mu_2$ to those in~\cite{Bah:2018lyv} as
\begin{equation}
\mu_1 = \sqrt{a_+} \hat{\mu}_1 \,,  \qquad\quad  \mu_2 = \sqrt{a_-} \hat{\mu}_2 \,.
\end{equation}
We then identify $\phi_1=\phi_+$, $\phi_2=\phi_-$ and
\begin{equation}
\p1 = 2m_+ = \kappa + \z \,, \qquad \quad \p2 = 2m_- = \kappa - \z \,,
\end{equation}
thus having $\sigma_1=\eta_+$, $\sigma_2=\eta_-$.
As consistency checks we have that $2(m_++m_-)=\p1+\p2=2\kappa$ and that the definition of $k_2$ in~\eqref{kar_6d-functions} agrees with~\eqref{match_k-vs-a}. 
Lastly, comparing the overall factors in the metric and the dilaton of the two solutions, we find
\begin{equation} \label{match_dilaton}
g^2 = \left(\frac32\right)^{3/2} k_8^{3/4} L_{\AdS_4}^{-2} \,,  \quad \qquad  \lambda^2 = \left(\frac32\right)^{-1/6} k_8^{1/4} \Fm^{-2/3} \,.
\end{equation}

The Romans mass~\eqref{kar_10d-F0} then reads $F_{(0)} = \frac{\Fm}{L_{\AdS_4}}$,
reducing to the expression in~\cite{Bah:2018lyv} for $L_{\AdS_4}=1$. Writing our four-form flux~\eqref{kar_10d-F4} in the variables used in~\cite{Bah:2018lyv}, we find 
\begin{equation}
\begin{split}
F_{(4)} &= \frac{H}{8} \left(\frac{2\mu_0}{3\Fm}\right)^{1/3} L_{\AdS_4}^3 \biggl\{ W \frac{\hat{\mu}_1 \hat{\mu}_2}{\mu_0} \, \dd\hat{\mu}_1 \wedge \dd\hat{\mu}_2 \wedge \eta_+ \wedge \eta_- \\
& - \vol{\Sigma_\mathrm{g}} \wedge \biggl[ \frac{m_+}{a_+} \, \dd\phi_- \wedge \bigl( 3\mu_0 \hat{\mu}_2 \, \dd\hat{\mu}_2 - 4a_- \hat{\mu}_2^2 \, \dd\mu_0 \bigr) \\
& + \frac{m_-}{a_-} \, \dd\phi_+ \wedge \bigl( 3\mu_0 \hat{\mu}_1 \, \dd\hat{\mu}_1 - 4a_+ \hat{\mu}_1^2 \, \dd\mu_0 \bigr) \biggr] \biggr\} \,,
\end{split}
\end{equation}
where we defined
\begin{equation}
W = 9H \mu_0^2 + 16H \bigl(a_+^3 \hat{\mu}_1^2 + a_-^3 \hat{\mu}_2^2\bigr) - 12 \,.
\end{equation}
This expression agrees with the $F_{(4)}$ given in~\cite{Bah:2018lyv} only partially\footnote{While the above $F_{(4)}$ satisfies the equations of motion and the Bianchi identity, we found that the four-form flux given in~\cite{Bah:2018lyv} is not closed.}. A couple of useful identities that we used for comparing the two solutions are 
\begin{equation}
\frac{m_+}{a_+} = -2\e^{2\nu} [1 - 2(a_+-a_-)] \,,  \qquad  \frac{m_-}{a_-} = -2\e^{2\nu} [1 - 2(a_--a_+)] \,.
\end{equation}

\section{$OSp(2|4)$ superalgebra from bilinears}
\label{app:osp}

Below we show that the $OSp(2|4)$ superalgebra can be reconstructed from the Killing spinors of our $\AdS_4\times\spindle$ solution, as expected. The fact that the Killing vectors of AdS can be constructed as bilinears in the Killing spinors is well known. Here we demonstrate that the $D=6$ solution captures part of the $R$-symmetry of the dual $d=3$, $\mc{N}=2$ SCFTs. Specifically, it includes the component of the $R$-symmetry along the spindle isometry. The full $R$-symmetry would arise from working with the Killing spinors of the uplifted ten-dimensional solution. 

Let us recall the relevant ingredients from section \ref{newlocal_section}.  The Killing spinors of the $D=6$ solution are a pair of symplectic-Majorana spinors, taking the form
\begin{equation} 
\epsilon^A = \vartheta_+ \otimes \eta^A_+ + \vartheta_- \otimes \eta^A_- \,,
\end{equation}
where $A=1,2$ is a symplectic index. The spinors $\vartheta_\pm$ are the chiral components of a Majorana Killing spinor $\vartheta=\vartheta(x^{\hat{\mu}})$ in $\AdS_4$, obeying 
\begin{equation}
\label{ads4kspinors}
\vartheta^*=\mc{B}_4\vartheta \,,  \qquad  \hat{\nabla}_{\hat{\mu}} \vartheta = \frac12 \gamma_{\hat{\mu}} \vartheta \,,
\end{equation}
thus we have $\vartheta =\vartheta_++\vartheta_-$ with $\gamma_5\vartheta_\pm=\pm\vartheta_\pm$. The spinors $\eta^A_\pm=\eta^A_\pm(y)$ are  Dirac spinors defined on the spindle.
Defining the following vector bilinear
\begin{equation}
\mathtt{K}^\mu = \ii \, \overline{\epsilon'^1} \Gamma^\mu \epsilon^1 \,,
\end{equation}
where $\epsilon^1$ and $\epsilon'^1$ are two Killing spinors,  straightforward manipulations lead to
\begin{equation} \label{killing-compact}
\mathtt{K}^{\hat{\mu}} = 2\ii\,m (\xi')^* \xi \, \bar{\vartheta}' \gamma^{\hat{\mu}} \vartheta \,,  \qquad
\mathtt{K}^y = 0 \,,  \qquad
\mathtt{K}^z = 2 (\xi')^* \xi \, \bar{\vartheta}' \vartheta \,. 
\end{equation}
From~\eqref{ads4kspinors} it immediately follows that 
\begin{equation}
\nabla_{(\hat{\mu}} (\bar{\vartheta}' \gamma_{\hat{\nu})} \vartheta) = 0 = \nabla_{\hat{\mu}} (\bar{\vartheta}' \vartheta) \, , 
\end{equation}
hence
\begin{equation}
\mathtt{K}^\mu \partial_\mu = 2 (\xi')^* \xi \bigl( \ii\,m \bar{\vartheta}' \gamma^{\hat{\mu}} \vartheta  \, \partial_{\hat{\mu}} + \bar{\vartheta}' \vartheta \, \partial_z \bigr)
\end{equation}
are Killing vectors for the $D=6$ solution.
One can check that $\overline{\epsilon'^2} \Gamma^\mu \epsilon^2 = -(\overline{\epsilon'^1} \Gamma^\mu \epsilon^1)^*$, while $\overline{\epsilon'^1} \Gamma^\mu \epsilon^2$, $\overline{\epsilon'^2} \Gamma^\mu \epsilon^1$ and $\overline{\epsilon'^A} \Gamma^7 \Gamma^\mu \epsilon^B$ are not Killing vectors, so there are no further vector bilinears to consider.

To make contact with the  $OSp(2|4)$ superalgebra we need to look in more detail into the structure of the Killing spinors in AdS$_4$.
Writing the unit radius $\AdS_4$ metric as 
\begin{equation}
\dd s_{\AdS_4}^2 =  u^2 \eta_{mn} \dd x^m \dd x^n + \frac{\dd u^2}{u^2}  \,,
\end{equation}
and defining the coordinates $x^{\hat{\mu}}=(t,x^1,x^2,u)\equiv (x^m,u)$, the Killing spinors on $\AdS_4$ can be written as~\cite{Lu:1998nu}
\begin{equation}
\epsilon_{\AdS_4} = \e^{\frac12(\ln u)\,\gamma_3} \biggl[ 1 + \frac{1}{2u} x^m \gamma_m (1 - \gamma_3) \biggr] \psi \,,
\end{equation}
with $\psi$ a constant spinor  and $\{\gamma_m,\gamma_n\}=u^2\eta_{mn}$. This can be conveniently split into two independent Killing spinors as $\epsilon_{\AdS_4} = \epsilon_+ + \epsilon_-$, where
\begin{equation}
\epsilon_+ = u^{1/2} \psi_+ \,,  \qquad \quad  \epsilon_- = u^{-1/2} \left( 1 + x^m \gamma_m \right) \psi_- \,,
\end{equation}
with
\begin{equation}
\psi_\pm \equiv \frac{I_4 \pm \gamma_3}{2} \psi  \qquad  \implies  \qquad  \gamma_3 \psi_\pm = \pm \psi_\pm \,.
\end{equation}
Note that  $\epsilon_\pm$ and $\psi_\pm$ are not chiral spinors, since $\gamma_3$ is not the chiral matrix. A Majorana condition can be consistently imposed on $\epsilon_{\AdS_4}$, which implies that $\psi$ and, in turn, $\psi_\pm$ are Majorana spinors. We can then identify $\vartheta$ with $\epsilon_{\AdS_4}$ and write the Killing vectors in~\eqref{killing-compact} as
\begin{equation} \label{killing-deco}
\begin{aligned}
\mathtt{K}^{\hat{\mu}} &= 2\ii\,m (\xi')^* \xi \bigl( \bar{\epsilon}'_+ \gamma^{\hat{\mu}} \epsilon_+ + \bar{\epsilon}'_+ \gamma^{\hat{\mu}} \epsilon_- + \bar{\epsilon}'_- \gamma^{\hat{\mu}} \epsilon_+ + \bar{\epsilon}'_- \gamma^{\hat{\mu}} \epsilon_- \bigr) \,, \\
\mathtt{K}^z &= 2 (\xi')^* \xi \bigl( \bar{\epsilon}'_+ \epsilon_+ + \bar{\epsilon}'_+ \epsilon_- + \bar{\epsilon}'_- \epsilon_+ + \bar{\epsilon}'_- \epsilon_- \bigr) \,.
\end{aligned}
\end{equation}
As standard, the two independent $\epsilon_+$ spinors are identified  with the real Poincar\'e supercharges~$Q$ and the $\epsilon_-$ with the real superconformal supercharges~$S$. For this reason we shall refer to the different contributions in~\eqref{killing-deco} as $\mathtt{K}^{\hat{\mu}}_{QQ}$, $\mathtt{K}^{\hat{\mu}}_{QS}$, $\mathtt{K}^{\hat{\mu}}_{SQ}$ and $\mathtt{K}^{\hat{\mu}}_{SS}$, respectively, and likewise for $\mathtt{K}^z$. Specifically, the components along the spindle are
\begin{equation}
\begin{aligned}
\mathtt{K}^z_{QQ} &= 0 \,,  \qquad  & \mathtt{K}^z_{QS} &= 2 (\xi')^* \xi \, \bar{\psi}'_+ \psi_- \,, \\
\mathtt{K}^z_{SS} &= 0 \,,  \qquad  & \mathtt{K}^z_{SQ} &= 2 (\xi')^* \xi \, \bar{\psi}'_- \psi_+ \,, \\
\end{aligned}
\end{equation}
and the components along $\AdS_4$ are
\begin{equation} \label{killing-para}
\begin{aligned}
\mathtt{K}^m_{QQ} &= 2\ii\,m (\xi')^* \xi \, u \, (\bar{\psi}'_+ \gamma^m \psi_+) \,,  \quad  & \mathtt{K}^u_{QQ} &= 0 \,, \\
\mathtt{K}^m_{QS} &= 2\ii\,m (\xi')^* \xi \, \bigl[ x^m (\bar{\psi}'_+ \psi_-) + x^n (\bar{\psi}'_+ \gamma^m_{\ \ n} \psi_-) \bigr] \,,  \quad  & \mathtt{K}^u_{QS} &= -2\ii\,m (\xi')^* \xi \, u \, \bar{\psi}'_+ \psi_- \,, \\
\mathtt{K}^m_{SQ} &= 2\ii\,m (\xi')^* \xi \, \bigl[ -x^m (\bar{\psi}'_- \psi_+) + x^n (\bar{\psi}'_- \gamma^m_{\ \ n} \psi_+) \bigr] \,,  \quad  & \mathtt{K}^u_{SQ} &= 2\ii\,m (\xi')^* \xi \, u \, \bar{\psi}'_- \psi_+ \,, \\
\mathtt{K}^m_{SS} &= (x \cdot x + u^{-2}) \, b^m - 2x^m (x \cdot b) \,,  \quad  & \mathtt{K}^u_{SS} &= 2u \, (x \cdot b) \,,
\end{aligned}
\end{equation}
with $b^m \equiv 2\ii\,m(\xi')^*\xi\,u\,(\bar{\psi}'_-\gamma^m\psi_-)$ and $x\cdot b=\eta_{mn} x^m b^n$. We now define the additional parameters
\begin{equation}
\label{parameters}
\begin{aligned}
a^m &\equiv 2\ii\,m (\xi')^* \xi \, u \, (\bar{\psi}'_+ \gamma^m \psi_+) \,, \\
\Lambda^m_{\ \ n} &\equiv 2\ii\,m (\xi')^* \xi \, (\bar{\psi}'_+ \gamma^m_{\ \ n} \psi_- + \bar{\psi}'_- \gamma^m_{\ \ n} \psi_+) \,, \\
\lambda &\equiv 2\ii\,m (\xi')^* \xi \, (\bar{\psi}'_+ \psi_- - \bar{\psi}'_- \psi_+) \,, \\
\mathtt{r} &\equiv 2 (\xi')^* \xi \, (\bar{\psi}'_+ \psi_- + \bar{\psi}'_- \psi_+) \,.
\end{aligned}
\end{equation}
It can be proven that in $D=4$, for any pair of generic Majorana spinors $\Psi_1$ and $\Psi_2$, we have
\begin{equation}
(\bar{\Psi}_1 \gamma^{\hat{\mu}_1} \ldots \gamma^{\hat{\mu}_n} \Psi_2)^* = -\bar{\Psi}_1 \gamma^{\hat{\mu}_1} \ldots \gamma^{\hat{\mu}_n} \Psi_2  \qquad\qquad  (n \ge 0) \,,
\end{equation}
namely any bilinear constructed with Majorana spinors is pure imaginary. It follows that the bilinears inside the parentheses in all  the parameters  are pure imaginary. This is consistent with the $\mc{N}=1$ AdS$_4$ superalgebra, which is obtained by formally setting $2m(\xi')^*\xi=1$ in $a^m$, $b^m$, $\Lambda^m{}_n$ and $\lambda$ and  $\mathtt{r}=0$. However, because we have $\mc{N}=2$, the parameters are complexified by $\xi$ and $\xi'$.
The Killing vector~$\mathtt{K}$ can then be expressed as
\begin{equation}
\mathtt{K} = \mathtt{K}_{QQ} + \mathtt{K}_{QS} + \mathtt{K}_{SQ} + \mathtt{K}_{SS} \,,
\end{equation}
with
\begin{equation}
\begin{aligned}
\mathtt{K}_{QQ} &= a^m \partial_m \equiv P \,, \\
\mathtt{K}_{SS} &= (x \cdot x + u^{-2}) \, b^m \partial_m - 2 (x \cdot b) (x^m \partial_m - u \, \partial_u) \equiv K \,, \\
\mathtt{K}_{QS} + \mathtt{K}_{SQ} &= \lambda (x^m \partial_m - u \, \partial_u) + \Lambda^m_{\ \ n} \, x^n \partial_m + \mathtt{r} \, \partial_z \equiv D + M + R \,,
\end{aligned}
\end{equation}
which  have the formal structure of the anticommutators of the  $OSp(2|4)$ superalgebra. Specifically, the left hand side of the equations above may be identified with the anticommutators $\{Q,Q\}$, $\{S,S\}$, and $\{Q,S\}$, respectively. On the right hand side we find the bosonic generators of the $SO(3,2)$ isometry of $\AdS_4$, namely translations~$P$, Lorentz transformations~$M$, dilatations~$D$, and special conformal transformations~$K$. Moreover, we find that $R\equiv\mathtt{r}\,\partial_z$ is the generator of the $R$-symmetry. The superalgebra may be completed by computing the spinorial Lie derivatives ${\cal L}_V \epsilon_+$, ${\cal L}_V\epsilon_-$ along the above Killing vectors.


\bibliographystyle{utphys}
\bibliography{biblio}

\providecommand{\href}[2]{#2}\begingroup\raggedright\begin{thebibliography}{10}

\bibitem{Karndumri:2015eta}
P.~Karndumri, ``{Twisted compactification of N = 2 5D SCFTs to three and two
  dimensions from F(4) gauged supergravity}'', {\em JHEP} {\bfseries 09} (2015)
  034, \href{http://arxiv.org/abs/1507.01515}{{\ttfamily arXiv:1507.01515
  [hep-th]}}.

\bibitem{Bah:2018lyv}
I.~Bah, A.~Passias, and P.~Weck, ``{Holographic duals of five-dimensional SCFTs
  on a Riemann surface}'', {\em JHEP} {\bfseries 01} (2019) 058,
  \href{http://arxiv.org/abs/1807.06031}{{\ttfamily arXiv:1807.06031
  [hep-th]}}.

\bibitem{Maldacena:2000mw}
J.~M. Maldacena and C.~Nunez, ``{Supergravity description of field theories on
  curved manifolds and a no go theorem}'', {\em Int. J. Mod. Phys. A}
  {\bfseries 16} (2001) 822--855,
  \href{http://arxiv.org/abs/hep-th/0007018}{{\ttfamily arXiv:hep-th/0007018}}.

\bibitem{Cacciatori:2009iz}
S.~L. Cacciatori and D.~Klemm, ``{Supersymmetric AdS(4) black holes and
  attractors}'', {\em JHEP} {\bfseries 01} (2010) 085,
  \href{http://arxiv.org/abs/0911.4926}{{\ttfamily arXiv:0911.4926 [hep-th]}}.

\bibitem{Benini:2013cda}
F.~Benini and N.~Bobev, ``{Two-dimensional SCFTs from wrapped branes and
  c-extremization}'', {\em JHEP} {\bfseries 06} (2013) 005,
  \href{http://arxiv.org/abs/1302.4451}{{\ttfamily arXiv:1302.4451 [hep-th]}}.

\bibitem{Bah:2012dg}
I.~Bah, C.~Beem, N.~Bobev, and B.~Wecht, ``{Four-Dimensional SCFTs from
  M5-Branes}'', {\em JHEP} {\bfseries 06} (2012) 005,
  \href{http://arxiv.org/abs/1203.0303}{{\ttfamily arXiv:1203.0303 [hep-th]}}.

\bibitem{Ferrero:2020twa}
P.~Ferrero, J.~P. Gauntlett, J.~M.~P. Ipi\~na, D.~Martelli, and J.~Sparks,
  ``{Accelerating black holes and spinning spindles}'', {\em Phys. Rev. D}
  {\bfseries 104} no.~4, (2021) 046007,
  \href{http://arxiv.org/abs/2012.08530}{{\ttfamily arXiv:2012.08530
  [hep-th]}}.

\bibitem{Ferrero:2020laf}
P.~Ferrero, J.~P. Gauntlett, J.~M. P\'erez Ipi\~na, D.~Martelli, and J.~Sparks,
  ``{D3-Branes Wrapped on a Spindle}'', {\em Phys. Rev. Lett.} {\bfseries 126}
  no.~11, (2021) 111601, \href{http://arxiv.org/abs/2011.10579}{{\ttfamily
  arXiv:2011.10579 [hep-th]}}.

\bibitem{Ferrero:2021wvk}
P.~Ferrero, J.~P. Gauntlett, D.~Martelli, and J.~Sparks, ``{M5-branes wrapped
  on a spindle}'', {\em JHEP} {\bfseries 11} (2021) 002,
  \href{http://arxiv.org/abs/2105.13344}{{\ttfamily arXiv:2105.13344
  [hep-th]}}.

\bibitem{Hosseini:2021fge}
S.~M. Hosseini, K.~Hristov, and A.~Zaffaroni, ``{Rotating multi-charge spindles
  and their microstates}'', {\em JHEP} {\bfseries 07} (2021) 182,
  \href{http://arxiv.org/abs/2104.11249}{{\ttfamily arXiv:2104.11249
  [hep-th]}}.

\bibitem{Boido:2021szx}
A.~Boido, J.~M.~P. Ipi\~na, and J.~Sparks, ``{Twisted D3-brane and M5-brane
  compactifications from multi-charge spindles}'', {\em JHEP} {\bfseries 07}
  (2021) 222, \href{http://arxiv.org/abs/2104.13287}{{\ttfamily
  arXiv:2104.13287 [hep-th]}}.

\bibitem{Ferrero:2021ovq}
P.~Ferrero, M.~Inglese, D.~Martelli, and J.~Sparks, ``{Multi-charge
  accelerating black holes and spinning spindles}'',
  \href{http://arxiv.org/abs/2109.14625}{{\ttfamily arXiv:2109.14625
  [hep-th]}}.

\bibitem{Couzens:2021rlk}
C.~Couzens, K.~Stemerdink, and D.~van~de Heisteeg, ``{M2-branes on Discs and
  Multi-Charged Spindles}'', \href{http://arxiv.org/abs/2110.00571}{{\ttfamily
  arXiv:2110.00571 [hep-th]}}.

\bibitem{Bah:2021hei}
I.~Bah, F.~Bonetti, R.~Minasian, and E.~Nardoni, ``{M5-brane sources,
  holography, and Argyres-Douglas theories}'', {\em JHEP} {\bfseries 11} (2021)
  140, \href{http://arxiv.org/abs/2106.01322}{{\ttfamily arXiv:2106.01322
  [hep-th]}}.

\bibitem{Couzens:2021tnv}
C.~Couzens, N.~T. Macpherson, and A.~Passias, ``{${\cal N}=(2,2)$ AdS$_3$ from
  D3-branes wrapped on Riemann surfaces}'',
  \href{http://arxiv.org/abs/2107.13562}{{\ttfamily arXiv:2107.13562
  [hep-th]}}.

\bibitem{Suh:2021ifj}
M.~Suh, ``{D3-branes and M5-branes wrapped on a topological disc}'',
  \href{http://arxiv.org/abs/2108.01105}{{\ttfamily arXiv:2108.01105
  [hep-th]}}.

\bibitem{Suh:2021aik}
M.~Suh, ``{D4-D8-branes wrapped on a manifold with non-constant curvature}'',
  \href{http://arxiv.org/abs/2108.08326}{{\ttfamily arXiv:2108.08326
  [hep-th]}}.

\bibitem{Suh:2021hef}
M.~Suh, ``{M2-branes wrapped on a topological disc}'',
  \href{http://arxiv.org/abs/2109.13278}{{\ttfamily arXiv:2109.13278
  [hep-th]}}.

\bibitem{Cassani:2021dwa}
D.~Cassani, J.~P. Gauntlett, D.~Martelli, and J.~Sparks, ``{Thermodynamics of
  accelerating and supersymmetric AdS4 black holes}'', {\em Phys. Rev. D}
  {\bfseries 104} no.~8, (2021) 086005,
  \href{http://arxiv.org/abs/2106.05571}{{\ttfamily arXiv:2106.05571
  [hep-th]}}.

\bibitem{Cabo-Bizet:2018ehj}
A.~Cabo-Bizet, D.~Cassani, D.~Martelli, and S.~Murthy, ``{Microscopic origin of
  the Bekenstein-Hawking entropy of supersymmetric AdS$_{5}$ black holes}'',
  {\em JHEP} {\bfseries 10} (2019) 062,
  \href{http://arxiv.org/abs/1810.11442}{{\ttfamily arXiv:1810.11442
  [hep-th]}}.

\bibitem{Hosseini:2019iad}
S.~M. Hosseini, K.~Hristov, and A.~Zaffaroni, ``{Gluing gravitational blocks
  for AdS black holes}'', {\em JHEP} {\bfseries 12} (2019) 168,
  \href{http://arxiv.org/abs/1909.10550}{{\ttfamily arXiv:1909.10550
  [hep-th]}}.

\bibitem{Benini:2015eyy}
F.~Benini, K.~Hristov, and A.~Zaffaroni, ``{Black hole microstates in AdS$_{4}$
  from supersymmetric localization}'', {\em JHEP} {\bfseries 05} (2016) 054,
  \href{http://arxiv.org/abs/1511.04085}{{\ttfamily arXiv:1511.04085
  [hep-th]}}.

\bibitem{Nian:2019pxj}
J.~Nian and L.~A. Pando~Zayas, ``{Microscopic entropy of rotating electrically
  charged AdS$_{4}$ black holes from field theory localization}'', {\em JHEP}
  {\bfseries 03} (2020) 081, \href{http://arxiv.org/abs/1909.07943}{{\ttfamily
  arXiv:1909.07943 [hep-th]}}.

\bibitem{Hristov:2019mqp}
K.~Hristov, S.~Katmadas, and C.~Toldo, ``{Matter-coupled supersymmetric
  Kerr-Newman-AdS$_4$ black holes}'', {\em Phys. Rev. D} {\bfseries 100} no.~6,
  (2019) 066016, \href{http://arxiv.org/abs/1907.05192}{{\ttfamily
  arXiv:1907.05192 [hep-th]}}.

\bibitem{BenettiGenolini:2019jdz}
P.~Benetti~Genolini, J.~M. Perez Ipi\~na, and J.~Sparks, ``{Localization of the
  action in AdS/CFT}'', {\em JHEP} {\bfseries 10} (2019) 252,
  \href{http://arxiv.org/abs/1906.11249}{{\ttfamily arXiv:1906.11249
  [hep-th]}}.

\bibitem{Ferrero:2021etw}
P.~Ferrero, J.~P. Gauntlett, and J.~Sparks, ``{Supersymmetric spindles}'',
  \href{http://arxiv.org/abs/2112.01543}{{\ttfamily arXiv:2112.01543
  [hep-th]}}.

\bibitem{Brandhuber:1999np}
A.~Brandhuber and Y.~Oz, ``{The D-4 - D-8 brane system and five-dimensional
  fixed points}'', {\em Phys. Lett. B} {\bfseries 460} (1999) 307--312,
  \href{http://arxiv.org/abs/hep-th/9905148}{{\ttfamily arXiv:hep-th/9905148}}.

\bibitem{Cvetic:1999xx}
M.~Cvetic, S.~S. Gubser, H.~Lu, and C.~N. Pope, ``{Symmetric potentials of
  gauged supergravities in diverse dimensions and Coulomb branch of gauge
  theories}'', {\em Phys. Rev. D} {\bfseries 62} (2000) 086003,
  \href{http://arxiv.org/abs/hep-th/9909121}{{\ttfamily arXiv:hep-th/9909121}}.

\bibitem{Romans:1985tw}
L.~J. Romans, ``{The F(4) Gauged Supergravity in Six-dimensions}'', {\em Nucl.
  Phys. B} {\bfseries 269} (1986) 691.

\bibitem{DAuria:2000afl}
R.~D'Auria, S.~Ferrara, and S.~Vaula, ``{Matter coupled F(4) supergravity and
  the AdS(6) / CFT(5) correspondence}'', {\em JHEP} {\bfseries 10} (2000) 013,
  \href{http://arxiv.org/abs/hep-th/0006107}{{\ttfamily arXiv:hep-th/0006107}}.

\bibitem{Hosseini:2018usu}
S.~M. Hosseini, K.~Hristov, A.~Passias, and A.~Zaffaroni, ``{6D attractors and
  black hole microstates}'', {\em JHEP} {\bfseries 12} (2018) 001,
  \href{http://arxiv.org/abs/1809.10685}{{\ttfamily arXiv:1809.10685
  [hep-th]}}.

\bibitem{Guarino:2015jca}
A.~Guarino, D.~L. Jafferis, and O.~Varela, ``{String Theory Origin of Dyonic
  N=8 Supergravity and Its Chern-Simons Duals}'', {\em Phys. Rev. Lett.}
  {\bfseries 115} no.~9, (2015) 091601,
  \href{http://arxiv.org/abs/1504.08009}{{\ttfamily arXiv:1504.08009
  [hep-th]}}.

\bibitem{Jafferis:2012iv}
D.~L. Jafferis and S.~S. Pufu, ``{Exact results for five-dimensional
  superconformal field theories with gravity duals}'', {\em JHEP} {\bfseries
  05} (2014) 032, \href{http://arxiv.org/abs/1207.4359}{{\ttfamily
  arXiv:1207.4359 [hep-th]}}.

\bibitem{Passias:2018zlm}
A.~Passias, D.~Prins, and A.~Tomasiello, ``{A massive class of $\mathcal{N} =
  2$ AdS$_4$ IIA solutions}'', {\em JHEP} {\bfseries 10} (2018) 071,
  \href{http://arxiv.org/abs/1805.03661}{{\ttfamily arXiv:1805.03661
  [hep-th]}}.

\bibitem{Crichigno:2020ouj}
P.~M. Crichigno and D.~Jain, ``{The 5d Superconformal Index at Large $N$ and
  Black Holes}'', {\em JHEP} {\bfseries 09} (2020) 124,
  \href{http://arxiv.org/abs/2005.00550}{{\ttfamily arXiv:2005.00550
  [hep-th]}}.

\bibitem{Crichigno:2018adf}
P.~M. Crichigno, D.~Jain, and B.~Willett, ``{5d Partition Functions with A
  Twist}'', {\em JHEP} {\bfseries 11} (2018) 058,
  \href{http://arxiv.org/abs/1808.06744}{{\ttfamily arXiv:1808.06744
  [hep-th]}}.

\bibitem{Lu:2003iv}
H.~Lu, C.~N. Pope, and J.~F. Vazquez-Poritz, ``{From AdS black holes to
  supersymmetric flux branes}'', {\em Nucl. Phys. B} {\bfseries 709} (2005)
  47--68, \href{http://arxiv.org/abs/hep-th/0307001}{{\ttfamily
  arXiv:hep-th/0307001}}.

\bibitem{Cvetic:1999un}
M.~Cvetic, H.~Lu, and C.~N. Pope, ``{Gauged six-dimensional supergravity from
  massive type IIA}'', {\em Phys. Rev. Lett.} {\bfseries 83} (1999) 5226--5229,
  \href{http://arxiv.org/abs/hep-th/9906221}{{\ttfamily arXiv:hep-th/9906221}}.

\bibitem{Seiberg:1996bd}
N.~Seiberg, ``{Five-dimensional SUSY field theories, nontrivial fixed points
  and string dynamics}'', {\em Phys. Lett. B} {\bfseries 388} (1996) 753--760,
  \href{http://arxiv.org/abs/hep-th/9608111}{{\ttfamily arXiv:hep-th/9608111}}.

\bibitem{Passias:2012vp}
A.~Passias, ``{A note on supersymmetric AdS$_6$ solutions of massive type IIA
  supergravity}'', {\em JHEP} {\bfseries 01} (2013) 113,
  \href{http://arxiv.org/abs/1209.3267}{{\ttfamily arXiv:1209.3267 [hep-th]}}.

\bibitem{Jafferis:2010un}
D.~L. Jafferis, ``{The Exact Superconformal R-Symmetry Extremizes Z}'', {\em
  JHEP} {\bfseries 05} (2012) 159,
  \href{http://arxiv.org/abs/1012.3210}{{\ttfamily arXiv:1012.3210 [hep-th]}}.

\bibitem{Hosseini:2021mnn}
S.~M. Hosseini, I.~Yaakov, and A.~Zaffaroni, ``{The joy of factorization at
  large $N$: five-dimensional indices and AdS black holes}'',
  \href{http://arxiv.org/abs/2111.03069}{{\ttfamily arXiv:2111.03069
  [hep-th]}}.

\bibitem{Hosseini:2018uzp}
S.~M. Hosseini, I.~Yaakov, and A.~Zaffaroni, ``{Topologically twisted indices
  in five dimensions and holography}'', {\em JHEP} {\bfseries 11} (2018) 119,
  \href{http://arxiv.org/abs/1808.06626}{{\ttfamily arXiv:1808.06626
  [hep-th]}}.

\bibitem{Bobev:2017uzs}
N.~Bobev and P.~M. Crichigno, ``{Universal RG Flows Across Dimensions and
  Holography}'', {\em JHEP} {\bfseries 12} (2017) 065,
  \href{http://arxiv.org/abs/1708.05052}{{\ttfamily arXiv:1708.05052
  [hep-th]}}.

\bibitem{Hosseini:2020mut}
S.~M. Hosseini and A.~Zaffaroni, ``{Universal AdS Black Holes in Theories with
  16 Supercharges and Their Microstates}'', {\em Phys. Rev. Lett.} {\bfseries
  126} no.~17, (2021) 171604, \href{http://arxiv.org/abs/2011.01249}{{\ttfamily
  arXiv:2011.01249 [hep-th]}}.

\bibitem{Hosseini:2020wag}
S.~M. Hosseini and K.~Hristov, ``{4d F(4) gauged supergravity and black holes
  of class $\mathcal{F}$}'', {\em JHEP} {\bfseries 02} (2021) 177,
  \href{http://arxiv.org/abs/2011.01943}{{\ttfamily arXiv:2011.01943
  [hep-th]}}.

\bibitem{Giri:2021xta}
S.~Giri, ``{Black holes with spindles at the horizon}'',
  \href{http://arxiv.org/abs/2112.04431}{{\ttfamily arXiv:2112.04431
  [hep-th]}}.

\bibitem{Alday:2015jsa}
L.~F. Alday, M.~Fluder, C.~M. Gregory, P.~Richmond, and J.~Sparks,
  ``{Supersymmetric solutions to Euclidean Romans supergravity}'', {\em JHEP}
  {\bfseries 02} (2016) 100, \href{http://arxiv.org/abs/1505.04641}{{\ttfamily
  arXiv:1505.04641 [hep-th]}}.

\bibitem{Genolini:2021urf}
P.~B. Genolini and P.~Richmond, ``{Supersymmetry of higher-derivative
  supergravity in AdS4 holography}'', {\em Phys. Rev. D} {\bfseries 104} no.~6,
  (2021) L061902, \href{http://arxiv.org/abs/2107.04590}{{\ttfamily
  arXiv:2107.04590 [hep-th]}}.

\bibitem{Couzens:2018wnk}
C.~Couzens, J.~P. Gauntlett, D.~Martelli, and J.~Sparks, ``{A geometric dual of
  $c$-extremization}'', {\em JHEP} {\bfseries 01} (2019) 212,
  \href{http://arxiv.org/abs/1810.11026}{{\ttfamily arXiv:1810.11026
  [hep-th]}}.

\bibitem{Hosseini:2019ddy}
S.~M. Hosseini and A.~Zaffaroni, ``{Geometry of $\mathcal{I}$-extremization and
  black holes microstates}'', {\em JHEP} {\bfseries 07} (2019) 174,
  \href{http://arxiv.org/abs/1904.04269}{{\ttfamily arXiv:1904.04269
  [hep-th]}}.

\bibitem{Gauntlett:2019roi}
J.~P. Gauntlett, D.~Martelli, and J.~Sparks, ``{Toric geometry and the dual of
  ${\cal I}$-extremization}'', {\em JHEP} {\bfseries 06} (2019) 140,
  \href{http://arxiv.org/abs/1904.04282}{{\ttfamily arXiv:1904.04282
  [hep-th]}}.

\bibitem{Gauntlett:2019pqg}
J.~P. Gauntlett, D.~Martelli, and J.~Sparks, ``{Fibred GK geometry and
  supersymmetric $AdS$ solutions}'', {\em JHEP} {\bfseries 11} (2019) 176,
  \href{http://arxiv.org/abs/1910.08078}{{\ttfamily arXiv:1910.08078
  [hep-th]}}.

\bibitem{Suh:2018szn}
M.~Suh, ``{Supersymmetric $AdS_6$ black holes from matter coupled $F(4)$ gauged
  supergravity}'', {\em JHEP} {\bfseries 02} (2019) 108,
  \href{http://arxiv.org/abs/1810.00675}{{\ttfamily arXiv:1810.00675
  [hep-th]}}.

\bibitem{Lu:1998nu}
H.~Lu, C.~N. Pope, and J.~Rahmfeld, ``{A Construction of Killing spinors on
  S**n}'', {\em J. Math. Phys.} {\bfseries 40} (1999) 4518--4526,
  \href{http://arxiv.org/abs/hep-th/9805151}{{\ttfamily arXiv:hep-th/9805151}}.

\end{thebibliography}\endgroup

\end{document}